\documentclass[12pt]{emulateapj}

\usepackage{amsmath,amssymb}  % Better maths support & more symbols
\usepackage{graphicx}
\usepackage{color}

\newcommand{\m}{$\rm{M}_{\rm 200b}$ }

\newcommand{\ie}{{\it i.e.}}

\newcommand{\msun}{\mbox{$M_{\odot}$}}

\newcommand{\sm}{$~{\rm M}_{\odot}$~}
\newcommand{\ncen}{\langle N_{\rm cen} \rangle}
\newcommand{\nsat}{\langle N_{\rm sat} \rangle}

\newcommand{\sigsm}{$\sigma_{\rm log M_*}$}
\newcommand{\sigmh}{$\sigma_{\rm log M_h}$}
\newcommand{\asat}{\alpha_{\rm sat}}

\def\fshmr{f^{-1}_{\textsc{shmr}}(M_\ast^{t_1})} 
\def\multidrizzle{{\tt MultiDrizzle}}

\slugcomment{ }
 
\shorttitle{The stellar-to-dark matter connection in COSMOS}
\shortauthors{A.\ Leauthaud et al.}

\begin{document}
  
%-------- TITLE  ---------------------

 \title{New constraints on the evolution of the stellar-to-dark matter
   connection: a combined analysis of galaxy-galaxy lensing,
   clustering, and stellar mass functions from z=0.2 to z=1}


%-------- AUTHORS  ---------------------


\author{Alexie Leauthaud\altaffilmark{1,2},
Jeremy Tinker\altaffilmark{3},
Kevin Bundy\altaffilmark{4},
Peter S. Behroozi\altaffilmark{5},
Richard Massey\altaffilmark{6},
Jason Rhodes\altaffilmark{7,8},
Matthew R. George\altaffilmark{4},
Jean-Paul Kneib\altaffilmark{9},
Andrew Benson\altaffilmark{7},
Risa H. Wechsler\altaffilmark{5},
Michael T. Busha\altaffilmark{5,10},
Peter Capak\altaffilmark{11},
Marina Cort\^{e}s\altaffilmark{1},
Olivier Ilbert\altaffilmark{9},
Anton M. Koekemoer\altaffilmark{12},
Oliver Le F\`{e}vre\altaffilmark{9},
Simon Lilly\altaffilmark{13},
Henry J. McCracken\altaffilmark{14},
Mara Salvato\altaffilmark{15},  
Tim Schrabback\altaffilmark{5,16},
Nick Scoville\altaffilmark{7},
Tristan Smith\altaffilmark{2},
James E. Taylor\altaffilmark{17}}

\submitted{Submitted to ApJ}
\email{asleauthaud@lbl.gov}

\altaffiltext{1}{Lawrence Berkeley National Lab, 1 Cyclotron Road,
  Berkeley, CA 94720, USA}

\altaffiltext{2}{Berkeley Center for Cosmological Physics, University
  of California, Berkeley, CA 94720, USA}

\altaffiltext{3}{Center for Cosmology and Particle Physics, Department of Physics, New York University}

\altaffiltext{4}{Department of Astronomy, University of California,
  Berkeley, CA 94720, USA}

\altaffiltext{5}{Kavli Institute for Particle Astrophysics and
  Cosmology; Physics Department, Stanford University, and SLAC
  National Accelerator Laboratory, Stanford CA 94305}

\altaffiltext{6}{Institute for Astronomy, Blackford Hill, Edinburgh
  EH9 3HJ UK}

\altaffiltext{7}{California Institute of Technology, MC 350-17, 1200
  East California Boulevard, Pasadena, CA 91125, USA}

\altaffiltext{8}{Jet Propulsion Laboratory, California Institute of Technology, Pasadena, CA 91109}

\altaffiltext{9}{LAM, CNRS-UNiv Aix-Marseille, 38 rue F. Joliot-Curis, 13013 Marseille, France}

\altaffiltext{10}{Institute for Theoretical Physics, Department of Physics, University of Zurich, CH-8057, Switzerland}

\altaffiltext{11}{Spitzer Science Center, 314-6 Caltech, 1201
  E. California Blvd. Pasadena, CA, 91125, USA}

\altaffiltext{12}{Space Telescope Science Institute, 3700 San Martin
  Drive, Baltimore, MD 21218, USA}

\altaffiltext{13}{Institute of Astronomy, Department of Physics, ETH Zurich, CH-8093, Switzerland}

\altaffiltext{14}{Institut d'Astrophysique de Paris, UMR 7095, 98 bis
  Boulevard Arago, 75014 Paris, France}

\altaffiltext{15}{SUPA, Institute for Astronomy, The University of Edinburgh, Royal Observatory, Edinburgh - EH9 3HJ, UK}

\altaffiltext{16}{Leiden Observatory, Leiden University, Niels Bohrweg 2, NL-2333 CA Leiden, The Netherlands}

\altaffiltext{17}{Department of Physics and Astronomy, University of
  Waterloo, 200 University Avenue West, Waterloo, Ontario, Canada N2L
  3G1}

%-------- ABSTRACT  ---------------------
  
\begin{abstract} Using data from the COSMOS survey, we perform the
  first joint analysis of galaxy-galaxy weak lensing, galaxy spatial
  clustering, and galaxy number densities. Carefully accounting for
  sample variance and for scatter between stellar and halo mass, we
  model all three observables simultaneously using a novel and
  self-consistent theoretical framework. Our results provide strong
  constraints on the shape and redshift evolution of the
  stellar-to-halo mass relation (SHMR) from $z=0.2$ to $z=1$. At low
  stellar mass, we find that halo mass scales as $M_h \propto
  M_*^{0.46}$ and that this scaling does not evolve significantly with
  redshift from $z=0.2$ to $z=1$. The slope of the SHMR rises sharply
  at $M_* > 5\times 10^{10}~{\rm M}_{\odot}$ and as a consequence, the
  stellar mass of a central galaxy becomes a poor tracer of its parent
  halo mass. We show that the dark-to-stellar ratio, $M_h/M_*$, varies
  from low to high masses, reaching a minimum of $M_h/M_*\sim 27$ at
  $M_*=4.5\times 10^{10}~{\rm M}_{\odot}$ and $M_h=1.2\times
  10^{12}~{\rm M}_{\odot}$. This minimum is important for models of
  galaxy formation because it marks the mass at which the accumulated
  stellar growth of the central galaxy has been the most efficient. We
  describe the SHMR at this minimum in terms of the ``pivot stellar
  mass'', $M_{*}^{\rm piv}$, the ``pivot halo mass'', $M_{h}^{\rm
    piv}$, and the ``pivot ratio'', $(M_h/M_*)^{\rm piv}$.  Thanks to
  a homogeneous analysis of a single data set spanning a large
  redshift range, we report the first detection of mass downsizing
  trends for both $M_{h}^{\rm piv}$ and $M_{*}^{\rm piv}$. The pivot
  stellar mass decreases from $M_{*}^{\rm piv}=5.75 \pm 0.13 \times
  10^{10}$\sm at $z=0.88$ to $M_{*}^{\rm piv}=3.55 \pm 0.17 \times
  10^{10}$\sm at $z=0.37$. Intriguingly, however, the corresponding
  evolution of $M_{h}^{\rm piv}$ leaves the pivot ratio constant with
  redshift at $(M_h/M_*)^{\rm piv}\sim 27$. We use simple arguments to
  show how this result raises the possibility that star formation
  quenching may ultimately depend on $M_h/M_*$ and not simply $M_h$,
  as is commonly assumed.  We show that simple models with such a
  dependence naturally lead to downsizing in the sites of star
  formation. Finally, we discuss the implications of our results in
  the context of popular quenching models, including disk
  instabilities and AGN feedback.
\end{abstract}
 
%-------- KEY WORDS  ---------------------

 
 

%-------- OBSERVATIONS  ---------------------

\altaffiltext{$\star$}{Based on observations with the NASA/ESA {\em
    Hubble Space Telescope}, obtained at the Space Telescope Science
  Institute, which is operated by AURA Inc, under NASA contract NAS
  5-26555; also based on data collected at : the Subaru Telescope,
  which is operated by the National Astronomical Observatory of Japan;
  the XMM-Newton, an ESA science mission with instruments and
  contributions directly funded by ESA Member States and NASA; the
  European Southern Observatory under Large Program 175.A-0839, Chile;
  Kitt Peak National Observatory, Cerro Tololo Inter-American
  Observatory, and the National Optical Astronomy Observatory, which
  are operated by the Association of Universities for Research in
  Astronomy, Inc.  (AURA) under cooperative agreement with the
  National Science Foundation; the National Radio Astronomy
  Observatory which is a facility of the National Science Foundation
  operated under cooperative agreement by Associated Universities, Inc
  and the Canada-France-Hawaii Telescope with MegaPrime/MegaCam
  operated as a joint project by the CFHT Corporation, CEA/DAPNIA, the
  National Research Council of Canada, the Canadian Astronomy Data
  Centre, the Centre National de la Recherche Scientifique de France,
  TERAPIX and the University of Hawaii.}

%%%%%%%%%%%%%%%%%%%%%%%%%%%%%%%%%%%%%%%%%%%%%%%%%%%%%%%%%%%%%%%%%%%%%%%%%%%%%%
%     INTRODUCTION
%%%%%%%%%%%%%%%%%%%%%%%%%%%%%%%%%%%%%%%%%%%%%%%%%%%%%%%%%%%%%%%%%%%%%%%%%%%%%%

\section{Introduction}

A fundamental goal in observational cosmology is to understand the
link between the luminous properties of galaxies and the dark matter
halos in which they reside. From an astrophysical perspective,
measurements of the relationship between dark matter halo mass ($M_h$)
and galaxy observables such as luminosity or stellar mass ($M_*$) are
critical for understanding how galaxy properties and their evolution
with time are shaped by the halos that host them.  Growing evidence
suggests that halos accumulate stellar mass with an efficiency
$\eta(M_h,z) \equiv (M_{*} / M_h)\times (\Omega_M / \Omega_b)$ that
depends strongly on halo mass, peaking at $M_h\sim 10^{12}M_{\sun}$
and declining towards lower and higher masses at $z\sim 0$
\citep[e.g.,][]{Mandelbaum:2006c,Conroy:2009,Moster:2010,
  Behroozi:2010,Guo:2010, More:2010}.  The stellar mass content is
determined not only by the past merging of smaller sub-components but
processes that, integrated over time, regulate the conversion of gas
into stars, including the rate at which fresh material is supplied to
the halo, feedback mechanisms from supernovae (SN), galactic winds,
and active galactic nuclei (AGN), and environmental effects such as
ram pressure stripping, just to name a few.  The global relationship
between halo mass and average stellar content---the stellar-to-halo
mass relation (SHMR)---probes the integrated outcome of these
processes and as such, provides clues to their physical nature and
constrains both semi-analytic models
\citep[e.g.,][]{Bower:2006,Croton:2006,Somerville:2008} and
hydrodynamical simulations
\citep[][]{Keres:2005,Keres:2009,Crain:2009,Brooks:2009,Gabor:2010,Agertz:2011}
that aim to disentangle the relative contributions of such mechanisms.

From a cosmological perspective, the SHMR is vital for determining how
galaxies trace dark matter. A complete picture of the manner in which
galaxies populate dark matter halos enhances the reconstruction of the
dark matter power spectrum from redshift surveys
\citep[][]{Sanchez:2008,Yoo:2009} and leads to improved constraints on
cosmological parameters \citep[][]{Yoo:2006,Zheng:2007a,Cacciato:2009}.

There are currently only two observational techniques capable of {\em
  directly} probing the dark matter halos of galaxies out to large
radii (above 50 kpc): galaxy-galaxy lensing \citep[e.g.,][]{
  Brainerd:1996,McKay:2001, Hoekstra:2004, Sheldon:2004,
  Mandelbaum:2006, Mandelbaum:2006c,
  Heymans:2006,Johnston:2007,Sheldon:2009,Leauthaud:2010} and the
kinematics of satellite galaxies \citep[][]{McKay:2002, Prada:2003,
  Brainerd:2003, van-den-Bosch:2004, Conroy:2007, Becker:2007,
  Norberg:2008, More:2009, More:2010}. The galaxy-galaxy lensing
technique (hereafter ``g-g lensing'') uses weak gravitational lensing
to probe the gravitational potential around foreground (``lens'')
galaxies.  The kinematic method uses satellite galaxies as test
particles that trace the local velocity field (and hence the local
gravitational potential).

Another popular, albeit more indirect method, to infer the galaxy-dark
matter connection is to measure the statistics of galaxy
clustering. The results are commonly interpreted using the halo
occupation distribution (HOD) model which describes the probability
distribution $P(N|M_h)$ that a halo of mass $M_h$ is host to N
galaxies above some threshold in luminosity or stellar-mass
\citep[e.g.,][]{Seljak:2000, Peacock:2000, Scoccimarro:2001,
  Berlind:2002,Bullock:2002, Zehavi:2002, Zehavi:2005, Zheng:2005,
  Zheng:2007,
  Tinker:2007,Wake:2011,Zehavi:2010,White:2011}. Variations on the HOD
approach include the conditional luminosity function $\Phi (L|M_h)dL$
which specifies the average number of galaxies of luminosity $L\pm
dL/2$ that reside in a halo of mass $M_h $ \citep[e.g.,][]{Yang:2003,
  van-den-Bosch:2003, Vale:2004, Cooray:2006, van-den-Bosch:2007,
  Vale:2008} and the conditional stellar mass function
$\Phi(M_{*}|M_h)dM_*$ which yields the average number of galaxies with
stellar masses in the range $M_{*}\pm dM_{*}/2$ as a function of host
halo mass $M_h$ \citep[e.g.,][]{Yang:2009,Moster:2010,Behroozi:2010}.

Finally, constraints on the SHMR have also been derived from the
so-called ``abundance matching'' technique, which assumes that there
is a monotonic correspondence between halo mass (or circular velocity)
and galaxy stellar mass (or luminosity)
\citep[e.g.,][]{Kravtsov:2004,Vale:2004,Tasitsiomi:2004,Vale:2006a,Conroy:2009,Drory:2009,Moster:2010,Behroozi:2010,Guo:2010}.

While individual applications of the techniques described above have
provided important insight, in \citet{Leauthaud:2011} (hereafter
``Paper I'') we take a further step by {\em combining} separate probes
into a self-consistent theoretical framework.  Specifically, our
method combines measurements of g-g lensing, galaxy clustering, and
the galaxy stellar mass function.  Beginning with the standard HOD
formalism, we make several modifications that enable us to a) extract
the parameters that determine the SHMR and to b) simultaneously fit
data from multiple probes while allowing for independent binning
schemes for each probe. The goal of the current paper is to apply the
methodology developed in Paper I to observations from the COSMOS
survey from $z=0.2$ to $z=1.0$.  This enables confident measurements
of the shape of the SHMR and its evolution with time, with important
implications for models of galaxy formation.

While stellar mass estimates are a key galaxy observable in this work,
it is important to highlight the uncertainties and sometimes unknown
systematic biases that affect them and, if not treated carefully, can
muddle attempts to compare results from disparate surveys (see
discussion in \citealt[][]{Behroozi:2010}).  One of the advantages of
the COSMOS data set is that evolutionary trends can be studied within
the sample using self-consistent stellar mass estimates, making our
conclusions more robust.

The layout of this paper is as follows. The data are described in
$\S$\ref{cosmos_survey} followed by the presentation of the g-g
lensing, clustering, and SMF measurements in
$\S$\ref{measurements}. In the interest of brevity, we only give a
short and necessarily incomplete review of the theoretical background
in $\S$\ref{theory}. We strongly encourage the reader to refer to
Paper I for a complete description of the theoretical foundations for
this work. Our main results are presented in
$\S$\ref{results}. Finally, we discuss the results and draw up our
conclusions in $\S$\ref{discussion} and $\S$\ref{conclusions}.

We assume a WMAP5 $\Lambda$CDM cosmology with $\Omega_{\rm m}=0.258$,
$\Omega_\Lambda=0.742$, $\Omega_{\rm b}h^2=0.02273$, $n_{\rm
  s}=0.963$, $\sigma_{8}=0.796$, $H_0=72$ km~s$^{-1}$~Mpc$^{-1}$
\citep[][]{Hinshaw:2009}. All distances are expressed in physical Mpc
units. The letter $M_h$ denotes halo mass in general whereas \m is
explicitly defined as $M_{200b}\equiv M(<r_{200b})=200\bar{\rho}
\frac{4}{3}\pi r_{200b}^3$ where $r_{200b}$ is the radius at which the
mean interior density is equal to 200 times the mean matter density
($\bar{\rho}$). Stellar mass is noted $M_{*}$ and has been derived
using a Chabrier Initial Mass Function (IMF). Stellar mass scales as
$1/H_0^2$. Halo mass scales as $1/H_0$. All magnitudes are given on
the AB system.

\section{Description of Data}\label{cosmos_survey}

The COSMOS survey \citep[][]{Scoville:2007} brings together a broad
array of panchromatic observations with imaging data from X-ray to
radio wavelengths and a large spectroscopic follow-up program
(zCOSMOS) with the VLT \citep{Lilly:2007}. In particular, the COSMOS
program has imaged the largest contiguous area (1.64 degrees$^2$) to
date with the {\it Hubble Space Telescope (HST)} using the {\it
  Advanced Camera for Surveys (ACS)} {\it Wide Field Channel (WFC)}
\citep[][]{Koekemoer:2007}.

\subsection{The ACS Lensing catalog}

The general methodology for the construction of the COSMOS ACS weak
lensing catalog and our shape measurement procedure are presented in
\citet{Leauthaud:2007} and \citet{Rhodes:2007}. In this section, we
present several updates to the pipeline that we have implemented since
those publications. In \citet{Leauthaud:2007} we used a parametric
correction for the effects of Charge Transfer Inefficiency (CTI) on
galaxy shape measurements. Instead, in this paper, we use a physically
motivated CTI correction scheme that operates on the raw data and
returns electrons to pixels from which they were unintentionally
dragged out during readout. This correction scheme has been shown by
\citet[][]{Massey:2010} to reduce the CTI trails by a factor of
$\sim$10 everywhere in the CCD and at all flux levels.

Following CTI correction in the raw images, image registration,
geometric distortion, sky subtraction, cosmic ray rejection and the
final combination of the dithered images are performed by the
\multidrizzle\ algorithm \citep{Koekemoer:2002}.  As described in
\citet{Rhodes:2007}, a finer pixel scale of $0.03\arcsec/$pix was used
for the final co-added images. The lensing source catalog is
constructed from 575 ACS/WFC tiles. Defects and diffraction spikes are
carefully removed, leaving a total of $1.2 \times 10^6$ objects to a
limiting magnitude of $\rm{I}_{F814W}=26.5$.

The next step is to measure the shapes of galaxies and to correct them
for the distortion induced by the time varying ACS PSF (Point Spread
Function) \citep[see][]{Rhodes:2007}. We continue to use a PSF model
based on physical parameters rather than arbitrary principal
components. In \citet{Leauthaud:2007}, we modelled the PSF as a
multivariate polynomial in x, y and focus. We now fit the PSF as a
function of x, y, focus and velocity aberration of the pointing
(recorded in CALACS headers as
``VAFACTOR''). \citet[][]{Schrabback:2010} found that VAFACTOR
partially correlates with the higher-order PSF variations, motivating
our use of this quantity. In the g-g lensing analysis presented here,
the weak lensing shear is azimuthally averaged.  Thus, any effects of
PSF anisotropy cancel to leading order. Therefore, our science
analysis is insensitive to subtle differences in the PSF modeling. We
confirmed this by repeating our analysis with the independently
obtained weak lensing catalog by \citet[][]{Schrabback:2010}, which
yields fully consistent results.

Finally, simulated images are used to derive the shear susceptibility
factors that are necessary in order to transform shape measurements
into unbiased shear estimators
\citep[][]{Leauthaud:2007}. Representing a number density of $66$
galaxies per arc-minute$^{2}$, the final COSMOS weak lensing catalog
contains $3.9 \times 10^5$ galaxies with accurate shape measurements.

\subsection{Photometric and Spectroscopic Redshifts}\label{photoz}

We use two updated versions (v1.8 dated from the 13$^{th}$ of July
2010 and v1.7 dated from the 1$^{st}$ of August 2009) of the
photometric redshifts (hereafter photo-$z$'s) presented in
\citet{Ilbert:2009} which have been computed with over 30 bands of
multi-wavelength data. In particular, deep $K_s$, $J$, and $u^{*}$
band data allow for a good photo-$z$ estimate at $z>1$ via the 4000\AA\
break which is increasingly shifted into the near infra-red
(IR). Further details regarding the data and the photometry can be
found in \citet{Capak:2007}.

The photo-$z$ catalog v1.8 has improved redshifts at $z>1$ compared to
v1.7. At $z<1$, the difference between the two catalogs is minor. We
use catalog v1.8 for g-g lensing measurements and v1.7 for the SMF and
galaxy clustering. Interchanging the two catalogs does not affect our
results.

Photo-$z$'s were estimated using a $\chi^2$ template fitting method
(Le Phare) and calibrated with large spectroscopic samples from
VLT-VIMOS \citep[][]{Lilly:2007,Ilbert:2009} and Keck-DEIMOS. The
dispersion in the photo-$z$'s as measured by comparing to the
spectroscopic redshifts is $\sigma_{\Delta z/(1+z_{\rm spec})}=0.007$
at $i^{+}_{AB}<22.5$ where $\Delta z =z_{\rm spec}-z_{\rm phot}$.  The
deep IR and IRAC (Infrared Array Camera on the Spitzer Space
Telescope) data enables the photo-$z$'s to be calculated even at
fainter magnitudes with a reasonable accuracy of $\sigma_{\Delta
  z/(1+z_{\rm spec})}=0.06$ at $i^{+}_{AB} \sim 24$.

\begin{figure}
\epsscale{1.3}
\plotone{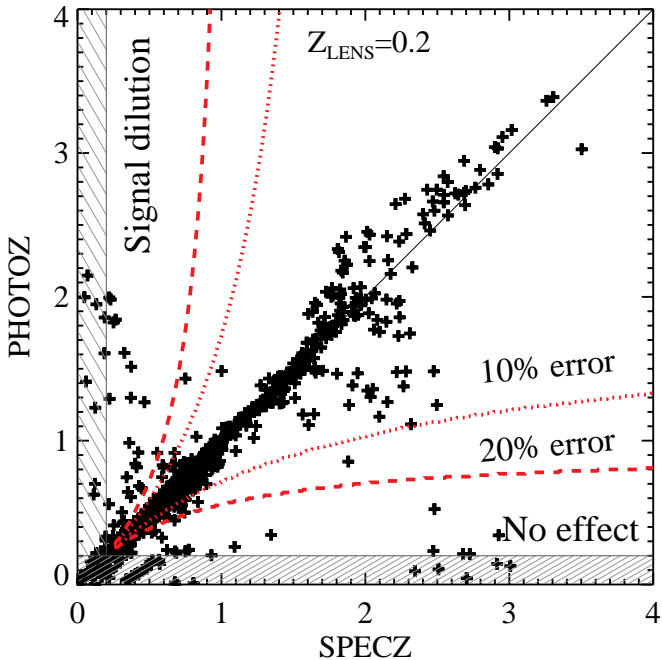}
\caption{The effects of photometric redshifts errors on galaxy-galaxy
  lensing signals. This figure illustrates the quality of the
  photometric redshifts for source galaxies in this paper by comparing
  them to a combined sample of 8812 spectroscopic redshifts from the
  zCOSMOS ``bright'' and ``faint'' programs for which we have applied
  the same selection as for source galaxies. We emphasize that since
  the spectroscopic sample does not go as faint as our source catalog,
  this figure is necessarily an underestimate of the true redshift
  errors for the source catalog -- a complete sample of faint galaxies
  with spectroscopic redshift would be necessary in order to fully
  test the photometric redshifts. There are three ways in which
  photometric redshift errors can impact g-g lensing signals. Firstly,
  any type of photometric error such that $z_{\rm phot}<z_{\rm lens}$
  will have no effect on the signal because such objects are not
  included in the background selection (bottom hashed
  region). Secondly, photometric errors such that $z_{\rm phot}>z_{\rm
    lens}$ and $z_{\rm spec}<z_{\rm lens}$ will lead to a signal
  dilution (left hashed region). Finally, photometric errors such that
  $z_{\rm phot}>z_{\rm lens}$ and $z_{\rm spec}>z_{\rm lens}$ but $z_{\rm
    phot}\neq z_{\rm spec}$ will lead to a bias in $\Delta\Sigma$
  because $\Sigma_{\rm crit}$ will be mis-estimated when transforming
  $\gamma$ into $\Delta\Sigma$. The dotted and dashed lines indicate
  where photo-$z$ errors lead to a $10\%$ and a $20\%$ error on
  $\Delta\Sigma$ for a lens located at z=0.2. As can be seen, g-g
  lensing signals are increasingly insensitive to photometric redshift
  errors for source galaxies at higher redshifts.}
\label{redshifts}
\end{figure}

Figure \ref{redshifts} compares the spectroscopic and photometric
redshifts of 8812 galaxies that belong to both the lensing source
catalog and the zCOSMOS ``bright'' or ``faint'' programs. This Figure
also illustrates the sensitivity of g-g lensing signals to photometric
redshift errors. As can be seen from Figure \ref{redshifts}, g-g
lensing signals are increasingly insensitive to photometric redshift
errors for source galaxies at higher redshifts.

\subsection{Stellar Mass Estimates}\label{stellar_masses}

Stellar masses are estimated using the Bayesian code described in
\citet{Bundy:2006} assuming a Chabrier IMF. Briefly, an observed
galaxy's spectral energy distribution (SED) and redshift is referenced
to a grid of models constructed using the \citet{bruzual:2003}
synthesis code.  The grid includes models that vary in age, star
formation history, dust content, and metallicity. The assumed dust
model is \citet[][]{Charlot:2000}. At each grid point, the probability
that the observed SED fits the model is calculated, and the
corresponding stellar mass to K-band luminosity ratio and stellar mass
is stored.  By marginalizing over all parameters in the grid, the
stellar mass probability distribution is obtained.  The median of this
distribution is taken as the stellar mass estimate, and the width
encodes the uncertainty due to degeneracies and uncertainties in the
model parameter space (also described as ``model error'' in $\S$
\ref{scatter_shmr}).  The final uncertainty on the stellar mass also
includes the K-band photometry uncertainty as well as the expected
error on the luminosity distance that results from the photo-$z$
uncertainty.  The typical final uncertainty is 0.1-0.2 dex (also see
$\S$ \ref{scatter_shmr} and Figure \ref{sm_error_kev}).  Systematic
uncertainties also come from the choice of stellar population
templates used to fit the data as well as the assumption of a
universal IMF.  To first order these will cause global offsets in the
mass estimates that will not affect comparisons {\em within} our
sample but may impact comparisons to work by other authors. Since the
primary goal of this paper is to study the redshift evolution of the
SHMR derived from COSMOS data alone, we do not include these
systematic uncertainties in our analysis and refer to
\citet{Conroy:2009a} and \citet{Behroozi:2010} for a broad discussion
of systematic errors in stellar mass estimates.

Following the approach in \citet{Bundy:2010}, we obtained PSF-matched
3\farcs0 diameter aperture photometry from the ground-based COSMOS
catalogs (filters $u^*, B_J, V_J, g^+, r^+, i^+, z^+, K_s$) described
in \citet{Capak:2007}, \citet{Ilbert:2009}, and
\citet{McCracken:2010}, after applying the photometric zero-point
offsets tabulated in \citet{Capak:2007}.  The depth in all bands
reaches at least 25th magnitude (AB) with the $K_s$-band limited to
$K_s < 24$.  Unlike \citet{Drory:2009}, we require $K_s$-band
detections for all galaxies in the sample.  We have found that the
mass estimates in \citet{Bundy:2010} agree with those of
\citet{Drory:2009} within the expected uncertainties (i.e., $<$ 0.2
dex).  The mass estimates used in this work are slightly different
from those in \citet{Bundy:2010} in that they are based on updated
redshift information (v1.7 of the photo-$z$ catalog and the latest
available spectroscopic redshifts as compiled by the COSMOS team) and
use a slightly different cosmology ($H_0=72$ km s$^{-1}$ Mpc$^{-1}$
instead of $H_0=70$ km s$^{-1}$ Mpc$^{-1}$).

The bins in our analysis are defined by using two estimates of the mass
completeness of the sample, as determined by the magnitude limits $K_s
< 24$ and $I_{\rm 814W} < 25$.  The first estimate is more
conservative and comes from estimating the observed magnitude of a
maximal $M_*/L$ stellar population model with solar metallicity, no
dust, and a $\tau = 0.5$ Gyr burst of star formation that occurred at
$z_{\rm form} = 5$.  As a function of redshift, the stellar mass of
such a population at the point where its observed $K_s$- and $I_{\rm
  814W}$-band flux falls below the magnitude limits defines the mass
completeness and roughly matches the 80\% completeness limits
determined when deeper samples are available \citep{Bundy:2006}.  This
redshift-dependent limit is plotted as the dashed line in Figure
\ref{sm_distribution}.  In practice, low mass galaxies exhibit more
star formation and therefore have lower $M_*/L$ ratios than the
passive template described above so we also define a second, more
liberal mass limit (i.e., lower) that corresponds to a star-forming
population.  This is plotted as the dotted line.  The majority of the
sample bins lie above the more conservative limit, but the lowest mass
bins are allowed to reach the star-forming mass limit under the
assumption that passive stellar populations (potentially missed) at
such low masses are extremely rare.

%%%%%%%%%%%%%%%%%%%%%%%%%%%%%%%%%%%%%%%%%%%%%%%%%%%%%%%%%%%%%%%%%%%%%%%%%%%%%%
%     MEASUREMENTS
%%%%%%%%%%%%%%%%%%%%%%%%%%%%%%%%%%%%%%%%%%%%%%%%%%%%%%%%%%%%%%%%%%%%%%%%%%%%%%

\section{Measurements}\label{measurements}

\subsection{Sample Selection}\label{sample selection}

We use two COSMOS galaxy catalogs: the Subaru (photo-$z$) catalog and the
ACS (lensing) catalog. The ACS catalog corresponds to the COSMOS area
that has been imaged with HST and covers a slightly smaller area than
the Subaru catalog. The ACS coverage of COSMOS is 1.64 degrees$^2$ and
the Subaru coverage of COSMOS is 2.3 degrees$^2$. The ACS catalog is
used for our g-g lensing measurements and for the SMFs. The Subaru
catalog is used to calculate the galaxy clustering.

Galaxies are selected with $K_s<24$ to ensure that stellar masses can
be computed for all galaxies in the sample. Note that for the g-g
lensing analysis, these cuts only apply to the foreground lens
sample. In addition to these cuts, we also reject galaxies that are in
masked areas where the photometry is deemed unreliable. For this, we
use the union of four masks in the B, V, ip, and zp bands
(``COSMOS.B.mask'', ``COSMOS.V.mask'', ``COSMOS.ip.mask'',
``COSMOS.zp.mask'').

For the ACS sample, star-galaxy separation is performed using the
morphological classifier described in \citet[][]{Leauthaud:2007}. For
the Subaru sample, we reject stars by imposing the condition that
$\chi_{\rm gal}^2-\chi_{\rm star}^2>0.5$ where $\chi_{\rm gal}^2$
represents the chi-square value of the best-fitting galaxy SED
template and $\chi_{\rm star}^2$ represents the chi-square value of
the best-fitting stellar SED template. In conjunction with the $K_s$
cut, this stellar selection agrees with our ACS based morphological
classification at the $97\%$ level. According to our ACS morphological
classifier: 2$\%$ of the objects in the final Subaru catalog are
classified as stars and 0.1$\%$ of galaxies are mis-classified as
stars.

\begin{figure*}[htp]
\epsscale{0.8}
\plotone{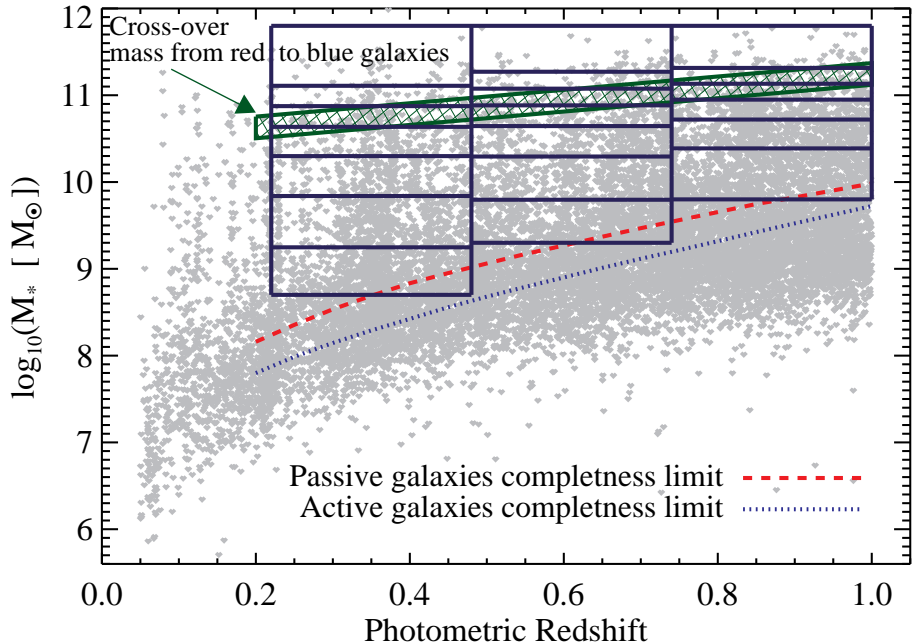}
\caption{Distribution of COSMOS galaxies as a function of redshift and
  stellar mass. The completeness limit for passive galaxies is shown
  by the red dashed line and for active galaxies by the blue dotted
  line (see $\S$ \ref{stellar_masses} for the derivation of the
  completeness limits). The green hashed region approximately shows
  the cross-over mass above which the number of red galaxies becomes
  larger than the number of blue galaxies. The vertical solid dark
  blue lines delineate the three redshift bins $z_1$, $z_2$, and $z_3$
  while the horizontal solid dark blue lines show the binning scheme
  for the g-g lensing measurements.}
\label{sm_distribution}
\end{figure*}

We compute the g-g lensing, galaxy clustering, and stellar mass in
three redshift intervals: $z_1=[0.22,0.48]$, $z_2=[0.48,0.74]$, and
$z_3=[0.74,1]$. For each redshift bin, we define a lower limit on the
stellar mass to ensure that all samples are complete in terms of
stellar mass. These cuts correspond to $M_*>10^{8.7}$\sm for $z_1$,
$M_*>10^{9.3}$\sm for $z_2$, and $M_*>10^{9.8}$\sm for $z_3$ (see
Table \ref{sample_selection}) and are defined by the stellar mass
completeness limit at the far edge of the redshift bin (see Figure
\ref{sm_distribution}). All galaxy samples used in this paper are
complete.

% **************** TABLE 1 ***************
% Number of Galaxies
\begin{deluxetable*}{cccccccc}
%  \tabletypesize{\scriptsize} 
\tablecolumns{5} \tablecaption{Characteristics of three redshift bins\label{sample_selection}} \tablewidth{0pt} 
\startdata
\hline 
\hline 
\\  [-1.5ex]
z$_{\rm min}$&z$_{\rm max}$& z median & ACS Volume&Subaru
Volume&Min $M_*$&N galaxies ACS\tablenotemark{a} & N galaxies Subaru\tablenotemark{a} \\
 & &   & 10$^6$ Mpc$^3$ & 10$^6$ Mpc$^3$ & & & \\ [1ex]
\hline\\  [-1.5ex]
0.22&0.48& 0.37 &0.88 &1.24 & 10$^{8.7}$ $M_{\odot}$ & 14956 & 20426 \\ 
0.48&0.74& 0.66 &2.03 &2.84 & 10$^{9.3}$ $M_{\odot}$ & 15103 & 20068\\ 
0.74&1.0 & 0.88 &3.13 &4.39 & 10$^{9.8}$ $M_{\odot}$ & 14387 & 18853\\ 
\enddata
\tablenotetext{a}{Numbers are quoted after stars have been removed,
  the selection $K_s<24$ as well as the lower limit stellar mass limit
  has been applied, and objects in masked areas have been removed from
  the catalog.}
\end{deluxetable*}

\subsection{Stellar Mass Functions}

The SMFs are calculated using the ACS catalog. As in
\citet{Bundy:2010}, we compute mass functions using the $V_{max}$
technique \citep{Schmidt:1968}.  We weight galaxies by the maximum
volume in which they would be detected within the $K_s$-band and
$I_{\rm 814W}$-band limits in a given redshift interval.  For each
host galaxy $i$ in the redshift interval $j$, the value of $V^i_{max}$
is given by the minimum redshift at which the galaxy would drop out of
the sample,

\begin{equation}
V^i_{max} = \int_{z_{low}}^{z_{high}} d \Omega \frac {dV}{dz} dz,
\end{equation}

\noindent where $d \Omega$ is the solid angle subtended by the survey area, and $dV/dz$ is the comoving
volume element.  The redshift limits are given as,

\begin{equation}
z_{high} = {\rm min}(z^j_{max}, z^j_{K_{lim}}, z^j_{I_{lim}}),
\end{equation}
\begin{equation}
z_{low} = z^j_{min},
\end{equation}

\noindent where the redshift interval, $j$, is defined by $[z^j_{min},
z^j_{max}]$ and $z^j_{K_{lim}}$ and $z^j_{I_{lim}}$ refer to the
redshift at which the galaxy would still be detected below the $K_s$-
and $I_{\rm 814W}$-band limits.  We use the best-fit SED template as
determined by the stellar mass estimator to calculate these values,
thereby accounting for the $k$-corrections necessary to compute
$V_{max}$ values (no evolutionary correction is applied).

\subsection{Galaxy Autocorrelation Function}\label{clustering_measure}

The galaxy clustering samples are defined by a series of stellar mass
thresholds rather than bins. For galaxy clustering, threshold samples
are the most straightforward to model in the HOD context. At small
scales, the signal-to-noise (S/N) in galaxy clustering is due mostly
to Poisson noise and at large scales it is subject to sample
(``cosmic'') variance. The optimal binning scheme for clustering can
be different than that for g-g lensing. For example, massive galaxies
produce the strongest shear signals, thus a relatively small sample is
required to produce a robust measurements relative to a simple
pair-counting statistic like the two-point correlation function
whereas the inverse would be true for low-mass galaxies. The threshold
samples we employ for clustering measurements are listed in Table
\ref{w_binning}. Except at the high mass end where galaxies become
scarce, the binning scheme employed for the clustering is a constant
0.5 dex in $\log_{10}(M_*)$.

% **************** TABLE 5 ***************
% Stellar mass bins for w(theta)
\begin{deluxetable*}{lcccccc}
%  \tabletypesize{\scriptsize} 
\tablecolumns{7} \tablecaption{Thresholds for the angular correlation function \label{w_binning}} \tablewidth{0pt} 
\startdata
\hline 
\hline
\\  [-1.5ex]
   & w bin1 & w bin2 & w bin3 & w bin4 & w bin5 & w bin6  \\ [1ex]
\hline\\  [-1.5ex]
minimum $\log_{10}(M_*)$, $z_1=[0.22,0.48]$ & 11.1 & 10.8 & 10.3 & 9.8
& 9.3 & 8.8 \\
minimum $\log_{10}(M_*)$, $z_2=[0.48,0.74]$ & 11.1 & 10.8 & 10.3 & 9.8
& 9.3 & none \\
minimum $\log_{10}(M_*)$, $z_3=[0.74,1.0]$  & 11.1 & 10.8 & 10.3 & 9.8 & none  & none \\ 
\enddata
\end{deluxetable*}

Since we do not need galaxy shape information to measure the angular
correlation function $w(\theta)$, we are not restricted to the ACS
coverage of the COSMOS field. We therefore use the COSMOS Subaru
catalog to calculate $w(\theta)$. For each threshold sample we measure
$w(\theta)$ using the well-known Landy-Szalay estimator of

\begin{equation}
\label{e.landy_szalay}
w(\theta) = \frac{DD - 2DR + RR}{RR},
\end{equation}

\noindent where $DD$ are the number of data-data pairs in a given bin
of angle $\theta$, $RR$ are the number of random-random pairs, and
$DR$ are the number of data-random pairs \citep[][]{Landy:1993}. Data
and random pairs are normalized by the total number of galaxies and
randoms, respectively. In all measurements we use $10^5$ randoms. The
distribution of the randoms are taken from a combination of four
COSMOS masks (B, V, ip, zp), thus mimicking the angular completeness
and geometry of the survey.

Because the volumes probed in each sample are small, our clustering
measurements are subject to the effect of the integral constraint (IC)
\citep[][]{Groth:1977}. Due to spatial fluctuations in the number
density of galaxies, the mean correlation function measured from an
ensemble of samples will be smaller than the correlation function
measured from a single contiguous sample of the same volume as the sum
of the ensemble sample. This attenuation of $w(\theta)$ becomes
relevant on angular scales significant with respect to the sample
size. We estimate the IC correction to our $w(\theta)$ measurements
through the use of mock galaxy distributions described in Paper I and
$\S$ \ref{theory}. In practice, we do not modify the measurements for
the IC but rather adjust the theoretical models to account for the
finite sample size.

Finally, in order to compute $w(\theta)$ from our model, we need to
know the normalized redshift distribution of the galaxy sample, $N(z)$
(see Equation 32 in Paper I). For this we use the probability
distribution functions of the photometric redshifts to estimate the
true $N(z)$ for our photo-$z$ slices. However, the COSMOS photo-$z$ errors
are accurate enough such that in tests we find a minimal effect when
using a flat top-hat $z$-bin for $N(z)$.

%%%%%%%%%%%% LENSING THEORY %%%%%%%%%%%%%%%%%%%%%%
\subsection{Galaxy-galaxy lensing: from galaxy shapes to $\Delta\Sigma$}

In the weak gravitational lensing limit, the observed shape
$\varepsilon_{\rm obs}$ of a source galaxy is directly related to the
lensing induced shear $\gamma$ according to

% Note : this is the calibrated gamma
\begin{equation}
  \varepsilon_{\rm obs} = \varepsilon_{\rm int}+\gamma,
\label{ellipticity_shear}
\end{equation}

\noindent where $\varepsilon_{\rm int}$ is the source galaxy's
intrinsic shape that would be observed in the absence of gravitational
lensing. In our notation, $\varepsilon_{\rm int}$, $\varepsilon_{\rm
  obs}$, and $\gamma$ are spin-2 tensors. The above relationship
indicates that galaxies would be ideal tracers of the distortions
caused by gravitational lensing if the intrinsic shape
$\varepsilon_{\rm int}$ of each source galaxy was known \textit{a
  priori}. However, lensing measurements exhibit an intrinsic
limitation, encoded in the width of the ellipticity distribution of
the galaxy population, noted here as $\sigma_{\rm int}$, and often
referred to as the ``intrinsic shape noise''. Because the intrinsic
shape noise \citep[of order $\sigma_{\rm int} \sim
0.27$,][]{Leauthaud:2007} is significantly larger than $\gamma$,
shears must be estimated by averaging over a large number of source
galaxies.

Throughout this paper, the gravitational shear is noted as $\gamma$
whereas $\tilde\gamma$ represents our estimator of $\gamma$. The
uncertainty in the shear estimator is a combination of unavoidable
intrinsic shape noise, $\sigma_{\rm int}=\sqrt{\langle
  \varepsilon_{\rm int}^2\rangle}$, and of shape measurement error,
$\sigma_{\rm meas}$:

\begin{equation}\label{eq:3}
  \sigma_{\rm \tilde\gamma}^2 = \sigma_{\rm int}^2+\sigma_{\rm meas}^2.
\end{equation}
 
We will refer to $\sigma_{\rm \tilde\gamma}$ as ``shape noise''
whereas $\sigma_{\rm int}$ will be called the ``intrinsic shape
noise''. The former includes shape measurement error and will vary
according to each specific data-set and shape measurement
method. Averaged over the whole COSMOS field, the weak lensing
distortions represent a negligible perturbation to Equation
\ref{eq:3}.

The derivation of our shear estimator is presented in
\citet[][]{Leauthaud:2007}. We employ the RRG method (see
\citealt{Rhodes:2000} for further details) for galaxy shape
measurements. Briefly, we form $\tilde\gamma$ from the PSF corrected
ellipticity according to

\begin{equation}
  \tilde\gamma=C \times \frac{\varepsilon_{\rm obs}}{G},
\end{equation}

\noindent where the shear susceptibility
factor\footnotemark[1]\footnotetext[1]{Not to be confused with
  Newton's constant which we have noted $G_{{ {\rm N}}}$.}, $G$, is
measured from moments of the global distribution of $\varepsilon_{\rm
  obs}$ and other, higher order shape parameters \citep[see Equation
28 in][]{Rhodes:2000}. Using a set of simulated images similar to
those of Shear TEsting Program
\citep[STEP;][]{Heymans:2006a,Massey:2007b} but tailored exclusively to
this data-set, we find that, in order to correctly measure the input
shear on COSMOS-like images, the RRG method requires an overall
calibration factor of $C=(0.86^{+0.07}_{-0.05})^{-1}$.

The shear signal induced by a given foreground mass distribution on a
background source galaxy will depend on the transverse proper distance
between the lens and the source and on the redshift configuration of
the lens-source system. A lens with a projected surface mass density,
$\Sigma(r)$, will create a shear that is proportional to the {\em
  surface mass density contrast}, $\Delta\Sigma(r)$:

\begin{equation}
  \Delta \Sigma(r)\equiv\overline{\Sigma}(< r)-\overline{\Sigma}(r)=\Sigma_{\rm crit}\times\gamma_t(r).
\label{dsigma}
\end{equation}

Here, $\overline{\Sigma}(< r)$ is the mean surface density within
proper radius $r$, $\overline{\Sigma}(r)$ is the azimuthally averaged
surface density at radius $r$
\citep[e.g,][]{Miralda-Escude:1991,Wilson:2001}, and $\gamma_t$ is the
tangentially projected shear. The geometry of the lens-source system
intervenes through the {\em critical surface mass
  density}\footnotemark[2]\footnotetext[2]{Note that some authors
  define the comoving critical surface mass density which has an extra
  factor of $(1+z)^{-2}$ with respect to ours due to the use of
  comoving instead of physical distances.}, $\Sigma_\mathrm{crit}$,
which depends on the angular diameter distances to the lens ($D_{\rm
  OL}$), to the source ($D_{\rm OS}$), and between the lens and source
($D_{\rm LS}$):

\begin{equation}
  \Sigma_\mathrm{crit} = \frac{c^2}{4\pi G_{{{\rm N}}}}\,
  \frac{D_\mathrm{OS}}{D_\mathrm{OL}\,D_\mathrm{LS}}\;,
\label{sigma_crit}
\end{equation}

\noindent where $G_{\rm N}$ represents Newton's constant. Hence, if
redshift information is available for every lens-source pair, each
estimate of $\gamma_{t}$ can be directly converted to an estimate of
$\Delta\Sigma(r)$ which is a more desirable quantity than $\gamma_{t}$
because it depends only on the mass distribution of the lens.

To measure $\Delta\Sigma(r)$ with high signal-to-noise, the lensing
signal must be stacked over many foreground lenses and background
sources. For every $i$th lens and $j$th source separated by a proper
distance $r_{ij}$, an estimator of the mean excess projected surface
mass density at $r_{ij}$ is computed according to:

\begin{equation}
  \Delta \tilde\Sigma_{ij}(r_{ij})=\tilde\gamma_{t,ij}\times \Sigma_{{\rm crit},ij},
\label{dsigma5}
\end{equation}

\noindent where $\tilde\gamma_{t,ij}$ is the tangential shear of the
source relative to the lens. The COSMOS photometric redshifts
described in $\S$\ref{photoz} are used to estimate $\Sigma_{{\rm
    crit},ij}$ for every lens-source pair. In order to optimize the
signal-to-noise, an inverse variance weighting scheme is employed when
$\Delta\Sigma_{ij}$ is summed over many lens-source pairs. Each
lens-source pair is attributed a weight that is equal to the estimated
variance of the measurement:

\begin{equation}
  w_{ij} = \frac{1}{ \left( \Sigma_{{\rm crit},ij} \times \sigma_{\tilde\gamma,ij} \right)^2}. 
\end{equation}

In this manner, faint small galaxies which have large measurement
errors are down-weighted with respect to sources that have well
measured shapes.

For the types of lenses studied in this paper, the signal-to-noise per
lens is not high enough to measure $\Delta\Sigma$ on an object by
object basis so instead we stack the signal over many lenses. For a
given sample of lenses, the total excess projected surface mass
density is the weighted sum over all lens-source pairs:

\begin{equation}
  \Delta\Sigma =  
  {\sum_{j=1}^{N_{Lens}} \sum_{i=1}^{N_{Source}} w_{ij} \times \tilde\gamma_{t,ij}\times \Sigma_{{\rm crit},ij}
    \over \sum_{j=1}^{N_{Lens}} \sum_{i=1}^{N_{Source}}w_{ij}} ~.
\label{ds_equation}
\end{equation}

\subsection{Galaxy-Galaxy Lensing Measurements}

We only give a brief outline of the overall methodology used to
compute the g-g lensing signals since this has already been presented
in detail in \citet[][]{Leauthaud:2010}. Foreground lens galaxies are
divided into three redshift samples and then are further binned by
stellar mass (see Figure \ref{sm_distribution} and Table
\ref{gg_binning}). For each lens sample, $\Delta\Sigma$ is computed
according to Equation \ref{ds_equation} from 25 kpc (physical
distance) to 1.5 Mpc in logarithmically spaced radial bins of 1.8
dex. In \citet[][]{Leauthaud:2010} we used a theoretical estimate of
the shape measurement error in order to derive the inverse variance
for each source galaxy. Instead, in this paper, the dispersion of each
shear component is measured directly from the data in bins of S/N and
magnitude. The measured shear dispersion is equal to the quadratic sum
of the intrinsic shape noise and of the shape measurement error. Our
empirical derivation of the shear dispersion varies from $\sigma_{\rm
  \tilde\gamma} \sim 0.25$ for bright galaxies with high S/N to
$\sigma_{\rm \tilde\gamma} \sim 0.4$ for faint galaxies with low
S/N. Overall, we find that the theoretical and the empirical schemes
yield very similar results with the latter method resulting in
slightly larger error bars because the theoretical scheme tends to
somewhat underestimate the shape measurement error for faint galaxies.

Photometric redshifts are used to derive $\Sigma_{\rm crit}$ for each
lens-source pair. The lower $68\%$ confidence bound on each source
redshift is used to select background galaxies. For each lens-source
pair, we demand that $z_{\rm source}-z_{\rm lens}>
\sigma_{68\%}(z_{\rm source})$ so as to minimize foreground
contamination. The g-g lensing signal is most sensitive to redshift
errors when $z_{\rm source}$ is only slightly larger then $z_{\rm
  lens}$ (see Figure \ref{redshifts}). For this reason, in addition to
the previous cut, we also implement a fixed cut so that $z_{\rm
  source}-z_{\rm lens}> 0.1$. Furthermore, in order to minimize the
effects of signal dilution caused by catastrophic errors, we also
reject all source galaxies with a secondary peak in the redshift
probability distribution function (\ie\ the parameter {\sc zp}$_2$ is
non zero in the \citealt{Ilbert:2009} catalog). This cut is aimed to
reduce the number of catastrophic errors in the source catalog. After
all cuts have been applied, the g-g lensing source catalog that we use
represents 35 galaxies per arcmin$^2$.

Finally, we re-compute all our g-g lensing signals using the
\citet[][]{Schrabback:2010} COSMOS shear catalog which has been
independently derived from ours. We find identical g-g lensing signals
from both shear catalogs, indicating that any relative shear
calibration differences between the two shear catalogs has no impact
on these results. This test provides an independent validation of our
g-g lensing results.

% **************** TABLE 3 ***************
% Stellar mass bins
\begin{deluxetable*}{lcccccccc}
%  \tabletypesize{\scriptsize} 
\tablecolumns{9} \tablecaption{Binning scheme for the g-g lensing \label{gg_binning}} \tablewidth{0pt} 
\startdata
\hline 
\hline 
\\  [-1.5ex]
  & & g-g bin1 & g-g bin2 & g-g bin3 & g-g bin4 & g-g bin5 & g-g bin6 & g-g bin7 \\ [1ex]
\hline\\  [-1.5ex]
$z_1=[0.22,0.48]$& min $\log_{10}(M_*)$ & 11.12 & 10.89 & 10.64 & 10.3 & 9.82 & 9.2 & 8.7 \\
                 & max $\log_{10}(M_*)$ & 12.0 & 11.12 & 10.89 & 10.64 & 10.3 & 9.8 & 9.2 \\
\hline
$z_2=[0.48,0.74]$& min $\log_{10}(M_*)$ & 11.29 & 11.05 & 10.88 & 10.65 & 10.3 & 9.8 & 9.3 \\
                 &max $\log_{10}(M_*)$ & 12.0 & 11.29 & 11.05 & 10.88 & 10.65 & 10.3 & 9.8 \\
\hline
$z_3=[0.74,1.0]$ & min $\log_{10}(M_*)$ & 11.35 & 11.16 & 10.97 & 10.74 & 10.39 & 9.8 & none \\
                 & max $\log_{10}(M_*)$ & 12.0 & 11.35 & 11.16 & 10.97 & 10.74 & 10.39 & none 
\enddata
\end{deluxetable*}

%%%%%%%%%%%%%%%%%%%%%%%%%%%%%%%%%%%%%%%%%%%%%%%%%%%%%%%%%%%%%%%%%%%%%%%%%%%%%%
%     THEORY
%%%%%%%%%%%%%%%%%%%%%%%%%%%%%%%%%%%%%%%%%%%%%%%%%%%%%%%%%%%%%%%%%%%%%%%%%%%%%%

\section{Theoretical Framework}\label{theory}

Paper I presents the general theoretical foundations that form the
backbone of this paper. In this section, we only give a brief, and
thus necessarily incomplete, review of the theoretical background and
strongly encourage the reader to refer to Paper I for further details.
We adopt the same model and notation as in Paper I.

Paper I describes an HOD-based model that can be used to analytically
predict the SMF, g-g lensing, and clustering signals. The key
component of this model is the SHMR which is modelled as a log-normal
probability distribution function with a log-normal
scatter\footnotemark[3]\footnotetext[3]{Scatter is quoted as the
  standard deviation of the logarithm base 10 of the stellar mass at
  fixed halo mass.} noted $\sigma_{\rm log M_{*}}$ and with a mean-log
relation denoted as $M_*=f_{\textsc{shmr}}(M_h)$.

For a given parameter set and cosmology, $f_{\textsc{shmr}}$ and
$\sigma_{\rm log M_{*}}$ can be used to determine the central and
satellite occupations functions, $\ncen$ and $\nsat$. These are used
in turn to predict the SMF, g-g lensing, and clustering signals.

\subsection{The stellar-to-halo mass relation}

\begin{figure*}[htb]
\epsscale{1.1}
\plotone{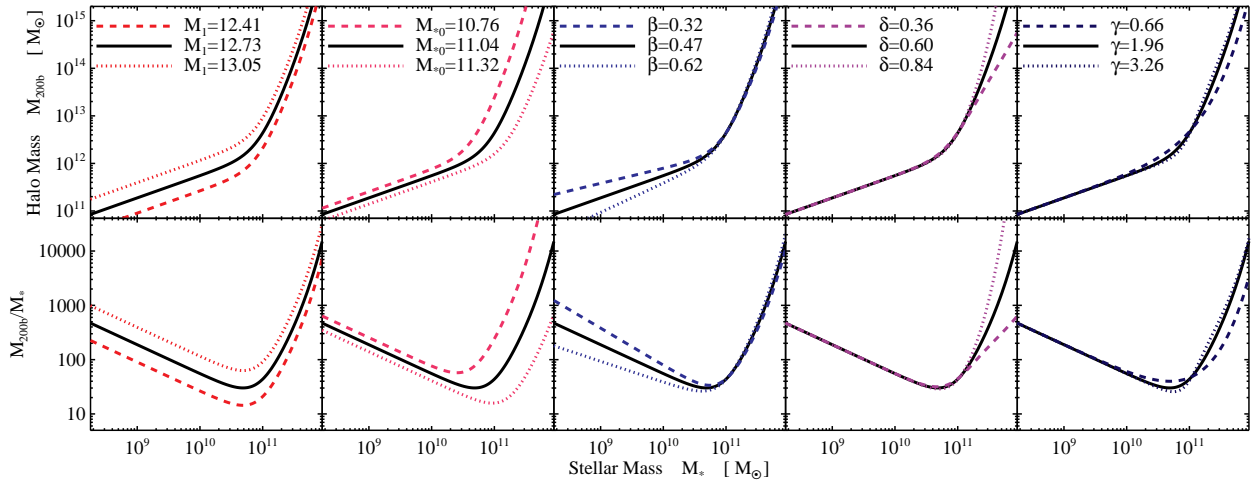}
\caption{Illustration of the influence of $M_{1}$, $M_{*,0}$, $\beta$,
  $\delta$ and $\gamma$ on the shape of the SHMR.}
\label{mh_ms_vary_p}
\end{figure*}

Following \citet{Behroozi:2010} (hereafter ``B10''),
$f_{\textsc{shmr}}(M_h)$ is mathematically defined via its
inverse function:

\begin{displaymath}
\log_{10}(f_{\textsc{shmr}}^{-1}(M_\ast)) =  \log_{10}(M_h) = \hspace{0.65\columnwidth}
 \end{displaymath}
 \vspace{-3ex}
\begin{equation}
 \label{shmr}
\quad \log_{10}(M_1) + \beta\,\log_{10}\left(\frac{M_\ast}{M_{\ast,0}}\right) +
 \frac{\left(\frac{M_\ast}{M_{\ast,0}}\right)^\delta}{1 + \left(\frac{M_{\ast}}{M_{\ast,0}}\right)^{-\gamma}} - \frac{1}{2},
\end{equation}

\noindent where $M_{1}$ is a characteristic halo mass, $M_{*,0}$ is a
characteristic stellar mass, $\beta$ is the faint end slope, and
$\delta$ and $\gamma$ control the massive end slope. We refer to B10
for a more detailed justification of this functional form. Briefly, we
expect that at least 4 parameters are required to model the SHMR: a
normalization, break, a faint end slope, and a bright end slope. In
addition, B10 have found that the SHMR displays sub-exponential
behaviour at large $M_*$. This is modelled by the $\delta$ parameter
which leads to a total of 5 parameters. Figure \ref{mh_ms_vary_p}
illustrates the influence of each parameter on the shape on the SHMR
and further details on the role of each parameter can be found in $\S$
2.1 of Paper I.

In contrast to B10, we do not parameterize the redshift evolution of
this functional form. Rather, we bin the data into three redshift bins
and check for redshift evolution in the parameters {\it a
  posteriori}. We also assume that Equation \ref{shmr} is only
relevant for central galaxies. Following the HOD ansatz, satellite
galaxies within groups and clusters occupy subhalos---bound,
virialized halos that are contained within the radius of a larger
halo. Abundance matching models like B10 assume that the halos and
subhalos of the same mass (or circular velocity) contain galaxies of
the same stellar mass, where the subhalo mass is taken as the mass at
the last time the galaxy was a central galaxy. Here we put no such
prior on the galaxy-subhalo connection. Rather, the halo occupation of
satellite galaxies is constrained by the data.

%vary_mhalo_ms.pro

\subsection{Scatter between stellar and halo mass}\label{scatter_shmr}

The measured scatter in stellar mass at fixed halo mass has an
intrinsic component (noted $\sigma_{\rm log M_{*}}^{\rm i}$), but also
includes a stellar mass measurement error due to redshift, photometry,
and modeling uncertainties (noted $\sigma_{\rm log M_{*}}^{\rm
  m}$). Ideally, we would measure both components but unfortunately we
can only constrain the quadratic sum of these two sources of
scatter. Nonetheless, given a model for $\sigma_{\rm log M_{*}}^{\rm
  m}$, we could in principle extract $\sigma_{\rm log M_{*}}^{\rm i}$
from \sigsm.

Previous work suggests that \sigsm~ is independent of halo mass. For
example, \citet[][]{Yang:2009} find that $\sigma_{\rm log M_{*}}=0.17$
dex and \citet{More:2009} find a scatter in luminosity at fixed halo
mass of $0.16 \pm 0.04$ dex. Both \citet[][]{Moster:2010} and B10 are
able to fit the SDSS (Sloan Digital Sky Survey) galaxy SMF assuming
$\sigma_{\rm log M_{*}}=0.15$ dex and $\sigma_{\rm log M_{*}}=0.175$
dex respectively. However, these results are derived with
spectroscopic samples of galaxies. In contrast to these surveys, we
expect a larger measurement error for the COSMOS stellar masses due to
the use of photometric redshifts. In addition, since photo-$z$ errors
increase for fainter galaxies, we might also expect that $\sigma_{\rm
  log M_{*}}^{\rm m}$ (and thus \sigsm) will depend on $M_*$.

To test if the assumption that \sigsm~ is constant has any impact on
our results, we implement two models for \sigsm. In the first case
(called ``\textsc{sig\_mod1}''), $\sigma_{\rm log M_{*}}$ is assumed
to be constant (this is our base-line model). In the second case
(called ``\textsc{sig\_mod2}''), we explicitly model $\sigma_{\rm log
  M_{*}}^{\rm m}$ to reflect stellar mass measurement errors. Note
that the goal of this exercise is not to perform a careful and
thorough error analysis, but simply to build a realistic enough model
to asses whether or not a $M_*$ dependent error has any strong impact
on our conclusions.

\begin{figure*}[htb]
\epsscale{1.1}
\plotone{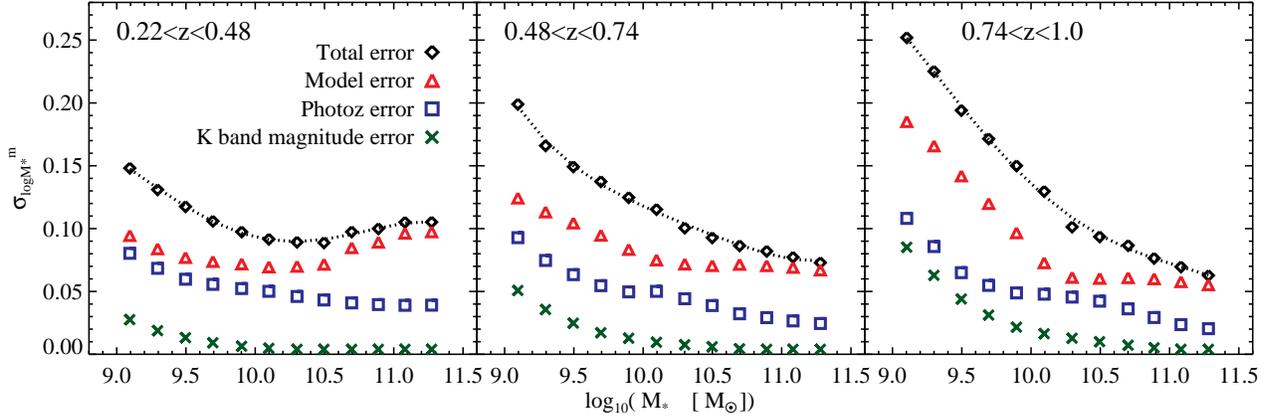}
\caption{The three sources of error that contribute to $\sigma_{\rm
    log M_{*}}^{\rm m}$ for the \textsc{sig\_mod2} model. At all
  redshifts, $\sigma_{\rm log M_{*}}^{\rm m}$ is dominated by model
  error followed by photo-$z$ error and finally by K-band magnitude
  error. The rise in the total error at high masses in the low-$z$ bin
  is caused by the model error term.  This is likely a result of
  degeneracies and template errors in the parameter space of models that provide
  reasonable fits to the early-type SEDs that dominate the high-mass
  end at low redshift.  Indeed, the estimated stellar mass PDFs at
  high mass in the low-$z$ bin tend to show more bi-modality.  The
  difference in error across the full mass range from this effect is
  small however, corresponding to only ~0.03 dex.}
\label{sm_error_kev}
\end{figure*}

For the \textsc{sig\_mod2} model, we consider three contributions to
the stellar mass error budget.  The first is called ``model error'':
this is measured by the 68\% confidence interval of the mass
probability distribution determined for each galaxy by the mass
estimator. It represents the range of model templates (each with its
own M/L ratio) that provide reasonable fits to the observed SED.  This
range is determined by the photometric uncertainty in the observed
SED, degeneracies in the grid of models used to fit the data, and how
well the grid of models represents the true parameter space of
observed galaxy populations as well as their
colors\footnotemark[4]\footnotetext[4]{Model template problems,
  especially in the restframe near-IR, and age-metallicity
  degeneracies may help explain the slight rise in the ``model error''
  contribution to the total stellar mass uncertainty for high mass
  galaxies at low redshift in Figure 4.}. The second term is the
photo-$z$ error, which derives from the uncertainty in the luminosity
distance owing to the error on a given photometric redshift.  The
final component is the photometric uncertainty from the observed
K-band magnitude, which translates into an uncertainty in luminosity
and therefore stellar mass. The total measurement error, $\sigma_{\rm
  log M_{*}}^{\rm m}$ is the sum in quadrature of these three sources
of error. The results are shown in Figure \ref{sm_error_kev} for the
three redshift bins.

%/Users/alexie/Work/HOD/Pro/study_redshift_errors_from_kev

As detailed in $\S$ \ref{results}, we find that our results are
largely unchanged, regardless of which form we adopt for \sigsm. This
can be explained as follows. Since the data are binned by $M_*$, the
observables are in fact sensitive to the scatter in halo mass at fixed
stellar mass, noted \sigmh. Given a model for the SHMR, \sigmh~can be
mathematically derived from \sigsm. Further details on the
mathematical connection between between \sigsm~ and \sigmh~ can be
found in Paper I. We find that the slope of the SHMR increases steeply
at $M_*>10^{11}$\sm so that \sigmh~ becomes quite large at the high
mass end. For example, $\sigma_{\rm log M_h}\sim 0.46$ dex at
$M_*=10^{11}$\sm and $\sigma_{\rm log M_h}\sim 0.7$ dex at
$M_*=10^{11.5}$ M$_{\odot}$. As a result, the data are particularly
sensitive to \sigsm~ at large $M_*$. Therefore, accounting for the
mass dependence of \sigsm~ has little impact on the results.

Finally, we note that our derived values for the parameters of the
SHMR should be independent of \sigsm. The observables (g-g lensing,
clustering, and SMF) do themselves depend on \sigsm but since we
account for \sigsm in the model, the extracted SHMR should reflect the
true underlying physical relationship between halo and stellar mass
(this however would not be the case if one were to neglect \sigsm). In
other terms, our model fully accounts for Eddington bias in all three
observables (also see discussion in Paper I).

\subsection{Parameters in the model}\label{parameters}

To model the central occupation function, we use five free parameters
($M_{1}, M_{*,0}, \beta, \delta, \gamma$) to model $f_{\textsc{shmr}}$
and we leave $\sigma_{\rm log M_{*}}$ as an additional free
parameter. As described in Paper I, the central occupation function
for a sample of galaxies more massive than $M_*^t$ is expressed as:

\begin{displaymath}
  \langle N_{\rm cen}(M_h|M_{*}^{t}) \rangle = \hspace{0.65\columnwidth}
\end{displaymath}
\begin{equation}
\label{ncen}
\frac{1}{2}\left[ 1-\mbox{erf}\left(\frac{\log_{10}(M_{*}^{t}) - \log_{10}(f_{\textsc{shmr}}(M_h)) }{\sqrt{2}\sigma_{\rm log M_{*}}} \right)\right].
\end{equation}

Our model also has an additional 5 parameters that are necessary to
model the satellite occupation function, $\nsat$. For a set of
galaxies more massive than threshold $M_*^t$, $\nsat$ is:

\begin{displaymath}
\langle N_{\rm sat}(M_h|M_{*}^{t}) \rangle = \hspace{0.65\columnwidth}
\end{displaymath}
\begin{equation}
\label{e.nsat}
\langle N_{\rm cen}(M_h|M_{*}^{t}) \rangle \left(\frac{f_{\textsc{shmr}}^{-1}(M_{*}^{t})}{M_{\rm sat}}\right)^{\asat} \exp\left(\frac{-M_{\rm cut}}{f_{\textsc{shmr}}^{-1}(M_{*}^{t})}\right).
\end{equation}

The free parameters that determine satellite occupation as a function
of stellar mass are $\beta_{\rm sat}$, B$_{\rm sat}$, $\beta_{\rm
  cut}$, B$_{\rm cut}$, and $\alpha_{\rm sat}$. The first two
parameters, $\beta_{\rm sat}$ and B$_{\rm sat}$, determine the
amplitude of $\nsat$. The second two parameters, $\beta_{\rm cut}$ and
B$_{\rm cut}$, set the scale of the exponential cut-off. These
parameters enter into $\nsat$ as:

\begin{equation}
\frac{M_{\rm sat}}{10^{12} M_{\odot}}= B_{\rm sat} \left(\frac{\fshmr}{10^{12} M_{\odot}}\right)^{\beta_{\rm sat}},
\end{equation}

and 

\begin{equation}
\label{mcut_eq}
\frac{M_{\rm cut}}{10^{12} M_{\odot}}= B_{\rm cut} \left(\frac{\fshmr}{10^{12} M_{\odot}}\right)^{\beta_{\rm cut}}.
\end{equation}

\noindent Finally, $\alpha_{\rm sat}$ represents the power-law slope of
the satellite mean occupation function. We set $\alpha_{\rm sat}=1$
for all samples which should be a good choice because the theoretical
expectation is that the number of sub-halos above a given mass scales
linearly with halo mass
\citep[][]{Kravtsov:2004,Conroy:2006,Moster:2010,Tinker:2010a}. Results
from group catalogs also indicate that $\alpha_{\rm sat} \sim 1$
\citep[][]{Collister:2005,Yang:2009}. Previous HOD analyses of
clustering results at varying redshifts also vary little from a value
of unity \citep[][]{Zehavi:2005, Zheng:2007,
  van-den-Bosch:2007,Tinker:2007, Zehavi:2010}.

In total, our model contains ten free parameters and one fixed
parameter ($\alpha_{\rm sat}$). A summary and description of these
parameters can be found in Table \ref{free_params} and also in Paper
I.

% **************** TABLE 1 ***************
% FREE PARAMETERS
\begin{deluxetable*}{lllll}
%  \tabletypesize{\scriptsize} 
\tablecolumns{5} \tablecaption{Parameters in model\label{free_params}} \tablewidth{0pt} 
\startdata
\hline 
\hline 
\\  [-1.5ex]
Parameter & Unit & Description & $\ncen$ or $\nsat$ & Free/fixed \\ [1ex]
\hline\\[-1.5ex]
$M_{1}$   &  $M_{\sun}$   & Characteristic halo mass in the SHMR &$\ncen$ &free \\ 
$M_{*,0}$  & $M_{\sun}$  & Characteristic stellar mass in the SHMR &$\ncen$ &free\\ 
$\beta$   & none & Faint end slope in the SHMR &$\ncen$ &free\\
$\delta$  & none & Controls massive end slope in the SHMR &$\ncen$&free\\
$\gamma$  & none & Controls the transition regime in the SHMR &$\ncen$&free\\
$\sigma_{\rm log M_{*}}$ & dex    & Log-normal scatter in stellar mass at fixed halo mass&$\ncen$ &free\\
$\beta_{\rm sat}$ & none   & Slope of the scaling of $M_{\rm sat}$ &$\nsat$ &free\\
$B_{\rm sat}$  &  none & Normalization of the scaling of $M_{\rm
  sat}$ &$\nsat$ &free\\
$\beta_{\rm cut}$ & none   & Slope of the scaling of $M_{\rm cut}$ &$\nsat$ &free\\
$B_{\rm cut}$  & none & Normalization of the scaling of $M_{\rm cut}$ &$\nsat$ &free\\
$\alpha_{\rm sat}$ & none    & Power-law slope of the satellite occupation function&$\nsat$ &fixed\\
\enddata
\end{deluxetable*}

\subsection{Covariance matrices}

The COSMOS survey covers a relatively small volume and therefore
sample variance effects must be taken into account. The volumes probed
in the three redshift bins are given in Table \ref{sample_selection}
and vary from $0.88\times 10^6$ Mpc$^3$ for the ACS region in $z_1$
($0.22<z<0.48$) to $4.39\times 10^6$ Mpc$^3$ for the Subaru region in
$z_3$ ($0.74<z<1.0$). The volumes sampled by COSMOS are too small to
obtain an accurate estimate of the sample variance from the data
itself.

A series of mock catalogs are used to calculate the co-variance
matrices for all three observables. Details regarding the construction
of these mocks can be found in Paper I. Briefly, COSMOS-like mocks are
created from a single simulation (named ``Consuelo'') 420 $h^{-1}$ Mpc
on a side, resolved with 1400$^3$ particles, and a particle mass of
1.87$\times 10^{9}$ $h^{-1}$
M$_{\odot}$\footnotemark[5]\footnotetext[5]{In this paragraph, numbers
  are quoted for $H_0=100$ h km~s$^{-1}$~Mpc$^{-1}$}. This simulation
can robustly resolve halos with masses above $\sim 10^{11}$ $h^{-1}$
M$_{\odot}$ and is part of the Las Damas
suite\footnotemark[6]\footnotetext[6]{Further details regarding this
  simulation can be found at {\tt
    http://lss.phy.vanderbilt.edu/lasdamas/simulations.html}} (McBride
et al. in prep).  We create mocks for the three redshift intervals:
$z_1=[0.22,0.48]$, $z_2=[0.48,0.74]$, and $z_3=[0.74,1]$. For each
redshift interval, we construct a series of mocks created from random
lines of sight through the simulation volume that have the same area
as COSMOS and the same comoving length for the given redshift
slice. This yields 405 independent mocks for the $z_1$ bin, 172 mocks
for the $z_2$ bin, and 109 mocks for the $z_3$ bin. For each redshift
bin, mocks are created from the simulation output at the median
redshift of the bin (see Table \ref{sample_selection}).

We converge on the mocks used for estimating errors using an iterative
method. To begin with, we find an initial best-fit model to the data,
without using any covariance matrices. This initial fit is used to
populate the mocks and to create a first set of covariance
matrices. We then re-fit the data using these co-variance matrices and
use the best-fit HOD models to create our final covariance
matrices. All results presented here use this final set of covariance
matrices.

%%%%%%%%%%%%%%%%%%%%%%%%%%%%%%%%%%%%%%%%%%%%%%%%%%%%%%%%%%%%%%%%%%%%%%%%%%%%%%
%     RESULTS
%%%%%%%%%%%%%%%%%%%%%%%%%%%%%%%%%%%%%%%%%%%%%%%%%%%%%%%%%%%%%%%%%%%%%%%%%%%%%%

\section{Results}\label{results}

We now present the results of fitting the model described in Paper I to
the observed COSMOS g-g lensing, clustering, and stellar mass
functions.

\subsection{Constraining the model}

In order to fit the model to the data, we minimize:

\begin{equation}
\chi^2_{\rm tot} = \chi^2_{\Phi_{\rm SMF}}+\sum_i\chi^2_{\Delta\Sigma,i}+\sum_j\chi^2_{w(\theta),j},
\end{equation}

\noindent where the sum over $i$ and $j$ indicates summation over the
different stellar mass bins and thresholds, respectively. For the SMF
and each stellar mass sample in $w(\theta)$ and $\Delta\Sigma$,
$\chi^2$ is calculated by:

\begin{equation}
\label{def_chi2}
\chi^2 = \sum_{n,l}\left(x_n-y_n\right)C^{-1}_{nl}\left(x_l-y_l\right),
\end{equation}

\noindent where $x_n$ is the model calculation of the quantity for
data point $n$, $y_n$ is the measurement, and $C^{-1}$ is the
inversion of the covariance matrix.

To obtain the posterior probability distributions for the parameter
set ($M_{1}$, $M_{*,0}$, $\beta$, $\delta$, $\gamma$, $\sigma_{\rm log
  M_{*}}$, $\beta_{\rm sat}$, $B_{\rm sat}$, $\beta_{\rm cut}$,
$B_{\rm cut}$) we implement a Markov Chain Monte Carlo (MCMC)
algorithm as follows:

\begin{enumerate}
\item We sample the region of interest around the best fit by
  initializing all chains close to the minimum $\chi^2$ solution with
  varying initial random spreads.
\item We apply conservative limits on all ten parameters. As shown in
  Figure \ref{z2_constraints}, the recovered posterior distributions
  are independent of these limits.
\item The covariance matrix is updated for the first 3000 elements in
  the chain, and then held fixed for the remainder of the chain.
\item We run 6 chains per case, with $\sim 40,000$ steps. We discard
  all elements before the covariance matrix is constant (the first
  3000), leaving $\sim 37,000$ elements per chain.
\item We use the {\tt GetDist} package provided with CosmoMC
  \citep{Lewis:2002} for computing convergence diagnostics. The chains
  are tested for convergence and mixing with the Gelman-Rubin
  criterion \citep[][]{Gelman:1992,Gilks:1996}. We impose a limit on
  the worst $R-1$ statistic of $R-1<0.01$ and observe that for all
  cases under study the mean and marginalized posterior distributions
  are well constrained.
\item We use {\tt GetDist} output statistics to estimate confidence
  limits on all parameters.
\end{enumerate}

\subsection{Fits to the data}

\begin{figure*}[p]
\epsscale{1.2}
\plotone{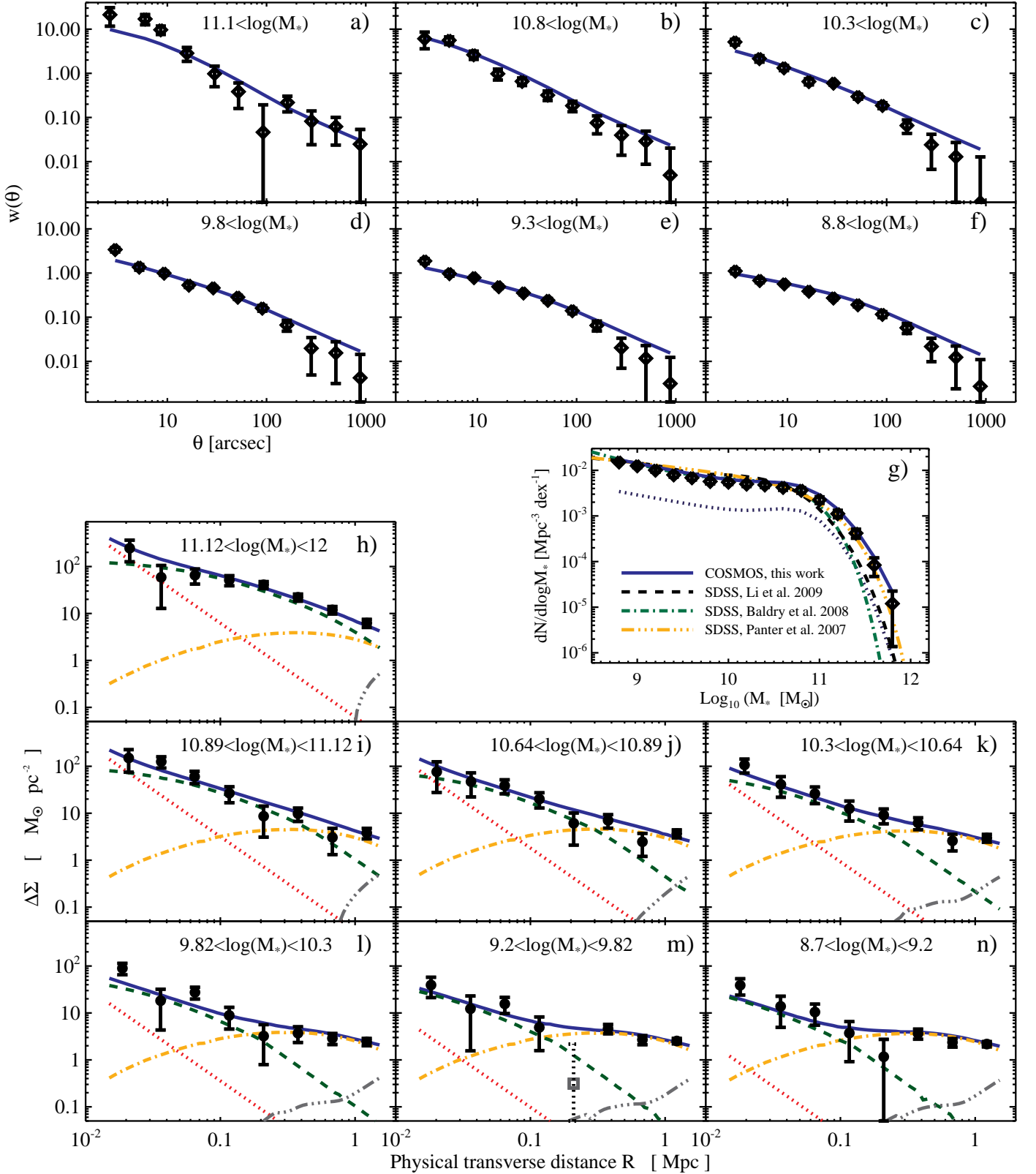}
\caption{ Best fit model for the $z_1$ ($0.22<z<0.48$) redshift bin
  (blue solid line). Panels a) to f): amplitude of the angular
  correlation function $w$ as a function of angular separation
  $\theta$ (in arc-seconds) in stellar mass thresholds. Note that in
  this redshift bin, the amplitude of $w$ at large separations is
  artificially deflated by integral constraint (but this is accounted
  for in the fitted model). Panel g): COSMOS SMF for $M_*>10^{8.7}$\sm
  (completeness limit for this redshift bin). For reference, in panel
  f), we also show the SDSS mass functions from \citet[][]{Li:2009}
  (triple schecter fit, black dashed line), \citet[][]{Baldry:2008}
  (green dash-dot line) and from \citet[]{Panter:2007} (orange,
  dash-dot-dot line). The dotted blue line in panel g) shows the SMF
  of satellite galaxies for the best fit model. Panels h) to n):
  galaxy-galaxy lensing signal in stellar mass bins. Note that in
  panel m), there is a negative data point represented by a grey
  square (this can occur due to noise and is not a concern). The
  lensing signal is decomposed into four components: the baryonic term
  (red dotted line), the central one-halo term (green dashed line),
  the satellite one-halo term (orange dash-dot), and the two-halo term
  (grey dash-dot-dot-dot line).}
\label{sm_z1}
\end{figure*}

\begin{figure*}[p]
\epsscale{1.2}
\plotone{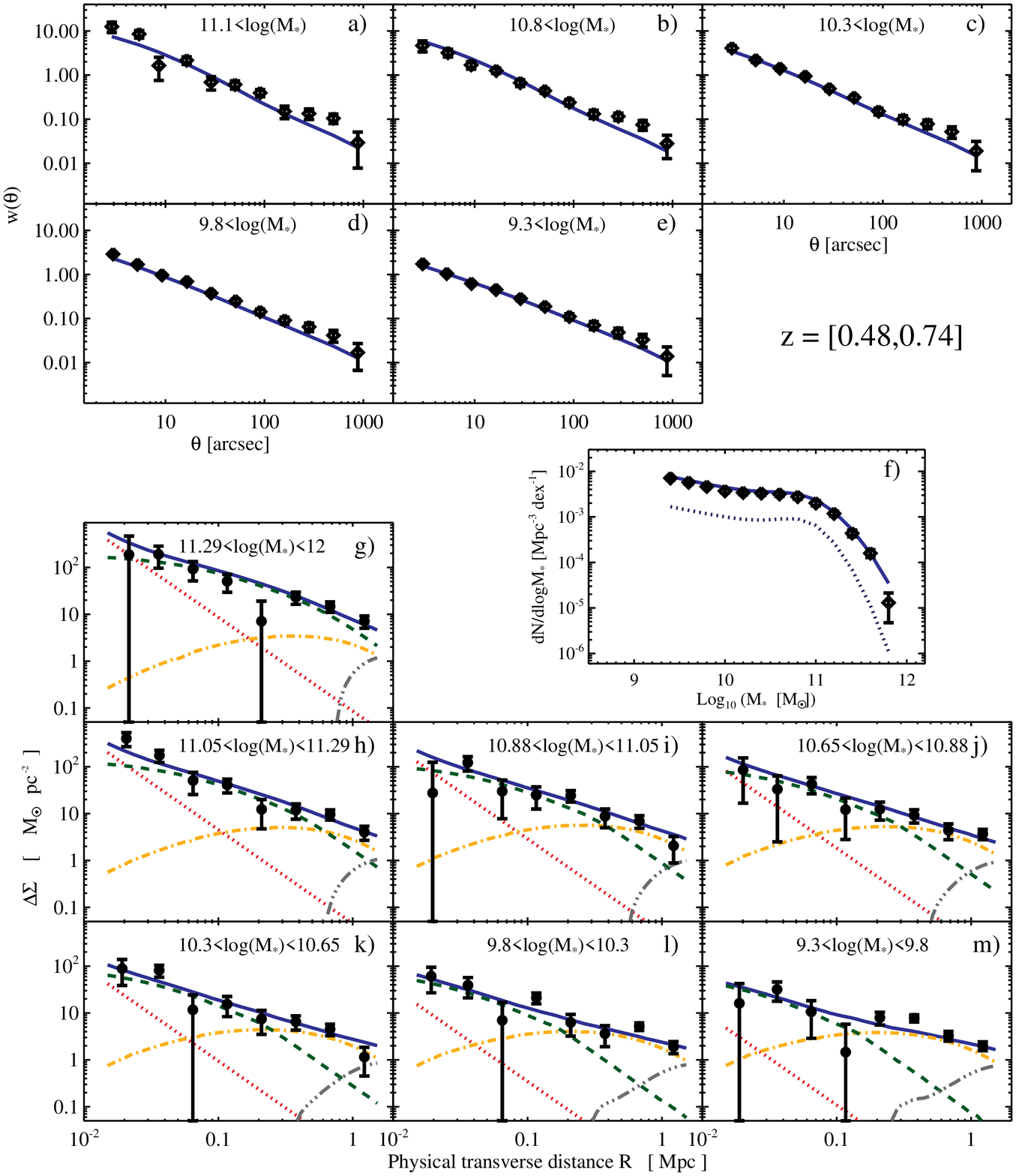}
\caption{Same as Figure \ref{sm_z1} but for the second redshift bin,
  $z_2$. Panels a) to e): angular correlation function. The angular
  correlation function is less affected by integral constraint in this
  redshift bin since the volume probed is 2.3 times larger than
  $z_1$. Panel f): COSMOS SMF for $M_*>10^{9.3}~{\rm M}_{\odot}$. The
  dotted blue line in panel f) shows the SMF of satellite galaxies for
  the best fit model. Panels g) to m): galaxy-galaxy lensing signal.}
\label{sm_z2}
\end{figure*}

\begin{figure*}[p]
\epsscale{1.2}
\plotone{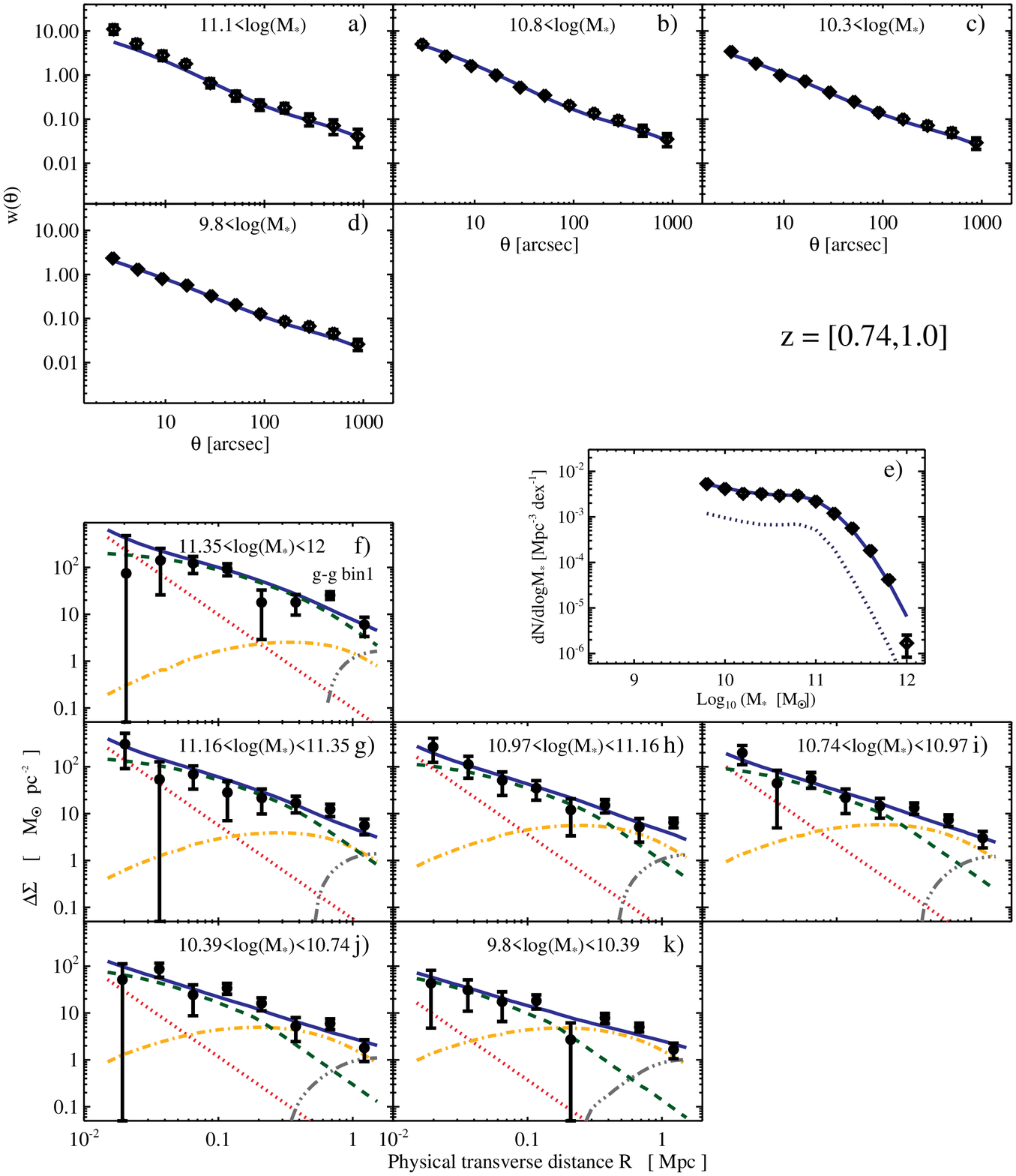}
\caption{Same as Figure \ref{sm_z1} but for the third redshift bin,
  $z_3$. Panels a) to d): angular correlation function. The angular
  correlation function is less affected by integral constraint in this
  redshift bin since the volume probed is 5 times larger than
  $z_1$. Panel e): COSMOS SMF for $M_*>10^{9.8}~{\rm M}_{\odot}$. The
  dotted blue line in panel e) shows the SMF of satellite galaxies for
  the best fit model. Panels f) to k): galaxy-galaxy lensing signal.}
\label{sm_z3}
\end{figure*}

Figures \ref{sm_z1}, \ref{sm_z2}, and \ref{sm_z3} show the best fit
models for each of the three redshift bins for the galaxy clustering,
the g-g lensing, and the SMF. The upper panels show the angular
correlation function $w(\theta)$ for stellar mass threshold
samples. The middle right panel shows the COSMOS SMF which is measured
for all galaxies with $M_*>10^{8.7}$\sm for $z_1$, $M_*>10^{9.3}$\sm
for $z_2$, and $M_*>10^{9.8}$\sm for $z_3$. The dotted blue line shows
the SMF of satellite galaxies from our model. The lower panels show
the g-g lensing signals for the stellar mass bins defined in Table
\ref{gg_binning}. When looking at Figures \ref{sm_z1}, \ref{sm_z2},
and \ref{sm_z3}, one must keep in mind that the data points are
correlated (see Paper I for the covariance matrices) and so it is
difficult to evaluate ``by eye'' whether or not the fits are
adequate. The reduced $\chi^2$ for the fits are 1.7, 1.6, and 1.9, for
$z_1$, $z_2$, and $z_3$ respectively.

%/Users/alexie/Work/HOD/Pro/doall_gg_cluster_10.pro

For our $z_1$ sample, we have compared the COSMOS mass function with
previously published mass functions from SDSS
\citep[][]{Li:2009,Baldry:2008,Panter:2007}. Because of the use of
photometric redshifts in COSMOS, we expect a larger stellar mass
measurement error which should cause Eddington bias and lead to an
inflated observed SMF at the high mass end in COSMOS compared to
SDSS. A more in-depth comparison of the mass functions is discussed
further on in $\S$ \ref{role_smf}.

\subsection{Parameter constraints and redshift evolution}

% **************** TABLE  ***************
% Best fit params
\begin{deluxetable}{llll}[htb]
\tablecolumns{4} \tablecaption{Best fit parameters for the 3 redshift bins \label{best_fit_params}} \tablewidth{0pt} 
\startdata
\hline 
\hline 
\\  [-1.5ex]
Parameter&$z_1=[0.22,0.48]$&$z_2=[0.48,0.74]$&$z_3=[0.74,1]$\\ [1ex]
\hline \\ 
SIG\_MOD1:   &  &  & \\[1.5ex]
\hline\\ %[-1.5ex]
$\log_{10}(M_{1})$  & $12.520\pm 0.037$ & $12.725\pm 0.032$ & $12.722\pm 0.027$ \\
$\log_{10}(M_{*,0})$  & $10.916\pm 0.020$ & $11.038\pm 0.019$ & $11.100\pm 0.018$ \\
$\beta$  & $0.457\pm 0.009$ & $0.466\pm 0.009$ & $0.470\pm 0.008$ \\
$\delta$  & $0.566\pm 0.086$ & $0.61\pm 0.13$ & $0.393\pm 0.088$ \\
$\gamma$  & $1.53\pm 0.18$ & $1.95\pm 0.25$ & $2.51\pm 0.25$ \\
$\sigma_{\rm log M_{*}}$  & $0.206\pm 0.031$ & $0.249\pm 0.019$ & $0.227\pm 0.020$ \\
B$_{\rm cut}$ & $1.47\pm 0.73$ & $1.65\pm 0.65$ & $2.46\pm 0.53$ \\
B$_{\rm sat}$  & $10.62\pm 0.87$ & $9.04\pm 0.81$ & $8.72\pm 0.53$ \\
$\beta_{\rm cut}$ & $-0.13\pm 0.28$ & $0.59\pm 0.28$ & $0.57\pm 0.20$ \\
$\beta_{\rm sat}$ & $0.859\pm 0.038$ & $0.740\pm 0.059$ & $0.863\pm 0.053$ \\
\hline \\ 
SIG\_MOD2:   &  &  & \\[1.5ex]
\hline\\ %[-1.5ex]
$\log_{10}(M_{1})$  & $12.518\pm 0.038$ & $12.724\pm 0.033$ & $12.726\pm 0.028$ \\
$\log_{10}(M_{*,0})$  & $10.917\pm 0.020$ & $11.038\pm 0.019$ & $11.100\pm 0.017$ \\
$\beta$  & $0.456\pm 0.009$ & $0.466\pm 0.009$ & $0.470\pm 0.008$ \\
$\delta$  & $0.582\pm 0.083$ & $0.62\pm 0.12$ & $0.47\pm 0.10$ \\
$\gamma$  & $1.48\pm 0.17$ & $1.93\pm 0.25$ & $2.38\pm 0.24$ \\
$\sigma_{\rm log M_{*}}^i$\tablenotemark{*}  & $0.192\pm 0.031$ & $0.245\pm 0.019$ & $0.220\pm 0.019$ \\
B$_{\rm cut}$  & $1.52\pm 0.79$ & $1.69\pm 0.65$ & $2.57\pm 0.56$ \\
B$_{\rm sat}$  & $10.69\pm 0.89$ & $9.01\pm 0.81$ & $8.66\pm 0.53$ \\
$\beta_{\rm cut}$  & $-0.11\pm 0.29$ & $0.60\pm 0.27$ & $0.58\pm 0.20$ \\
$\beta_{\rm sat}$ & $0.860\pm 0.039$ & $0.740\pm 0.059$ & $0.863\pm 0.053$ \\
\enddata
\tablenotetext{*}{In the SIG\_MOD2 case we fit for $\sigma_{\rm log
    M_{*}}^i$ whereas in the SIG\_MOD1 case we fit for $\sigma_{\rm
    log M_{*}}$.}  \tablenotetext{}{Note: the quoted errors are always
  marginalized over all other parameters.}
\end{deluxetable}

\begin{figure*}[htb]
\epsscale{1.1}
\plotone{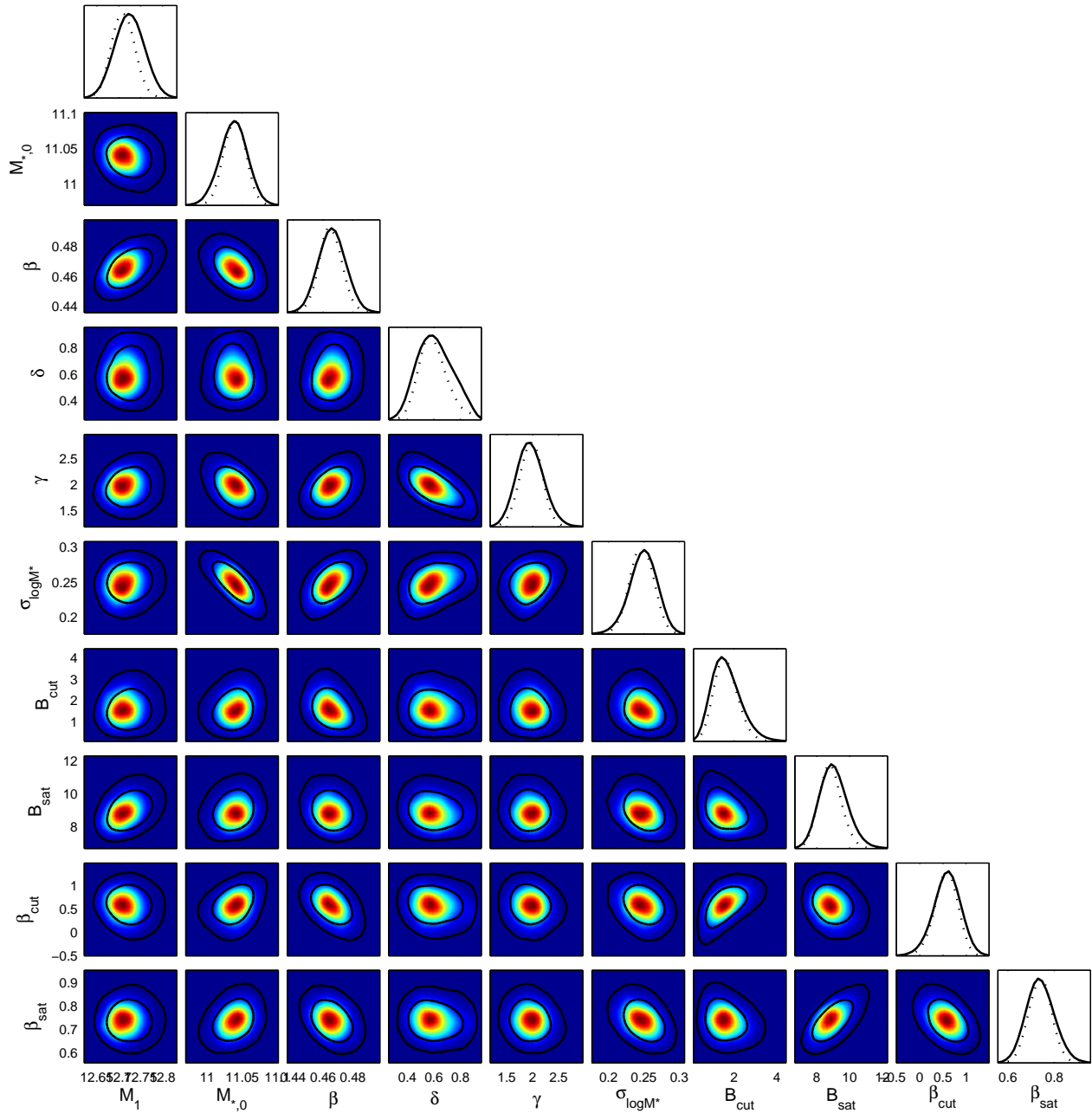}
\caption{One and two dimensional joint-mean and marginalized
  distributions for all ten parameters for the $z_2$ bin. Dotted lines
  in the 1D distributions (and shaded contours in the 2D planes) show
  mean likelihoods of samples and solid lines show marginalized
  probabilities. In the 2D planes, blue to red denotes the region of
  lowest to highest likelihood density. Solid contours in the 2D plane
  represent the 68\% (1$\sigma$) and 95\% (2$\sigma$) confidence
  regions.}
\label{z2_constraints}
\end{figure*}

Table \ref{best_fit_params} gives the best fit values from {\tt
  GetDist} for all ten parameters and for the three redshift
bins. Figure \ref{z2_constraints} shows the one dimensional and two
dimensional joint marginalized constraints on the parameters for the
$z_2$ bin. All of the parameters have well behaved, uni-modal
distributions. In the interest of brevity, we have not included
equivalent Figures for $z_1$ and $z_3$ but they are similar to Figure
\ref{z2_constraints}. All parameters are well constrained in the three
redshift bins.

Table \ref{best_fit_params} lists the best fit values for
\textsc{sig\_mod1} where we have assumed that \sigsm~ is constant and
for \textsc{sig\_mod2} where we have explicitly accounted for stellar
mass dependent errors. We find no strong difference in our results,
regardless of which model we adopt. We conclude that our results are
robust to the effects of mass dependent scatter. In the
\textsc{sig\_mod2} case, we model $\sigma_{\rm log M_{*}}^{\rm m}$ and
assume that $\sigma_{\rm log M_{*}}$ is the sum in quadrature of
$\sigma_{\rm log M_{*}}^{\rm i}$ (which is assumed to be constant) and
$\sigma_{\rm log M_{*}}^{\rm m}$. Therefore, the quantity that we
actually fit for in the \textsc{sig\_mod2} case is $\sigma_{\rm log
  M_{*}}^{\rm i}$ (whereas in \textsc{sig\_mod1} we fit for
$\sigma_{\rm log M_{*}}$). This is why, as expected, the best-fit
scatter in Table \ref{best_fit_params} is slightly lower for
\textsc{sig\_mod2} compared to \textsc{sig\_mod1}. Note that we are
not claiming to actually extract meaningful values for the intrinsic
scatter in stellar mass at fixed halo mass. To do so would require a
more thorough error analysis which is beyond the scope of this
paper. Overall, our conclusions regarding $\sigma_{\rm log M_{*}}$ are
twofold. Firstly, we can safely assume that $\sigma_{\rm log M_{*}}$
is constant and ignore any mass dependent effects induced by
measurement error. This is due to the fact that all three observables
are primarily sensitive to the effects of $\sigma_{\rm log M_{*}}$ at
large $M_*$ where the slope of the SHMR increases sharply. Similar
conclusions were reached by B10 who find that the effects of scatter
are insignificant below $M_*=10^{10.5}$\sm where the slope of the SHMR
is not steep enough to have a significant impact. Secondly, we find
that $\sigma_{\rm log M_{*}} \sim 0.23 \pm 0.03$ dex, in broad
agreement with previous results. For example, B10 estimate that
$\sigma_{\rm log M_{*}}^{m}=0.07$ dex and $\sigma_{\rm log
  M_{*}}^{i}=0.16$ dex, resulting in a total scatter of $\sigma_{\rm
  log M_{*}}=0.175$ dex (their total scatter is lower than ours as
expected because we have a larger $\sigma_{\rm log M_{*}}^{m}$
component).

\begin{figure*}[htb]
\epsscale{1.2}
\plotone{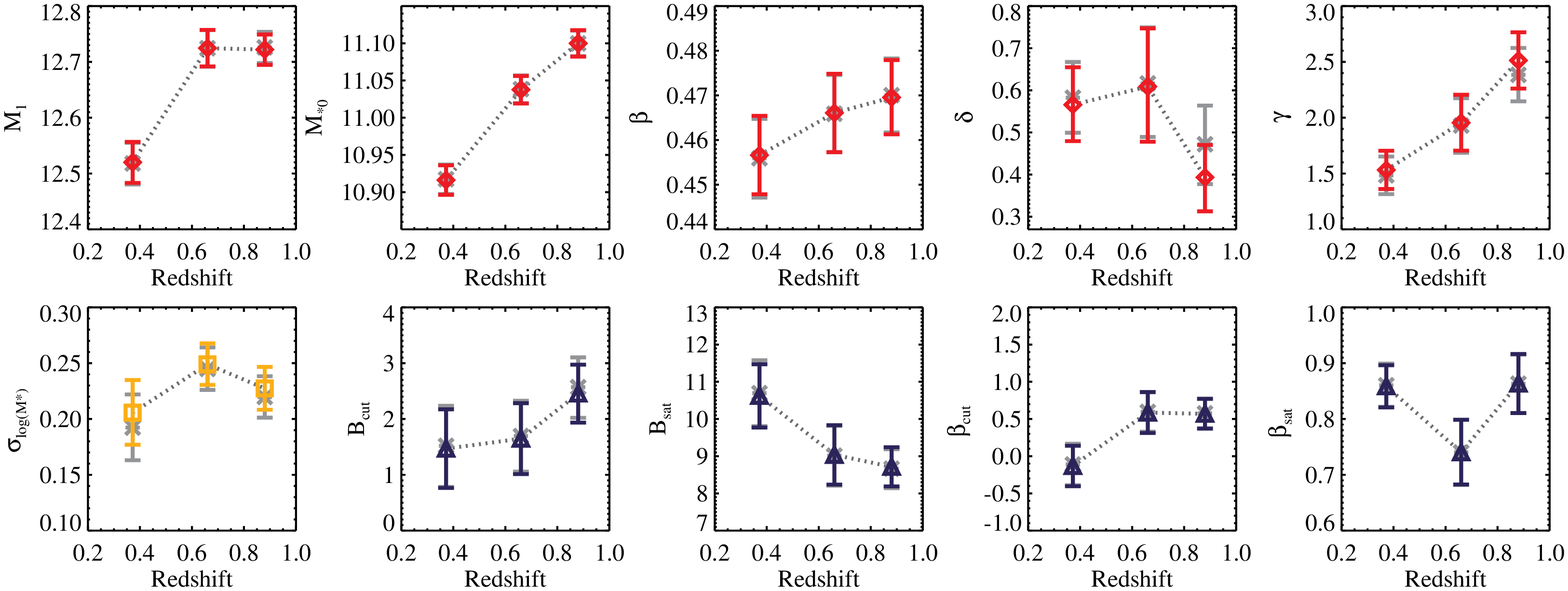}
\caption{Redshift evolution in the ten parameters. Red diamonds
  indicate evolution in the five parameters that describe the SHMR
  ($M_{1}$, $M_{*,0}$, $\beta$, $\delta$, $\gamma$). Orange squares
  show the scatter in stellar mass at fixed halo mass ($\sigma_{\rm
    log M_{*}}$). Blue triangles show the remaining four parameters
  that describe the satellite occupation function ($\beta_{\rm sat}$,
  $B_{\rm sat}$, $\beta_{\rm cut}$, $B_{\rm cut}$). Grey data points shows
  the best fit parameters when we assume a stellar mass dependent
  scatter due to photometric redshifts (the \textsc{sig\_mod2} model).}
\label{param_z_evol}
\end{figure*}

Figure \ref{param_z_evol} shows the measured redshift evolution for
all ten parameters. Previous work on this topic has been limited by
systematic differences in stellar mass estimates between low and
high-z surveys which can be of order $0.25$ dex (excluding IMF
uncertainties) according to B10. We stress that the results in this
paper have been derived in a homogeneous fashion from high to low
redshift. Our conclusions regarding redshift evolution should thus be
robust in this respect.

The most striking result from Figure \ref{param_z_evol} is the
redshift evolution in the two parameters $M_1$ and $M_{*,0}$. This is
one of the major results of this paper which we will discuss in more
detail in the following section. Apart from these two parameters,
there is marginal evidence for some evolution in $\gamma$ and
$B_{\rm sat}$. No strong evolution is detected in the remaining
six parameters. It is interesting to note that $\beta$ (which controls
the low mass slope of the SHMR) remains constant at $\beta \sim
0.46$. We will provide an interpretation of this result in the
discussion section.

% This figure comes from get_dist
%/Users/alexie/Work/HOD/Chains_Marina_4paper/Results

% Using the param constraint file from Marina
% plot_param_evolution

\subsection{The SHMR from $z=0$ to $z=1$}

\begin{figure*}[htb]
\epsscale{1.2}
\plotone{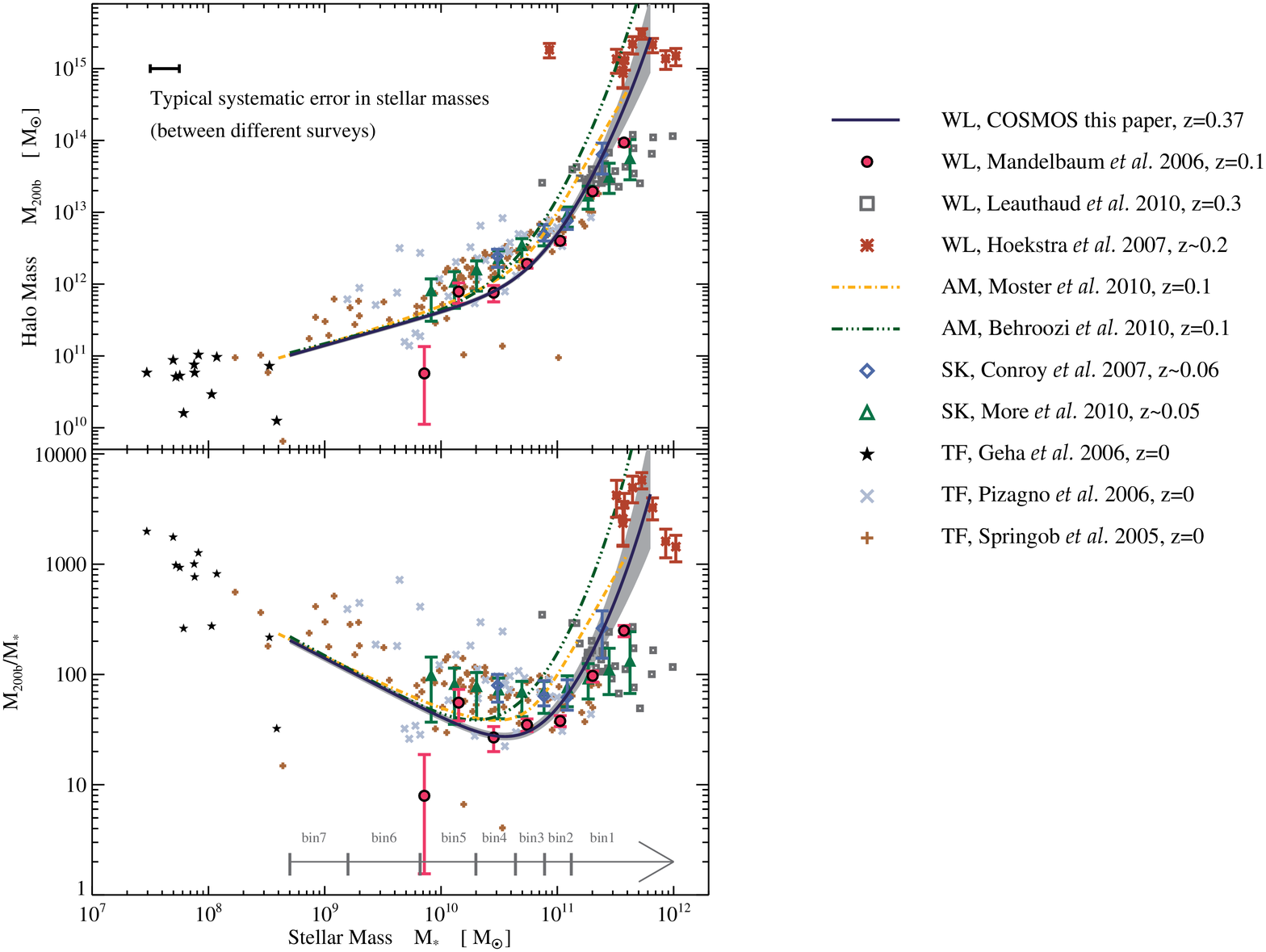}
\caption{{\em Top panel}: Inferred SHMR in the $z_1$ redshift bin
  compared to other low redshift measurements from weak lensing (WL),
  abundance matching (AM), satellite kinematics (SK), and the Tully
  Fisher relation (TF). The COSMOS $z_1$ results are shown by the
  solid dark blue line and the shaded grey region represents the one
  sigma error on the SHMR. This SHMR represents
  $\langle\log_{10}(M_*(M_h))\rangle$. With the exception of
  \citet[][]{Conroy:2007}, all data points either use or have been
  converted to this same averaging system . Overall, there is a broad
  agreement between various probes. Detailed comparisons between
  various data-sets however, are limited by systematic differences in
  stellar mass estimates due to varying assumptions (e.g., star
  formation histories, extinction laws, stellar population
  models). {\em Bottom panel}: Dark-to-stellar mass ratio as a function of
  stellar mass. We observe a clear variation in $M_{200b}/M_*$ with
  $M_{200b}/M_*$ reaching a minimum of $M_h/M_*\sim 27$ at $M_*\sim
  4.5\times 10^{10}~{\rm M}_{\odot}$ and $M_{200b} \sim 1.2\times
  10^{12}~{\rm M}_{\odot}$. The dark-to-stellar mass ratio rises
  sharply at $M_*> 5\times 10^{10}~{\rm M}_{\odot}$ so that a cluster
  of halo mass $M_{200b}\sim 10^{15}~{\rm M}_{\odot}$ will reach a
  ratio of $M_{200b}/M_* \sim 2000$. Note that this ratio only refers
  to the ratio between the halo mass and the stellar mass of the {\em
    central galaxy}. For example, in the case of clusters, we are
  comparing the ratio of the cluster halo mass to stellar mass of the
  central Brightest Cluster Galaxy (BCG).}
\label{shmr_z1}
\end{figure*}

Figure \ref{shmr_z1} shows the best fit SHMR for $z_1$ compared to a
variety of low redshift constraints from weak lensing
\citep[][]{Mandelbaum:2006c,Leauthaud:2010,Hoekstra:2007}, abundance
matching \citep[][]{Moster:2010,Behroozi:2010}, satellite kinematics
\citep[][]{Conroy:2007,More:2010}, and the Tully-Fisher relation
\citep[][]{Geha:2006,Pizagno:2006,Springob:2005,Blanton:2008} (see
$\S$ \ref{previous_work} for a more in-depth comparison). Most of the
data are in broad agreement and show clear evidence for a variation in
the dark-to-stellar mass ratio with a minimum of $M_h/M_*\sim 27$ at
$M_* \sim 4.5\times 10^{10}~{\rm M}_{\odot}$ and $M_h \sim 1.2\times
10^{12}~{\rm M}_{\odot}$. As demonstrated by B10, however, meaningful
and detailed comparisons between various data-sets are hampered by
systematic uncertainties in stellar mass estimates. For this reason,
we will mainly focus in what follows on the conclusions that can be
drawn by inter-comparing the three COSMOS redshift bins.

%mhalo_ms.pro

At low stellar mass, $M_h$ scales as $M_h \propto (M_*)^{0.46}$ and
this scaling does not evolve significantly with redshift from $z=0.2$
to $z=1$. At high stellar mass, the SHMR rises sharply at $M_* >
10^{10.5}$\sm as a result of which \sigmh~(the scatter in halo mass at
fixed stellar mass) also increases and $M_*$ is clearly no longer a
good tracer of halo mass. For example, a galaxy with $M_* \sim
10^{11.3}$\sm may be the central galaxy of group with $M_h \sim
10^{13}-10^{14}~{\rm M}_{\odot}$ or may also be the central galaxy of
a cluster with $M_h>10^{15}~{\rm M}_{\odot}$.

A quantity that is of particular interest is the mass at which the
ratio $M_h/M_*$ reaches a minimum. This minimum is noteworthy for
models of galaxy formation because it marks the mass at which the
accumulated stellar growth of the central galaxy has been the most
efficient. We describe the SHMR at this minimum in terms of the
``pivot stellar mass'', $M_{*}^{\rm piv}$, the ``pivot halo mass'',
$M_{h}^{\rm piv}$, and the ``pivot ratio'', $(M_h/M_*)^{\rm
  piv}$. Note that $M_{*}^{\rm piv}$ and $M_{h}^{\rm piv}$ are not
simply equal to $M_{1}$ and $M_{*,0}$. Indeed, the mathematical
formulation of the SHMR is such the pivot masses depends on all five
parameters. The three parameters that have the strongest effect on the
pivot masses are $M_{1}$, $M_{*,0}$, and $\gamma$ (see Paper I).

\begin{figure*}[htb]
\epsscale{1.2}
\plotone{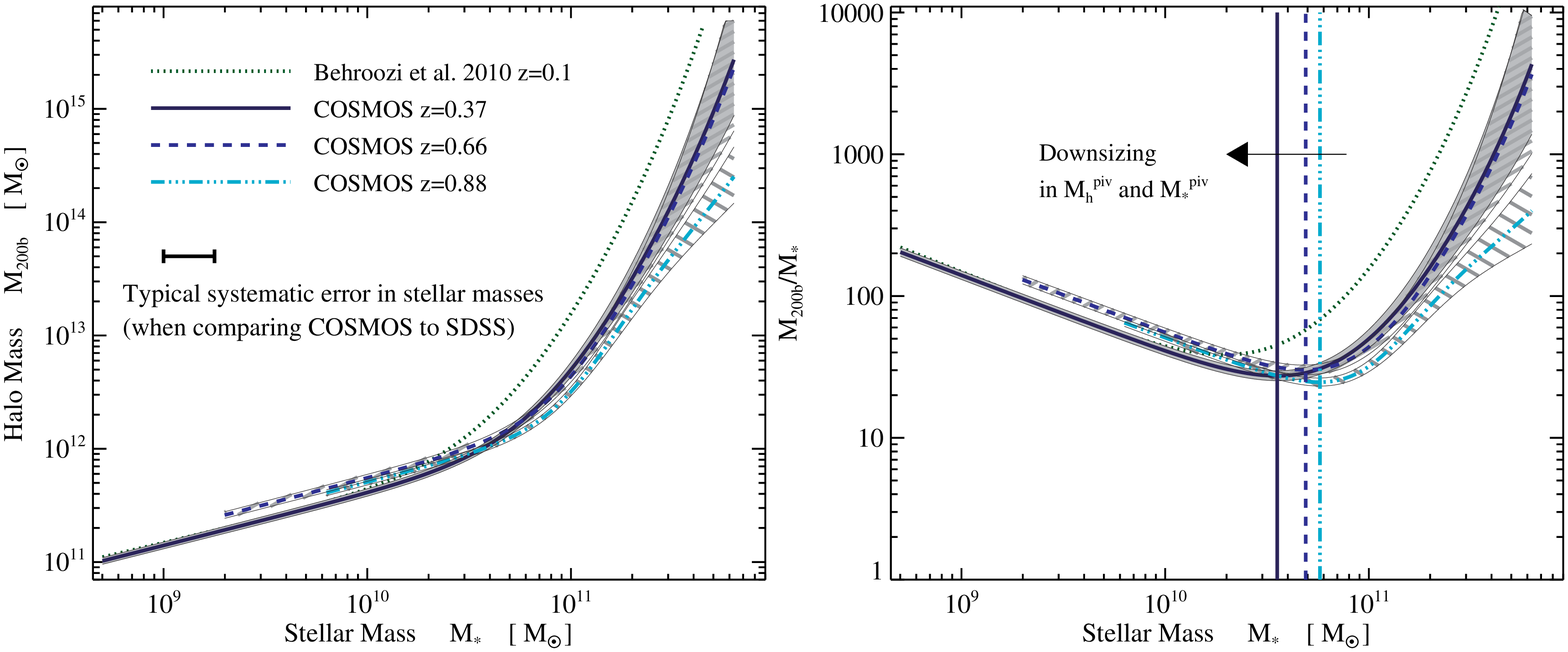}
\caption{{\em Left panel}: Redshift evolution in the SHMR. Because the
  COSMOS results have been derived in a homogeneous fashion, we can
  inter-compare the three COSMOS redshift bins. The low mass slope of
  the SHMR remains constant with $\beta\sim 0.46$. We detect a clear
  evolution in $M_*^{\rm piv}$ and $M_h^{\rm piv}$ which are
  respectively the stellar and halo mass at which $M_{200b}/M_{*}$
  reaches a minimum (see right panel).  At high masses, there is an
  interesting hint that the amplitude of the SHMR is decreasing at
  higher redshifts, but we lack the statistics to make a clear
  detection, mainly due to the small volume probed by COSMOS. We have
  also plotted the SHMR reported by B10. However, caution must be
  taken when making a direct comparisons between COSMOS and B10
  because our stellar masses have been derived under different
  assumptions. According to B10, the level of systematic uncertainty
  in stellar masses is of order $0.25$ dex. Given this 0.25 dex
  systematic uncertainty, we are in broad agreement with B10 but a
  direct comparison would require a more homogeneous analysis between
  COSMOS and SDSS. {\em Right panel}: redshift evolution in
  $M_{200b}/M_{*}$. Both $M_*^{\rm piv}$ and $M_h^{\rm piv}$ exhibit
  downsizing trends, decreasing at later epochs. This effect is shown
  more explicitly in Figure \ref{downsizing}.}
\label{shmr_z_evol}
\end{figure*}

Figure \ref{shmr_z_evol} shows the redshift evolution of the
SHMR. Three points are of particular interest in this Figure. First,
we detect no strong redshift evolution in the low mass slope of the
SHMR ($M_* <10^{10.2}$ M$_{\odot}$). Indeed, as highlighted in the
previous section already, $\beta$ is remarkably constant out to
$z=1$. In Paper I we have shown that $\beta$ regulates the faint end
slope of the SMF so in other terms, we could also state that the faint
end slope of the SMF shows remarkably little redshift evolution. We do
however find that the amplitude of the low-mass ($M_* <10^{10.2}$
M$_{\odot}$) SHMR was higher at earlier times so that galaxies at
fixed stellar mass live in more massive halos at earlier epochs. We
will return to this result in the discussion section.

\begin{figure*}[htb]
\epsscale{1.2}
\plotone{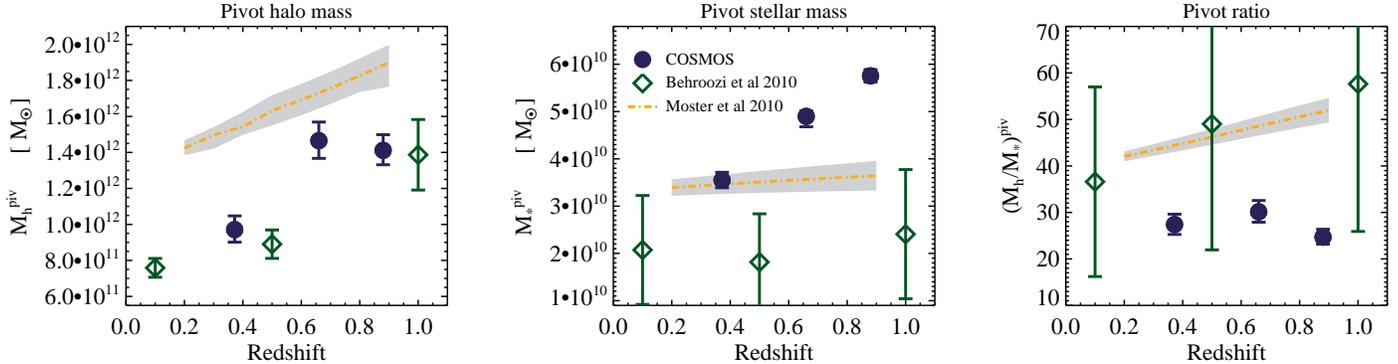}
\caption{Redshift evolution in the pivot halo mass, the pivot stellar
  mass, and the pivot ratio. Dark blue circles show the results from
  this paper. We detect a striking evolution in $M_*^{\rm piv}$ with
  $M_*^{\rm piv}$ decreasing at later epochs. This downsizing in
  $M_*^{\rm piv}$ is accompanied by a downsizing in $M_h^{\rm piv}$
  with $M_h^{\rm piv}$ also decreasing at later epochs.  We detect no
  strong evolution in the pivot ratio which remains constant at
  $(M_h/M_*)^{\rm piv}\sim 27$.  Green diamonds: comparison with B10.
  Yellow dash-dot line: comparison with \citet[][]{Moster:2010} (from
  Table 7 of their paper). The downsizing trend in $M_h^{\rm piv}$
  (which is not sensitive to systematic errors associated with $M_*$)
  is detected by all three studies. The normalization and evolution of
  $M_h^{\rm piv}$ is similar for COSMOS and for B10. However, the
  normalization of $M_h^{\rm piv}$ from \citet[][]{Moster:2010} is
  higher than our prediction. This could be due to differences in the
  adopted parametric form of the SHMR. The three studies show
  different results for the evolution of $M_*^{\rm piv}$. We stress
  that stellar mass estimates computed under different assumptions may
  present relative systematic errors of order 0.25 dex. This error
  will affect quantitative comparisons for the normalization $M_*^{\rm
    piv}$ (and by consequence, also $(M_h/M_*)^{\rm piv}$).  Since our
  results have been derived homogeneously from $z=0.2$ to $z=1$, our
  conclusions regarding the evolutionary trends of $M_*^{\rm piv}$
  should be robust. In contrast, B10 and \citet[][]{Moster:2010} infer
  evolutionary trends from a combination of SDSS data and other
  distinct surveys at higher redshifts. Therefore, their results have
  large systematic uncertainties with respect to $M_*^{\rm piv}$.
  These systematic errors are reflected in the B10 error bars (B10
  account for systematic errors and sample variance).  The
  \citet[][]{Moster:2010} results do not account for systematic error
  or sample variance from mock catalogs. Adding these two sources of
  error to the \citet[][]{Moster:2010} results would lead to errors on
  the pivot quantities of roughly the same order as B10.  Given these
  considerations: both the B10 and the \citet[][]{Moster:2010} results
  (provided larger errors are included) are consistent with our
  detection of an evolving $M_*^{\rm piv}$ and a constant
  $(M_h/M_*)^{\rm piv}$.}
\label{downsizing}
\end{figure*}

Second, we detect a strong redshift evolution in $M_{*}^{\rm piv}$ and
$M_{h}^{\rm piv}$ which is shown more explicitly in Figure
\ref{downsizing}. The detected evolution is such that both the halo
mass and the stellar mass for which the accumulated stellar growth of
the central galaxy has been the most efficient is smaller at late
times than at earlier times. This trend is a manifestation of at least
one meaning of the term ``downsizing''
\citep[][]{Cowie:1996,Brinchmann:2000,Juneau:2005}. Originally, the
term downsizing referred to the notion that the maximum K-band
luminosity of galaxies above a specific star formation rate threshold
decreases with time \citep[][]{Cowie:1996}. Since then, downsizing has
been widely employed to more generally describe the behaviour in which
a certain parameter that regulates galaxy formation decreases with
time (for recent discussions on downsizing see \citealt{Fontanot:2009}
and \citealt{Conroy:2009}). Our results show strong evidence for
downsizing in both the pivot halo mass and the pivot stellar mass. We
have already remarked in the previous section that a strong evolution
in $M_1$ and $M_{*,0}$ is seen in Figure \ref{param_z_evol}. Although
these two parameters are not strictly equal to $M_{*}^{\rm piv}$ and
$M_{h}^{\rm piv}$, they do have a strong impact on the location of the
pivot masses. Thus, the evolution seen in $M_1$ and $M_{*,0}$ is
directly related to the observed downsizing behaviour in the pivot
masses that is apparent in Figures \ref{shmr_z_evol} and
\ref{downsizing}. The pivot stellar mass evolves from $M_{*}^{\rm
  piv}=5.75 \pm 0.13 \times 10^{10}$\sm at $z=0.88$ to $M_{*}^{\rm
  piv}=3.55 \pm 0.17 \times 10^{10}$\sm at $z=0.37$ with an evolution
detected at 10$\sigma$. We note that all errors have been derived by
marginalizing over all other parameters.

The evolution in $M_{*}^{\rm piv}$ varies smoothly in the three
redshift bins, however the evolution in $M_{h}^{\rm piv}$ is less
smooth, in particular in the $z_2$ bin. We suggest that $M_{h}^{\rm
  piv}$ is more sensitive to sample variance than $M_{*}^{\rm
  piv}$. Indeed, the first order effect of sample variance is to
change the normalization of the SMF (see Paper I); this will directly
affect $M_1$ and thus $M_{h}^{\rm piv}$. In summary: $M_{h}^{\rm piv}$
is sensitive to sample variance between redshift bins whereas
$M_{*}^{\rm piv}$ is sensitive to systematic errors in stellar mass
measurements between redshift bins.

Finally, at high masses ($M_* >10^{11}$ M$_{\odot}$) there is an
interesting hint that the amplitude of the SHMR is decreasing at
higher redshifts, but we lack the statistics for a clear detection,
mainly due to the small volume probed by COSMOS.

%mhalo_ms.pro

%mhalo_ms.pro

% mhalo_ms.pro
\begin{figure*}[htb]
\epsscale{1.2}
\plotone{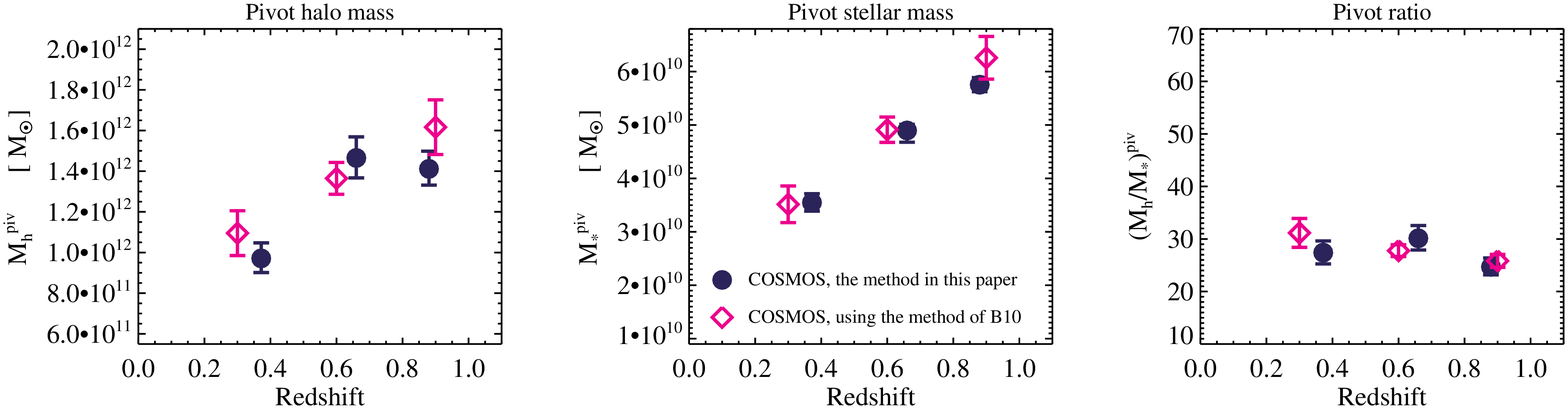}
\caption{An independent validation of the evolution of the pivot
  masses using the abundance matching method of B10. Blue circles show
  our detected evolution of the pivot quantities. Magenta diamonds
  show the evolution of the pivot quantities when the abundance
  matching method of B10 is applied to the COSMOS SMFs. We find that
  the analysis of B10 applied to the COSMOS results fully agrees with
  our claims concerning the evolution of the pivot quantities. We
  conclude that the detected downsizing behaviour of $M_{h}^{\rm piv}$
  and $M_*^{\rm piv}$ is robust to the methodology that is employed.}
\label{downsizing2}
\end{figure*}

\subsection{Comparison with previous work}\label{previous_work}

Figure \ref{shmr_z1} compares our $z_1$ SHMR to previous work on this
topic at low redshift. The general picture that emerges from Figure
\ref{shmr_z1} is one of remarkable broad agreement between various
methods on the overall shape of the SHMR. In detail however,
meaningful comparison between various surveys are severely limited by
systematic differences in stellar mass estimates ($\sim$0.25 dex
between different surveys). For this reason, we mainly focus on
qualitative comparisons in this section. All results have been
adjusted to our assumed value of $H_0=72$ km~s$^{-1}$~Mpc$^{-1}$ and
unless stated otherwise, halo masses are converted to $M_{200b}$
assuming a NFW profile and a \citet{Munoz-Cuartas:2010}
mass-concentration relation for a WMAP5 cosmology when necessary. All
results quoted here assume either a Kroupa or a Chabrier IMF. Since
the systematic shift in $M_*$ between these two IMFs is small ($\sim$
0.05 dex), we do not adjust for this difference. We also do not
attempt to correct for differences in the assumed cosmological model.

\subsubsection{Comparison with previous work: low redshift}\label{previous_work_low_z}

% --- Mandelbaum ---
\citet[][]{Mandelbaum:2006} have used g-g lensing in the SDSS to
measure halo masses for lens galaxies at $z\sim 0.1$. Since
\citet[][]{Mandelbaum:2006} only present their results as a function
of both $M_*$ and color, the data points in Figure \ref{shmr_z1} have
been re-computed as a function on $M_*$ only (Rachel Mandelbaum, priv
comm). Except perhaps for one data point at low $M_*$, these results
are in good agreement with ours.

% COSMOS groups
At high masses, an alternative method to probe the central SHMR is to
directly compare the halo masses of groups and clusters of galaxies to
the masses of their central galaxies. Since we are primarily
interested in $\Phi_c(M_{*}|M_h)$, it is critical, as much as
possible, to use {\em halo mass} selected samples of groups and
clusters for this type of comparison. Using samples selected on the
basis of the stellar mass of the central galaxy, for example, would
result in biased conclusions. In \citet[][]{Leauthaud:2010} we
presented a sample of X-ray groups ($M_{200b}\sim 10^{13}-10^{14}~{\rm
  M}_{\odot}$) in COSMOS for which we have calibrated the relationship
between halo mass and X-ray luminosity (L$_{\rm X}$) using g-g
lensing. The expected scatter in halo mass at fixed L$_{\rm X}$ is of
order 0.13 dex so the sample presented in \citet[][]{Leauthaud:2010}
is halo mass selected to a good approximation. In parallel, George et
al. (in prep) have constructed an algorithm to identify the central
galaxies of these groups and have used the weak lensing signal itself
to optimize the algorithm by maximizing the weak lensing signal at
small radial separations from the central galaxy. The grey squares in
Figure \ref{shmr_z1} report the stellar mass of the central galaxy
versus $M_h$ for groups at $0.22<z<0.48$ and with a high quality
flag. These data points are directly comparable to ours since we have
used exactly the same stellar masses and confirm that our results are
consistent with \citet[][]{Leauthaud:2010}.

% Hoekstra
We present a similar exercise for a sample of X-ray luminous clusters
(A68, A209, A267, A383, A963, A1689, A1763, A2218, A2390, A2219) from
\citet{Hoekstra:2007} with weak lensing masses from
\citet{Mahdavi:2008}. The central galaxies of these clusters have been
studied in detail by \citet[][]{Bildfell:2008}. Using the same stellar
mass code and assumptions as in this paper, we have computed stellar
masses for the central cluster galaxies using a compilation of optical
data provided by Chris Bildfell. The results are shown by the red
asterisk points in Figure \ref{shmr_z1}. Unfortunately, these mass
estimates are based on just two optical bands (B-band and R-band) and
as such will have larger uncertainties than the COSMOS stellar masses
used in this paper which are constrained with many more filters and
normalized to a near-IR luminosity.  We estimate that an additional
0.25 dex stellar mass uncertainty should be included when interpreting
these data points, which may account for their scatter in stellar mass
seen in Figure \ref{shmr_z1}.  With this cautionary note, plus the
additional caveat that this sample is not as homogeneously selected as
the groups from \citet[][]{Leauthaud:2010}, and that COSMOS is too
small to probe over-densities of these masses, the results are
nonetheless in good agreement with the extrapolation of our $z_1$
SHMR.

Both \citet[][]{Moster:2010} and B10 have presented constraints on the
SHMR and its redshift evolution by using the abundance matching
technique. A more detailed comparison with their work is presented in
$\S$ \ref{mos_comp} and $\S$ \ref{beh_comp}

% Conroy
\citet[][]{Conroy:2007} have used the kinematics of satellite galaxies
in SDSS and DEEP2 (Davis 2003) to probe the SHMR at $z\sim 0.06$ and
at $z\sim0.8$. Their low redshift results are shown by the blue
diamonds in Figure \ref{shmr_z1}. A 30$\%$ downward correction to halo
masses has been applied due to incompleteness effects as described in
their paper. In addition to systematic differences in stellar mass
estimates, direct comparisons with our results are further complicated
by the fact that our model describes $\langle
\log_{10}(M_*(M_h))\rangle$ whereas the \citet[][]{Conroy:2007}
results represent $\langle M_h(M_*)\rangle$. The two averaging systems
will yield different results: $\langle M_h(M_*)\rangle$ will be
increasingly biased low with respect to $\langle
\log_{10}(M_*(M_h))\rangle$ with increasing $\sigma_{\log M_*}$ and
for steeper values of the slope of the SHMR.

\citet[][]{More:2010} have used SDSS data to probe the halo masses of
$\sim 3900$ central galaxies in the range $0.02\leq z \leq 0.072$
using the kinematics of satellite galaxies. In their paper,
\citet[][]{More:2010} have analyzed red and blue galaxies
separately. For Figure \ref{shmr_z1} we have asked the authors to
provide the data for all central galaxies as a function of $M_*$,
irrespective of color, and also to convert their results to reflect
the mean log-relation ($\langle \log_{10}(M_*(M_h))\rangle$ as opposed
to $\langle M_h(M_*)\rangle$). Overall, there is some disagreement
between our results and \citet[][]{More:2010} regarding of the general
shape of the SHMR. Indeed, our results display a more strongly varying
power-law index compared to \citet[][]{More:2010}. This disagreement
is perhaps more apparent in the lower panel of Figure
\ref{shmr_z1}. Indeed, the \citet[][]{More:2010} results display a
fairly broad minimum in $M_{200b}/M_*$ whereas our results predict a
more strongly varying $M_{200b}/M_*$ ratio that reaches a minimum at
$M_*\sim 4.5\times 10^{10}~{\rm M}_{\odot}$. \citet[][]{More:2010}
have suggested that satellite kinematics may yield halo masses that
may be systematically higher by a factor of 2-3 than other methods at
low $M_*$. Lowering the \citet[][]{More:2010} results at low $M_*$
would certainly bring their results into better agreement with ours in
terms of the shape of the SHMR. \citet[][]{More:2010} also provide
estimates for $\sigma_{\rm log M_{*}}$. They find $\sigma_{\rm log
  M_{*}}=0.19_{-0.03}^{0.03}$ for red centrals and $\sigma_{\rm log
  M_{*}}=0.15_{-0.07}^{0.12}$ for blue centrals. Both of these values
are in broad agreement with our estimate of $\sigma_{\rm log
  M_{*}}\sim 0.23$ dex.
 
In the $z_1$ redshift bin, the COSMOS results are limited by
completeness to $M_*>10^{8.7}~M_{\odot}$. Nonetheless, it is of
interest to see how our results extrapolate to galaxies of even lower
stellar masses, even though measurements of $M_h$ for such low mass
galaxies are fraught with difficulties and for the most part limited
to the local volume. \citet[][]{Blanton:2008} have presented an effort
to address the very low mass SHMR (see their Figure 12) using
measurements of the maximum circular velocities from H$_{\rm I}$ disks
around isolated nearby dwarf galaxies. Since \citet[][]{Blanton:2008}
have applied criteria to specifically select isolated galaxies,
their sample should be dominated by central galaxies and so comparable
to our Figure \ref{shmr_z1}. Upon request, the authors provided us
with the full data set from Figure 12 in \citet[][]{Blanton:2008}
which we have reproduced in Figure \ref{shmr_z1}, including additional
data from \citet[][]{Springob:2005} and \citet[][]{Pizagno:2007} based
on H$_{\rm I}$ and H$\alpha$ rotation curves respectively. This data
compilation is restricted to galaxies that are isolated and with axis
ratios $b/a<0.5$ in order to minimize inclination uncertainties and
extinction corrections. Halo masses in \citet[][]{Blanton:2008} have
been estimated by assuming that the optical circular velocity, $V_{\rm
  opt}$, is equal to $V_{\rm max}$, the maximum circular velocity for
a NFW halo. $V_{\rm max}$ is then converted to the viral velocity,
$V_{200}$, using N-body calibrations from
\citet[][]{Bullock:2001}. For galaxy mass halos, $V_{\rm max}/V_{200}
\simeq 1.1-1.2$ under the assumption of no adiabatic contraction of
the dark matter due to galaxy formation. When incorporating adiabatic
contraction into Tully-Fisher analyses, \cite{Gnedin:2007} find a
factor of $\sim 2.5$ decrease in the inferred halo mass at fixed
stellar mass. Such a correction would put the Tully-Fisher constraints
into better agreement with our results.

To first order, there is relatively good agreement between our SHMR
and the data from \citet[][]{Blanton:2008}, albeit with a much larger
scatter in the Tully-Fisher based SHMR than predicted by our
results. In particular, the dwarf galaxy data points from the
\citet[][]{Geha:2006} sample are in good agreement with the
extrapolation of our SHMR to lower masses. At
$10^9~M_{\odot}<M_*<10^{11}~M_{\odot}$, however, there may be some
indication that the halo-masses inferred by \citet[][]{Blanton:2008}
are too high on average compared to our results with the possible
implication that the $V_{\rm opt}/V_{\rm 200}$ is larger than
$1.1-1.2$. The $V_{\rm opt}/V_{\rm 200}$ ratio contains information
about the relative importance of baryons versus dark matter on galaxy
scales: $V_{\rm opt}/V_{\rm 200} \geq 1.1-1.2$ would imply that the
baryons have modified the dark matter profile in the very inner halo
regions. \citet[][]{Dutton:2010} have used the
\citet[][]{Pizagno:2007} data in combination with a compilation of
prior work on the SHMR to place constraints on $V_{\rm opt}/V_{\rm
  200}$ for late type galaxies (the requirement that the
\citealt{Pizagno:2007} galaxies have sufficiently extended H$\alpha$
emission to yield a useful rotation curve implies that this is
primarily a late type sample). They find that $V_{\rm opt} \sim V_{\rm
  200}$, however, the normalization of the SHMR that they employ is
more similar to the \citet[][]{More:2010} results than to
ours. Therefore, a similar analysis as \citet[][]{Dutton:2010} but
applied to the \citet[][]{Pizagno:2007} data in combination with our
results would yield a higher $V_{\rm opt}/V_{\rm 200}$.

The agreement between the extrapolation of our SHMR to lower masses
and the \citet[][]{Geha:2006} sample is encouraging. In fact,
\citet{Busha:2010} found that a similar scaling continues to work down
to the faintest satellites of the Milky Way. However, Figure
\ref{shmr_z1} clearly reveals a lack of data at $M_*<10^9$\sm due to
the stellar mass completeness limits of current optical and IR
surveys. Pushing the SHMR down to $10^8~{\rm
  M}_{\odot}<M_*<10^{9}~{\rm M}_{\odot}$ using techniques such as
described in this paper is clearly an exciting avenue to explore and
will be facilitated by upcoming very deep optical and IR surveys such
at UltraVista and the Hyper Suprime Cam (HSC) survey on the Subaru
telescope\footnotemark[7]\footnotetext[7]{For example, the UltraVISTA
  survey of the COSMOS field will obtain IR imaging to Y=26.7, J=26.6,
  H=26.1, K$_s$=25.6, pushing low redshift stellar mass completeness
  limits to below $M_*=10^8$ M$_{\odot}$ at $z<0.3$. The HSC
  intermediate layer survey will cover 20 deg$^2$ to g=28.6, r=28.1,
  i=27.7, z=27.1, and Y=26.6}.

\subsubsection{Comparison with previous work: high redshift}

% Kroupa IMF. Virial masses. h=0.7.
% code to convert results: compare_catherine.pro
\citet[][]{Heymans:2006} have used g-g lensing to estimate halo masses
for a sample of 626 galaxies with $M_*>10^{10.5}$ M$_{\odot}$ and with
$0.2<z<0.8$ from the 0.25 deg$^2$ HST/GEMS survey
\citep[][]{Rix:2004}. \citet[][]{Heymans:2006} find $M_{\rm vir}/M_*
=53_{-16}^{+13}$ at a mean stellar mass of $M_*=7.2\times 10^{10}$
M$_{\odot}$. Converting their result to our assumed value of $H_0$ and
to $M_{200b}$ yields $M_{200b}/M_*\sim 58_{-17}^{+14}$. In a similar
mass and redshift range, our results produce $M_{200b}/M_*\sim 34$. We
note, however, that a careful comparison between our work and
\citet[][]{Heymans:2006} is limited by several differences in the way
the analyses have been performed. Firstly, \citet[][]{Heymans:2006}
fit a NFW profile to the g-g lensing signal and so the masses that
they measure will reflect the mean halo mass at fixed stellar mass,
which is different than our averaging system. Secondly,
\citet[][]{Heymans:2006} do not account for the contribution of
satellite galaxies to the g-g lensing signal and so they will tend to
overestimate halo masses. The lens sample of \citet[][]{Heymans:2006}
is roughly similar to our g-g bin4 in the $z_2$ redshift range: the
contribution to the g-g lensing signal from satellites for this sample
is shown in panel j of Figure \ref{sm_z2}. Given these two caveats our
results are in fairly good agreement.

\subsubsection{Comparison with Moster et al.}\label{mos_comp}

\citet[][]{Moster:2010} derive constraints on the redshift evolution
of the SHMR by abundance matching to the SDSS SMF of
\citet[][]{Panter:2007} at low redshift and to mass functions from the
MUNICS survey (detection limit K$\sim$19.5, area 0.28 deg$^2$)
\citep[][]{Drory:2004} and the GOODS-MUSIC sample (detection limit
K$\sim$23.5, area 143.2 arcmin$^2$) \citep[][]{Fontana:2006} at high
redshift. There are two main differences between
\citet[][]{Moster:2010} and our work. Firstly, we adopt the functional
form advocated by B10 for the SHMR which is sub-exponential at high
$M_*$. In contrast, the \citet[][]{Moster:2010} parametrization
asymptotes to a power-law at high $M_*$. According to B10, such a
parametrization may be problematic. Indeed, because the logarithmic
slope of the SHMR increases with increasing $M_*$, the best-fit
power-law for high mass galaxies will depend on the upper limit in the
available data for the SMF. Secondly, the \citet[][]{Moster:2010}
errors do not reflect possible systematic errors in stellar mass
estimates between \citet[][]{Panter:2007}, \citet[][]{Drory:2004}, and
\citet[][]{Fontana:2006}.

Although we find the same qualitative behaviour as
\citet[][]{Moster:2010}: $M_h/M_*$ decreases to a minimum at $M_h \sim
10^{12} M_{\odot}$ and then rises at higher masses, interestingly, our
results differ regarding the evolutionary trends of the SHMR. The two
parameters for which our conclusions differ in particular are
$M_*^{\rm piv}$ and $(M_h/M_*)^{\rm piv}$. In Figure \ref{downsizing}
(yellow dash-dot line) we show the evolution of $M_h^{\rm piv}$,
$M_*^{\rm piv}$, and $(M_h/M_*)^{\rm piv}$, as inferred from Table 7
in \citet[][]{Moster:2010}.

Our results agree with \citet[][]{Moster:2010} in terms of the
qualitative downsizing trend seen for $M_h^{\rm piv}$. However, it is
interesting to note that our measurements differ with respect to the
normalization of $M_h^{\rm piv}$. The exact origin of this discrepancy
remains unclear. In light of the results of B10, we hypothesize that
this discrepancy may be caused by the difference in the assumed
parametric form of the SHMR. In any case, further investigation
regarding the source of this discrepancy, though beyond the scope of
this paper, is clearly warranted.

Our conclusions differ with respect to $M_*^{\rm piv}$ and
$(M_h/M_*)^{\rm piv}$. Whereas our results suggest that $M_*^{\rm
  piv}$ increases with redshift and that $(M_h/M_*)^{\rm piv}$ remains
constant, in contrast, the \citet[][]{Moster:2010} results imply that
$M_*^{\rm piv}$ is constant with redshift and that instead,
$(M_h/M_*)^{\rm piv}$ increases with redshift. We hypothesize that
this discrepancy is simply due to the fact that the errors in
\citet[][]{Moster:2010} are likely to be under-estimated. Indeed,
accounting for sample variance with mocks as well as for systematic
differences in the relative stellar masses between
\citet[][]{Panter:2007}, \citet[][]{Drory:2004},
\citet[][]{Fontana:2006} would lead to similar errors as B10 (Figure
\ref{downsizing}, green diamonds). Indeed, the evolution that we
detect in $M_*^{\rm piv}$ is $\sim$0.21 dex from $z=0.37$ to $z=0.88$
which is similar to the expected error in stellar mass estimates
between different surveys. This could perhaps also explain why we
reach similar conclusions regarding the evolution of $M_h$ (which
should be less affected by systematic errors associated with $M_*$)
but not $M_*^{\rm piv}$ or $(M_h/M_*)^{\rm piv}$.

\subsubsection{Comparison with Behroozi et al.}\label{beh_comp}

The closest comparison with our work is B10 since we employ the same
functional form for the SHMR, the same halo mass function from
\citet[][]{Tinker:2008}, and we both account for the effect of scatter
in the SHMR and for sample variance in the data using mock catalogs.

B10 derive constraints on the redshift evolution of the SHMR by
abundance matching to the SDSS SMF of \citet[][]{Li:2009} at low
redshift and to mass functions from the FIDEL Legacy Project in the
extended Groth strip at higher redshifts
\citep[][]{Perez-Gonzalez:2008}. As a result of the fact that B10
combine data from distinct surveys, their systematic uncertainties on
the evolution of the SHMR are fairly large. We also note that there
are differences between the stellar mass estimates used in B10 and in
this paper (see $\S$ \ref{role_smf}) which will lead to normalization
differences in $M_*^{\rm piv}$ and $(M_h/M_*)^{\rm piv}$ for example.

Another difference between B10 and our work is the treatment of
satellite galaxies. In our model, the SHMR only applies to central
galaxies and satellites are modelled via $\nsat$. Indeed, we require a
more sophisticated treatment of satellites in order to fit the
clustering and the g-g lensing for which the satellite term plays a
larger role than in the SMF (the satellite term is sub-dominant at all
scales for the SMF). In contrast, B10 assume that the SHMR applies
also to satellite galaxies, on condition that the ``halo mass'' for
satellite galaxies is defined as the halo mass at the epoch when
satellites were accreted onto their parent halos (the ``infall mass'',
$M_{\rm infall}$). Thus, there could be subtle differences between the
two methods due to the treatment of satellite galaxies
\citep[for example, see discussion in][]{Neistein:2011}.

Figure \ref{downsizing} (green diamonds) shows the prediction from B10
for the pivot quantities. Our results are in striking agreement with
B10 with respect to the evolution of $M_{h}^{\rm piv}$. The errors
from B10 are larger for $M_*^{\rm piv}$ and $(M_h/M_*)^{\rm
  piv}$. Thus our results agree with B10 in terms of qualitative
evolutionary trends for $M_*^{\rm piv}$ and $(M_h/M_*)^{\rm
  piv}$. There is a normalization difference for $M_*^{\rm piv}$
between B10 and our results. However, this normalization offset is not
unexpected given systematic differences due to varying assumptions for
stellar mass estimates. We address this issue further below.

\subsubsection{Testing the redshift evolution of the pivot
  masses}\label{piv_mass}

Given the striking downsizing signal that we detect for the pivot
masses, we would like to check that this result is not an artifact of
the method we are using. As such we have also re-analysed the three
COSMOS stellar mass functions using the abundance matching method of
B10.  In this test, the method followed here is identical to that of
B10 but with different constraints on the stellar mass function.
Specifically, we use the three cosmos SMFs to constrain the redshift
evolution of the SHMR and do not include any SDSS data. The results
are shown in Figure \ref{downsizing2}. We find that the downsizing
signal for $M_{h}^{\rm piv}$ and $M_*^{\rm piv}$ is clearly detected
in the COSMOS data using the methods of B10. This provides an
independent test on our detected evolution of the pivot quantities and
suggests that the detected downsizing signal is robust to the
methodology that is employed.

This test also raises the interesting question of how the method used
in this paper (an HOD-based model that includes fits to clustering and
g-g lensing) compares to the method of B10 (abundance matching using
only the SMF). As can be seen in Figure \ref{downsizing2}, we find
that the two methods yield very similar results for the pivot
quantities. We do however find subtle differences in the actual SHMRs
between the two methods. Tracking down the cause of the exact
differences, although a very interesting question in itself, is beyond
the scope of this paper and we defer this study to follow-up work. For
the purposes of this paper, we will simply emphasize that the analysis
of B10 applied to the COSMOS results fully agrees with our claims
concerning the evolution of the pivot quantities.

%%%%%%%%%%%%%%%%%%%%%%%%%%%%%%%%%%%%%%%%%%%%%%%%%%%%%%%%%%%%%%%%%%%%%%%%%%%%%%

\subsection{The role of the stellar mass function}\label{role_smf}

In our analysis, the errors on the SMF are small compared to the
clustering and the lensing. It is always the case that a measurement
of a one-point statistic from a given set of data is more precise than
a measurement of a two-point (or higher) statistic. Thus, the SMF
plays an important role in constraining our parameter set. Therefore,
we investigate the SMF in further detail in this section, and in
particular, we show a more in-depth comparison with SDSS mass
functions.

\begin{figure*}[htb]
\epsscale{1.13}
\plotone{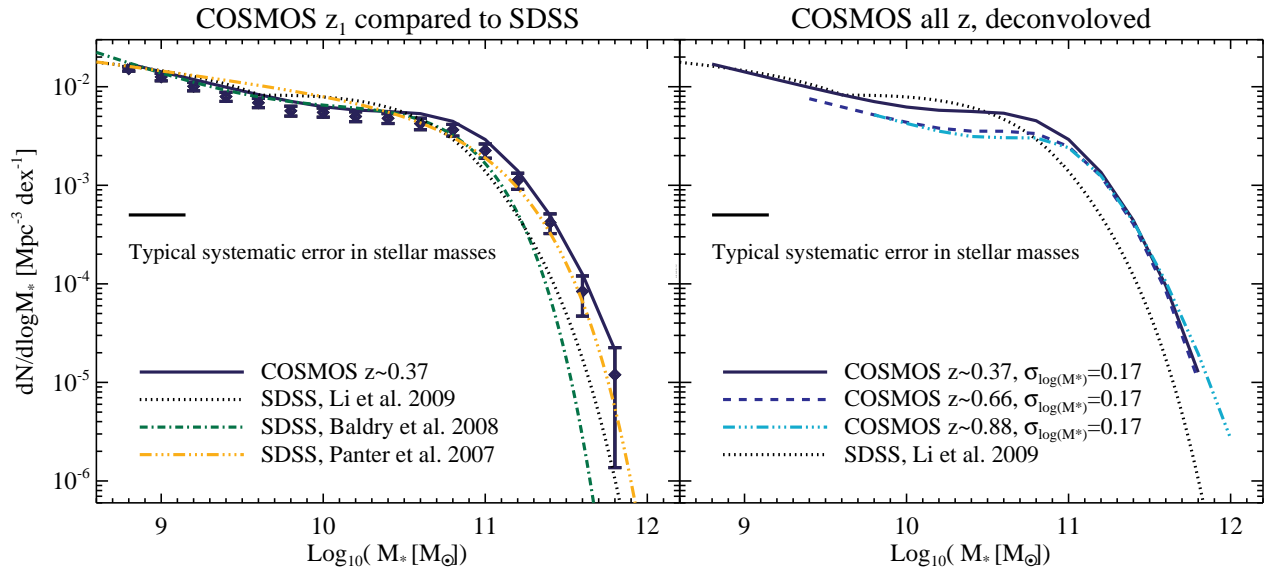}
\caption{Comparison of the COSMOS stellar mass functions to the SDSS
  mass functions of \citet[][]{Panter:2007}, \citet[][]{Baldry:2008},
  and \citet[][]{Li:2009}. In the left panel we show the COSMOS $z_1$
  mass function. The error bars represent the expected sample variance
  in COSMOS. In the right panel we show the mass functions for all
  three COSMOS mass functions, de-convolved to a common $\sigma_{\rm
    log M_{*}}$ of 0.17 dex. The differences between the COSMOS mass
  functions and the SDSS mass functions cannot be explained by
  Eddington bias and are therefore more likely to be due, for
  example, to differences in assumptions regarding stellar population
  and/or dust models.}
\label{compare_smf}
\end{figure*}

Figure \ref{compare_smf} shows the COSMOS mass functions compared to
various SDSS mass functions that have been commonly employed in the
literature \citep[][]{Panter:2007,Baldry:2008, Li:2009}. The main
difference that we may expect between the COSMOS mass functions and
the SDSS ones (besides sample variance and systematic error) is that
the high end of the mass function may be inflated due to a larger
value of $\sigma_{\rm log M_{*}}$ in COSMOS. To gauge how much the
COSMOS mass functions are affected by Eddington bias compared to SDSS,
we use our model to predict the COSMOS mass functions, de-convolved to
the expected scatter for SDSS ($\sigma_{\rm log M_{*}}\sim0.17$
dex). The results are shown in the right hand panel of Figure
\ref{compare_smf}. We find that the difference in scatter is not
significant enough to explain the differences between the COSMOS and
SDSS mass functions. It is more likely that the differences are due,
for example, to varying assumptions regarding stellar population and
dust models.
 
The difference between COSMOS and \citet[][]{Li:2009} corresponds
roughly ``by eye'' to a ``left/right'' shift along the X-axis
($\log_{10}(M_*^{\rm li}) \sim \log_{10}(M_*^{\rm cosmos})-0.2$). This
difference is within the estimated systematic uncertainties (0.25 dex
according to B10). However, this type of systematic shift will be
reflected directly in the SHMR (Figures \ref{shmr_z1} and
\ref{shmr_z_evol}) by a ``left/right'' shift along the X-axis and will
also affect the normalization of $M_*^{\rm piv}$ and $(M_h/M_*)^{\rm
  piv}$. This $\sim$0.2 dex shift would bring the normalization of B10
into closer agreement with our results. Tracking down the exact source
of this systematic shift is beyond the scope of this paper but it
could be associated with differences in the assumed dust
model\footnotemark[8]\footnotetext[8]{Our stellar masses used the
  \citet[][]{Charlot:2000} dust model whereas \citet[][]{Li:2009} use
  \citet{Blanton:2007}.} for example.

We conclude that in order to use the low-$z$ SDSS data as a $z\sim 0$
anchor to study the redshift evolution of the SHMR, a homogeneous
analysis of both the SDSS and the COSMOS data is critical. This will
be the focus of a future paper.

% compare_smf_funct_z.pro

\subsection{The total galaxy stellar content as a function of halo mass}\label{section_tot_sm_content}

\begin{figure*}[htb]
\epsscale{1.2}
\plotone{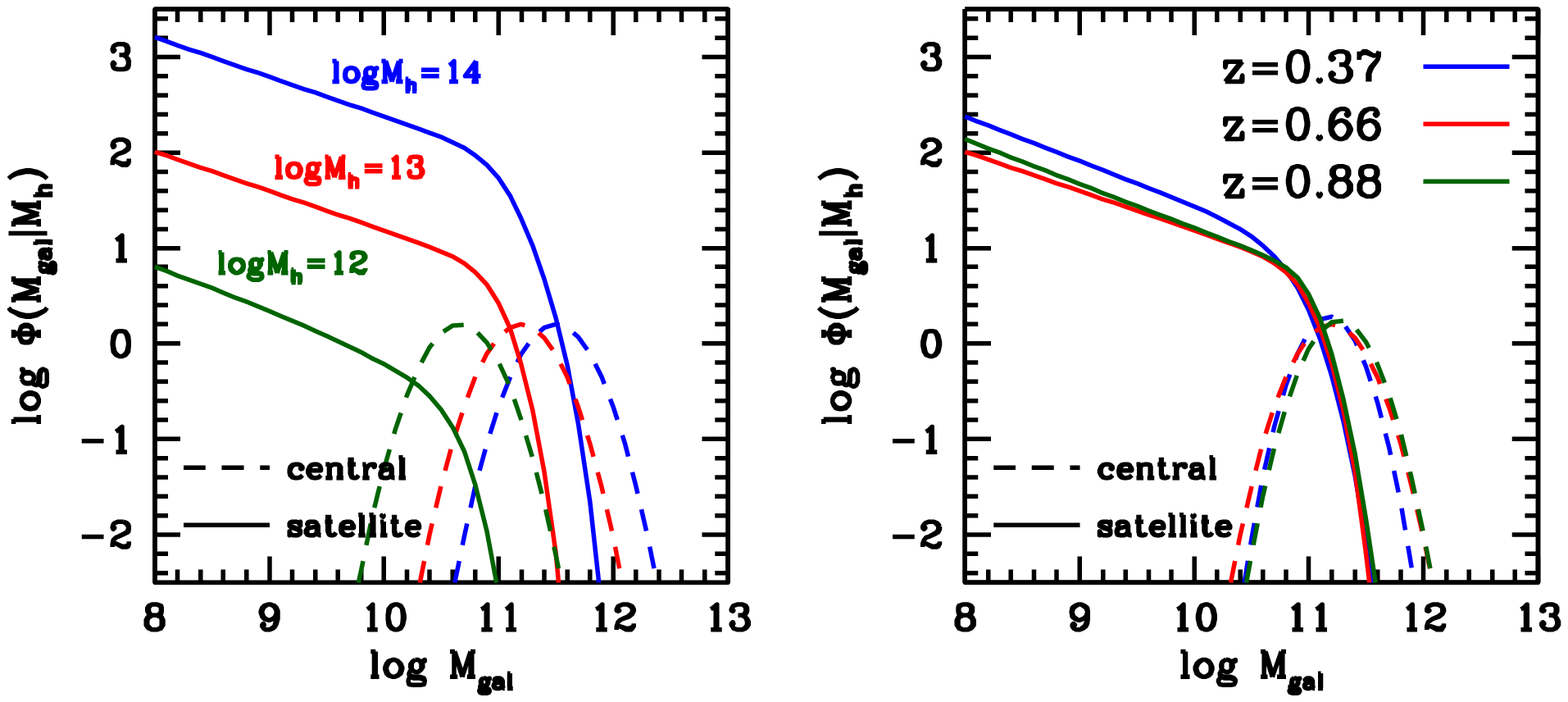}
\vspace{-9.5cm}
\caption{{\em Left panel}: The conditional stellar mass functions for
  central and satellite galaxies for three different halos
  masses. Results are for $z=0.66$. Solid curves represent
  $\Phi_s(M_*|M_h)$ while dashed curves represent
  $\Phi_c(M_*|M_h)$. The curves are normalized such that the total
  area is the mean number of galaxies at that halo mass. {\em Right
    panel}: Conditional stellar mass functions for halos of
  $M_h=10^{13}$ M$_\odot$ for our three redshift bins.}
\label{csmf}
\end{figure*}

Figure \ref{csmf} shows the conditional stellar mass functions, for
various halo masses and redshifts from our best fit model. We can use
these functions to calculate the {\em total} amount of stellar
material locked up in galaxies as a function of halo mass, noted
hereafter $M_*^{\rm tot}$ (see Equation 16 in Paper I). Investigating
$M_*^{\rm tot}$ is of interest because it reveals the efficiency with
which dark matter halos accumulate stellar mass from the combined
effects of in-situ star formation and accretion via merging.

One might worry that calculating the total amount of stellar material
locked up in satellite galaxies requires extrapolating our model
beyond the lower and upper stellar mass bounds for which our model has
been calibrated. As discussed in $\S$ \ref{previous_work_low_z}, the
extrapolation of our model is in good agreement with results from
\citet[][]{Blanton:2008} at low $M_*$ and with
\citet[][]{Hoekstra:2007} at high $M_*$. Thus, to first order, this
extrapolation does not appear unreasonable. Let us therefore make the
assumption that the extrapolation of our SHMR is not wildly
incorrect. We will now investigate which mass range of satellite
galaxies contribute most to $M_*^{\rm tot}$.

From Figure \ref{csmf}, it is clear that satellite galaxies are a
subdominant component of the total stellar mass at $M_h=10^{12}$
M$_{\odot}$. Our present concern is therefore only relevant for
$M_h>10^{12}$ M$_{\odot}$. Let us consider the total stellar mass
associated with satellite galaxies as a function of $M_h$ in a fixed
stellar mass bin: $M_{*}^{\rm tot,sat}(M_h|M_*^{t1},M_*^{t2})$. As
shown in Paper I, the expression for $M_{*}^{\rm
  tot,sat}(M_h|M_*^{t1},M_*^{t2})$ is given by:

\begin{equation}
M_{*}^{\rm tot,sat}(M_h|M_*^{t1},M_*^{t2})=\int_{M_*^{t1}}^{M_*^{t2}}\Phi_s(M_*|M_h)M_*{\rm d}M_* .
\label{tot_msat_1}
\end{equation}

We have tested how $M_{*}^{\rm tot,sat}(M_h|M_*^{t1},M_*^{t2})$ varies
with the integral limits, $M_*^{t1}$ and $M_*^{t2}$. We find that at
fixed halo mass, most of the stellar mass associated with satellite
galaxies arises from a relatively narrow range in stellar mass. In
particular, for halos with $M_h>10^{12}$ M$_{\odot}$, the bulk of
$M_*^{\rm tot,sat}$ is built from satellite galaxies in the range
$10^{10} M_{\odot}<M_{*}<10^{11} M_{\odot}$. Therefore, provided that
the extrapolation of our model is not wildly incorrect, the bulk of
$M_*^{\rm tot,sat}$ arises from satellites that are within the tested
limits of our model.

\begin{figure*}[htb]
\epsscale{1.2}
\plotone{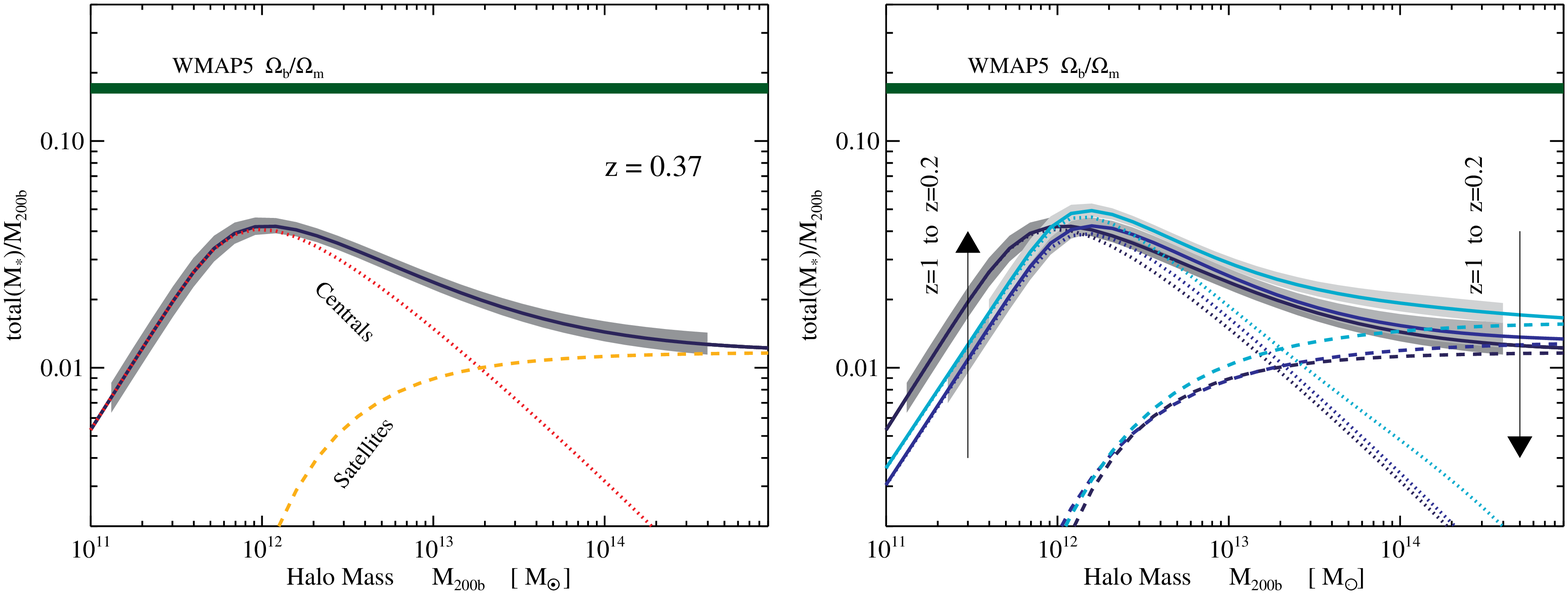}
\caption{Total stellar content locked up in galaxies as a function of
  halo mass compared to the cosmic baryon fraction measured by the
  Wilkinson Microwave Probe \citep[WMAP5;
  $f_b=\frac{\Omega_b}{\Omega_m}=0.171\pm
  0.009$;][]{Dunkley:2009}. {\em Left panel}: Our prediction from the
  $z_1$ bin ($z\sim 0.37$). {\em Right panel}: Our three redshift
  bins. $z_1$ is shown by the solid dark blue line, $z_2$ is shown by
  the blue line, and $z_3$ is shown by the turquoise line. Dotted
  lines show the contribution to $M_*^{\rm tot}$ from the central
  galaxy and dashed lines show the contribution from satellite
  galaxies.  Shaded regions represent the errors on $M_*^{\rm
    tot}/M_{200b}$. $M_*^{\rm tot}$ is dominated by the central galaxy
  at $M_h < 2\times 10^{13}$ M$_{\odot}$ and by satellites at $M_h >
  2\times 10^{13}$ M$_{\odot}$.}
\label{tot_sm_content}
\end{figure*}

Having underlined this caveat, we have calculated $M_*^{\rm tot}$
using the best fit parameters for each of the three redshift bins and
the results are shown in Figure \ref{tot_sm_content}. This Figure will
be discussed in detail in the following section.

% plot_total_sm_from_hod_with_errors.pro

% plot_total_sm_from_hod_with_errors.pro

%%%%%%%%%%%%%%%%%%%%%%%%%%%%%%%%%%%%%%%%%%%%%%%%%%%%%%%%%%%%%%%%%%%%%%%%%%%%%%
%     DISCUSSION
%%%%%%%%%%%%%%%%%%%%%%%%%%%%%%%%%%%%%%%%%%%%%%%%%%%%%%%%%%%%%%%%%%%%%%%%%%%%%%

\section{Discussion}\label{discussion}

Using a self-consistent framework to simultaneously fit the g-g
lensing, spatial clustering, and number densities of galaxies in
COSMOS, we have obtained a robust characterization of the evolving
relationship between stellar mass and halo mass over two orders of
magnitude in $M_*$.  The nature of this relationship, shown in Figure
\ref{tot_sm_content}, is not only a byproduct of cosmic mass assembly
but is also shaped by the physical processes that drive galaxy
formation, ultimately providing valuable constraints on both.  In this
section we begin by discussing various processes that shape the form
of $M_*/M_h$ versus $M_h$. We will then introduce a simple framework
for interpreting evolution in this relation by considering the
relative growth of stellar mass as compared to the growth of dark
matter halos. Finally, we will discuss the observed evolution of the
pivot quantities and we will show how a constant pivot ratio may imply
that the mechanism responsible for the shut-down of star formation in
massive galaxies may have a physical dependence on $M_*/M_h$.

\subsection{The total stellar mass content of dark matter halos}

Figure \ref{tot_sm_content} separates the stellar content of the
average dark matter halo into a contribution from the central galaxy
and a contribution from the sum of satellite galaxies.  Central
galaxies show a $M_*/M_h$ ratio that rises steeply to a maximum at
$M_h\sim 10^{12}$ M$_{\odot}$ before decreasing somewhat more
gradually in halos of higher mass.  The fact that halos above this
mass scale (at the redshifts considered) have cooling times longer
than their dynamical times has been invoked by modelers for some time
to help explain why cooling and star formation shut down at the
highest masses, with some refinement due to the presence of so-called
``cold-mode accretion''
\citep[][]{Birnboim:2003,Keres:2005,Birnboim:2007,Cattaneo:2006}.  We
will return to the evolution of this mass scale at a later point in
the discussion.

Central galaxies strongly dominate the total stellar mass content at
$M_h \lesssim 2 \times 10^{13}$ M$_{\odot}$ (``the central dominated
regime''), including at the peak mass, $M_h \sim 10^{12}$ M$_{\odot}$,
while the stellar mass in satellites dominates at $M_h \gtrsim 2
\times 10^{13}$ M$_{\odot}$ (``the satellite dominated regime'').

The transition between the two regimes is driven by the steep decline
in $M_{*}^{\rm cen}/M_h$ at $M_h > 10^{12}$ M$_{\odot}$. This decline
occurs as the contribution from satellites begins to rise. One might
then naturally ask if central galaxies in group-scale halos experience
stunted growth simply because stellar mass is accumulating within the
halo in the form of satellite galaxies, instead of merging onto the
central galaxy. Figure \ref{tot_sm_content} reveals that this is not
the case. Indeed, the solid line in this Figure demonstrates that {\em
  the total stellar mass fraction of halos declines at $M_h>10^{12}$
  M$_{\odot}$}. Thus, even if all satellite galaxies were allowed to
rapidly coalesce at the center of the potential well, the central
galaxies of group-scale halos would still have lower $M_*/M_h$ ratios
than those in halos of $M_h\sim 10^{12}$ M$_{\odot}$. Thus, we
conclude that dark matter halos globally decline in the efficiency by
which they accumulate stellar mass at $M_h>10^{12}$ M$_{\odot}$.

We note that in massive halos, the intra-cluster light (ICL) (not
accounted for in this analysis) is estimated to contribute an
additional 20--30\% of the total stellar mass
\citep[][]{Feldmeier:2004,Zibetti:2005,Gonzalez:2005,Krick:2006}. Adding
this to the satellite component in Figure \ref{tot_sm_content} does
little to bridge the factor of 2--4 gap in $M_*^{\rm tot}/M_h$ between
the satellite component in high-mass halos and centrals at the peak
mass.

\subsection{The role of galaxy mergers in determining the shape of
  $M_*^{\rm tot}/M_h$}

The majority of the total mass in an average dark matter halo is built
from halo-halo mergers with mass ratios above 1:10
\citep[e.g.,][]{Hopkins:2010}. At halo masses below $10^{12}$
M$_{\odot}$, the steep rise in $M_*^{\rm tot}/M_h$ with $M_h$ implies
that the typical {\em stellar} mass ratio of galaxy mergers will be
less than the typical mass ratio of the dark matter halos hosting
these galaxies.  In other words, major halo mergers are minor galaxy
mergers in this regime.  Thus, the accumulation of stellar mass
through the effects of merging will be limited compared to the growth
in total mass of such halos.  The steep rise of $M_*^{\rm tot}/M_h$
must therefore reflect the greater importance of star formation at
masses below $M_h \sim 10^{12}$ M$_{\odot}$ over assembly from galaxy
mergers \citep{Bundy:2009}. Similar conclusions have also been reached
by \citet[][]{Conroy:2009} (see their Figures 2 and 3 in particular).

Simple arguments suggest, however, that once halos grow past the pivot
mass and in the absence of significant star formation, $M_*^{\rm
  tot}/M_h$ should dip below the peak value since these halos can only
grow by merging with halos with lower values of $M_*^{\rm tot}/M_h$.
At slightly higher mass, the decline in $M_*^{\rm tot}/M_h$ now means
that stellar mass ratios are enhanced with respect to halo mass
ratios, and the trend must reverse again.  This repeating pattern
should cause a flattening of $M_*^{\rm tot}/M_h$ above the pivot mass.
While this behavior is certainly apparent in Figure
\ref{tot_sm_content}, our different redshift bins also reveal that at
fixed mass among high-mass halos ($M_h > 4 \times 10^{13}$), the total
stellar mass content {\em declines} at later epochs. We speculate that
this trend could arise from the smooth accretion of dark matter, which
brings no new stellar mass, and amounts to as much as 40\% of the
growth of dark matter halos \citep[e.g.,][]{Fakhouri:2010}.  One way
to test this hypothesis would be to populate a $z=0.88$ N-body
simulation with our $z_3$ HOD and evolve the subhalo and halo
populations to $z=0$, assuming no star-formation. This would reveal
the amount of stars that are acquired through mergers in this redshift
range \citep[][]{Zentner:2005}.  We note that the destruction of
satellites and a growing ICL component could also contribute to this
trend given that the ICL at $z=0$ could make up 20--30\% of the
stellar content of massive halos, roughly the amount by which
$M_*^{\rm tot}/M_h$ declines over our redshift range.

\subsection{The pivot quantities and the quenching of star formation
  in central galaxies}

The location at which halos reach their maximum accumulated stellar
mass efficiency is encoded by the pivot mass quantities, $M_{*}^{\rm
  piv}$, $M_{h}^{\rm piv}$, and $(M_h/M_*)^{\rm piv}$. While we
observe downsizing trends for both $M_{*}^{\rm piv}$ and $M_{h}^{\rm
  piv}$, the co-evolution of these two parameters leaves
$(M_h/M_*)^{\rm piv}$ constant with redshift. This can be seen in the
right panel of Figure \ref{tot_sm_content}: the pivot ratio evolves
very little over our redshift range, while the mass scale of the peak
($M_{h}^{\rm piv}$) does evolve downward by nearly a factor of
2. Given the low satellite content of halos at these masses, merging
is not likely to play a dominant role in driving growth in $M_*$ below
the pivot peak, arguing instead that the regulation of star formation
is key to understanding this behavior. The physical process that sets
the pivot peak at $M_h=10^{12}$ M$_{\odot}$, and drives the subsequent
decline in $M_{*}^{\rm cen}/M_h$ at higher halo masses, must be linked
to the shut-down of star formation in central galaxies. In the
reminder of this discussion, we will focus on interpreting the
downsizing behaviour of the pivot quantities in this context.

\subsection{A simple model for interpreting evolution in $M_*/M_h$
  versus $M_h$}

A complete and comprehensive interpretation of our results requires
modelling and accounting for dark matter accretion histories, galaxy
merger rates, and star formation rates (SFR) as a function of redshift
(for example, see \citealt{Conroy:2009}). Nonetheless, we will
introduce some toy models based on simple arguments to provide a first
interpretation of our results. Our goal here is to evaluate our
results in the context of other observations and theoretical work on
galaxy formation models and to set the stage for a more detailed
treatment in subsequent work. We begin with a general treatment of
evolution in the SHMR and then will focus on applying this treatment
to interpret the evolution we observe in the pivot quantities.

\begin{figure}[htb]
\epsscale{1.2}
\plotone{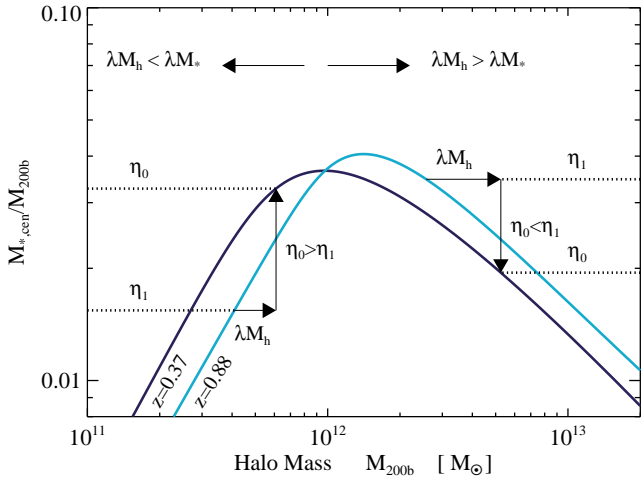}
\caption{Schematic illustration of our results in terms of galaxy mass
  assembly versus halo growth. The turquoise line represents our $z_3$
  result and the dark blue line represents our $z_1$ result. From the
  relative positions of these two curves we can infer that from $z\sim
  0.88$ to $z \sim 0.37$ and at $M_h<10^{12}$ M$_{\odot}$, the stellar
  mass of the central galaxy has experienced a stronger growth in
  proportion to the growth of the dark matter. On the contrary, at
  $M_h>10^{12}$ M$_{\odot}$, the stellar mass of the central galaxy
  has experienced a more mild growth in proportion to the dark
  matter.}
\label{illustration_eta}
\end{figure}

The physical basis for evolution in $M_*/M_h$ versus $M_h$ must be
considered carefully because the stellar mass and associated halo mass
of a galaxy can evolve independently, depending on the mass scale
involved and on processes including merging, smooth (diffuse) dark
matter accretion, star formation, and even tidal stripping.  The sum
of these processes on the growth of $M_h$ and $M_*$ shapes the
behavior of $M_*/M_h$ versus $M_h$ in different ways, as illustrated
by the schematic diagram in Figure \ref{illustration_eta}. Here we let
$\eta_1$ represent the ratio $M_*/M_h$ at redshift $z_{\rm high}$ and
$\eta_0$ represent the equivalent ratio at redshift $z_{\rm low}$ with
$z_{\rm high}>z_{\rm low}$.  We further consider a dark matter halo of
mass $M_h$ that has grown by a relative factor of $\lambda _{M_h}$
from $z_{\rm high}$ to $z_{\rm low}$.  We can write that $M_h(z_{\rm
  low})=\lambda_{M_h}\times M_h(z_{\rm high})$.  Characterizing growth
in the stellar mass of the central galaxy (although similar arguments
apply to $M_*^{\rm tot}$) by a factor of $\lambda_{M_*}$, we can
simply write that:

\begin{equation}
\lambda_{M_*} = \frac{\eta_0}{\eta_1}\times\lambda_{M_h}.
\end{equation}

If $\eta_0>\eta_1$, we infer that $\lambda_{M_*}>\lambda_{M_h}$
and that the stellar mass has experienced a stronger relative amount
of growth compared to that of the dark matter from $z_{\rm high}$ to
$z_{\rm low}$. If on the contrary, $\eta_0<\eta_1$ then the relative
growth of the stellar mass is less than the dark matter.  Note that
this schematic view demonstrates that identical curves for $M_*/M_h$
versus $M_h$ at different redshifts do not necessarily indicate a
lack of evolution, since $\lambda_{M_h}$ is always greater than zero.

% make_eta_illustration_figure.pro

We now apply this intuitive framework to the evolution observed in
Figure \ref{tot_sm_content}.  Considering values of $\eta_1$ and
$\eta_0$ applied to the total stellar mass curves in Figure
\ref{tot_sm_content}, we see that below $M_h = 10^{12}$, the
fractional growth in stellar mass outweighs the growth in halo mass.
This reflects the greater importance of star formation at low masses
over assembly from galaxy mergers, the same conclusion reached above
by simply considering the shape of the SHMR.  This evolutionary trend
reverses above $M_h = 10^{12}$ M$_{\odot}$, consistent with the notion
that star formation is largely shut down in centrals above this mass.

\subsection{Understanding the evolution of the pivot quantities}

We now apply these simple arguments to the evolution in the pivot
quantities. We focus only on central galaxies, neglecting the minor
contribution from satellites near the pivot mass. The aim here is to
explore several simple models for the quenching of star formation and
to investigate which models might reproduce the observed evolution of
the pivot quantities, namely, a pivot halo and stellar mass that
decrease at later epochs (downsizing) but leave the pivot ratio
constant.

\begin{figure*}[htb]
  \epsscale{1.15} \plotone{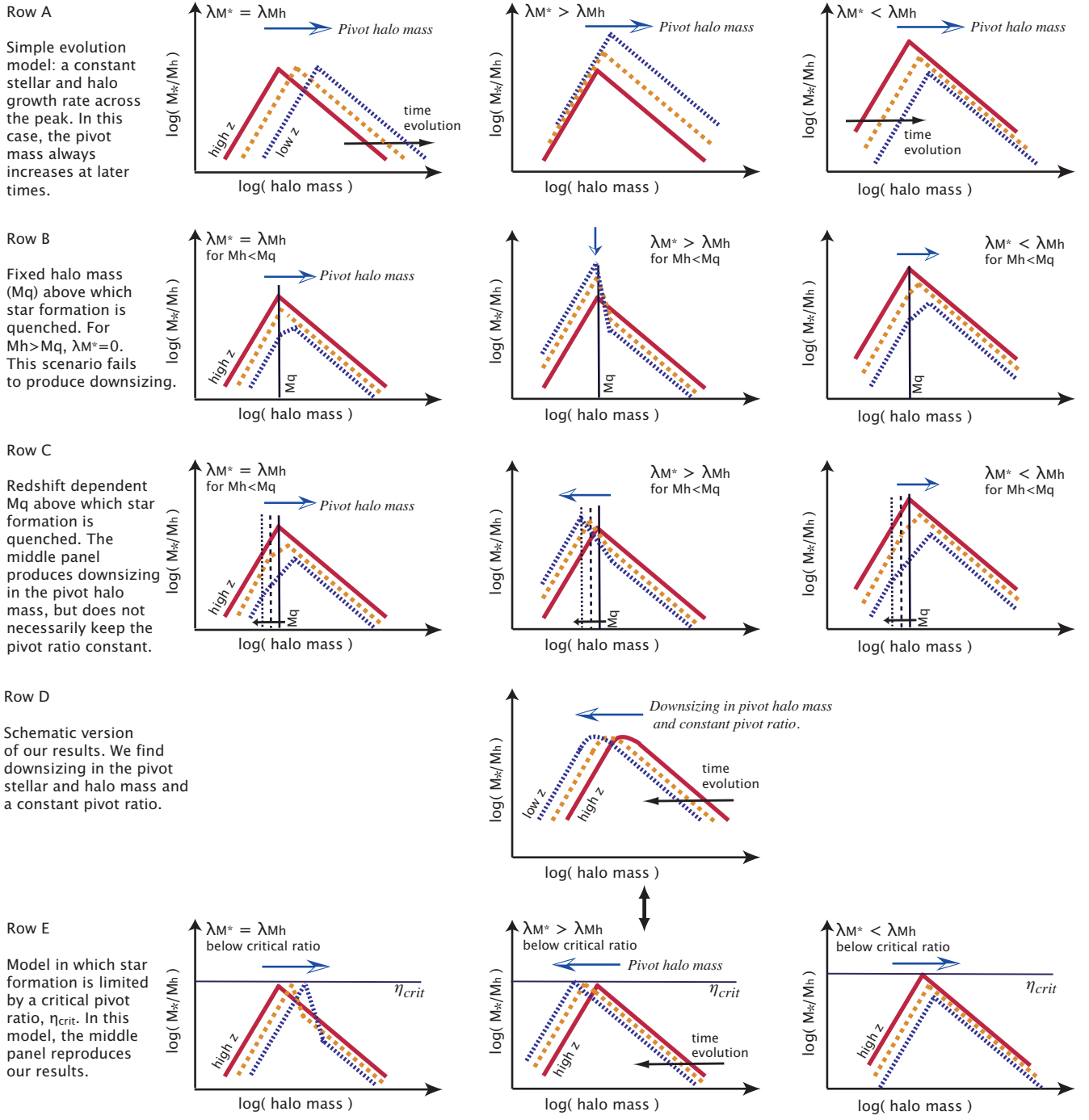}
  \caption{A schematic (and simplistic picture) of how $M_*/M_h$ (for
    central galaxies) varies with redshift for different quenching
    models and for various prescriptions for stellar growth and halo
    growth (parametrized here by $\lambda_{M_*}$ and $\lambda_{M_h}$
    respectively). Our $z_3$ (z=0.88) relation is represented by the
    solid red line. In this picture, we are not interested in
    understanding why the high redshift relation has the particular
    form that is observed, but simply in predicting roughly what the
    evolution of this relation should look like given various
    quenching models. The orange dashed line and the blue dotted line
    show how we expect the SHMR to evolve with time. In Row A we
    consider a model without any quenching and where $\lambda_{M_*}$
    and $\lambda_{M_h}$ are constant with redshift and halo mass. In
    this model, the pivot halo mass will increase at later epochs
    which is not what we observe. In Row B, we consider a model in
    which star formation is quenched above a fixed halo mass,
    $M_q$. This model fails to produce the observed downsizing
    behaviour for $M_{h}^{\rm piv}$. In Row C, we consider a model
    where $M_q$ decreases at lower redshifts. In this case when
    $\lambda_{M_*}>\lambda_{M_h}$ below $M_q$ (Row C, middle panel),
    $M_{h}^{\rm piv}$ follows a downsizing trend. However, in this
    scenario, a pivot ratio that is constant with redshift requires a
    fine tuning between $\lambda_{M_*}$ and the rate at which $M_q$
    declines. In Row E, instead of assuming that star formation is
    quenched at a fixed halo mass, we now assume that star formation
    is quenched at a fixed critical $M_*/M_h$ ratio ($\eta_{\rm
      crit}$). Given this assumption and if
    $\lambda_{M_*}>\lambda_{M_h}$ below $\eta_{\rm crit}$ (a
    reasonable assumption given estimates for halo growth and star
    formation at these scales), we can qualitatively reproduce our
    main results (compare the middle panel of Row D to the middel
    panel of Row E).}
\label{ms_mh_evol_illustration}
\end{figure*}

Shown schematically in Figure \ref{ms_mh_evol_illustration}, we
consider how our high redshift ($z=0.88$) SHMR relation would evolve
towards lower redshifts under several prescriptions for stellar and
halo growth.  We begin with no assumptions about the SFR but adopt a
halo growth rate ($\lambda_{M_h}$) that is roughly constant over the
mass range spanned by the peak. This assumption is well justified by
dark matter mass accretion rates derived from N-body simulations
\citep[][]{Wechsler:2002,McBride:2009,Fakhouri:2010a}. For example,
\citet{Fakhouri:2010a} find that $\dot{M_h}/M_h$ is only weakly
dependent on halo mass with $\dot{M_h}/M_h\propto M_h^{0.1}$.

% Fig made in illustrator

We now consider several different quenching models and investigate
their impact on the redshift evolution of the pivot quantities.

\begin{itemize}
\item No quenching model: to begin with, we consider a model with no
  quenching of star formation and in which the stellar growth rate
  ($\lambda_{M_*}$) is constant over the halo mass range spanned by
  the peak (Row A in Figure \ref{ms_mh_evol_illustration}). We
  consider the redshift evolution of the SHMR for three values of
  $\lambda_{M_*}$ defined with respect to $\lambda_{M_h}$.  In all
  three cases, this model leads to an increase in $M_{h}^{\rm piv}$
  with time (contrary to what we observe). We can therefore conclude
  that $\lambda_{M_*}$ must vary with $M_h$, not a surprise given the
  expectation that the SFR shuts down above $M_{h}= 10^{12}$
  M$_{\odot}$.
\item Fixed halo mass for quenching: we next consider a model in Row B
  in which star formation is quenched as galaxies cross above a fixed
  halo mass, $M_q$.  However, this instantaneous quenching model fails
  to reproduce (not surprisingly) the downward evolution of
  $M_{h}^{\rm piv}$ and also yields evolution in the pivot ratio,
  which is not detected.
\item Redshift dependent halo mass for quenching: Row C shows a model
  in which $M_q$ shifts downward with time. This model leads to
  downsizing in the pivot halo mass if $\lambda_{M_*}>\lambda_{M_h}$
  below $M_q$ (Row C, middle panel). However, in order to keep the
  pivot ratio fixed in this scenario, the growth rate,
  $\lambda_{M_*}$, must be tuned with respect to the rate at which
  $M_q$ declines.  This would require a fortuitous coincidence, but
  obviously cannot be dismissed as an explanation.
\item Critical $M_*/M_h$ ratio for quenching: we finally consider an
  alternative scenario in which star formation is limited by a
  critical mass ratio, $\eta_{\rm crit} \equiv M_*/M_h \approx 0.04$
  (Row E). In this model, we can qualitatively reproduce our main
  results, including the downsizing trends in the pivot stellar and
  halo mass, and, by construction, a constant pivot ratio set by
  $\eta_{\rm crit}$ (compare the middle panel of Row D to the middle
  panel of Row E). However, in order to produce downsizing behaviour
  in this model, $\lambda_{M_*}$must be larger than $\lambda_{M_h}$
  below $\eta_{\rm crit}$. It is important to note that if
  $\lambda_{M_*}\leq \lambda_{M_h}$ below $\eta_{\rm crit}$, this
  model would fail to produce downsizing.
\end{itemize}
 
The model explored in Row E seems a promising and simple mechanism
that can explain the observed evolution of the pivot quantities. We
now test if observations are consistent with the requirement that
$\lambda_{M_*}>\lambda_{M_h}$ below $M_h\sim10^{12}$\msun. A halo of
mass $M_h \sim 2\times 10^{11}$\msun~ grows by a factor of
$\lambda_{M_h}\sim 1.4$ from $z=0.88$ to $z=0.37$
\citep{Fakhouri:2010a}. Star formation rates as a function of $M_*$
and redshift have recently been measured by \citet{Noeske:2007},
\citet{Cowie:2008}, and \citet[][]{Gilbank:2010}. These measurements
indicate that galaxies of mass $M_*\sim 10^{10}$\msun~ grow by roughly
a factor of $\lambda_{M_*}=2-3$ from $z=0.88$ to $z=0.37$. Therefore,
at $z<1$ and for $M_h<M_h^{\rm piv}$, current estimates for halo
growth coupled with estimates for stellar growth are indeed consistent
with our observation that stellar mass has experienced a stronger
relative amount of growth compared to that of the dark matter
($\lambda_{M_h}<\lambda_{M_*}$ below $\eta_{\rm crit}$).

We further note that the relatively weak dependence of $\lambda_{M_*}$
on $M_h$ required to maintain a constant slope in $M_*/M_h$ versus
$M_h$ with redshift (again for $M_h < M_{h}^{\rm piv}$) is implied by
the weak SSFR-$M_*$ relation observed for galaxies with $M_*<M_*^{\rm
  piv}$ (a typical power-law fit gives SSFR$ \sim M_*^{0.4}$). This
could explain why $\beta$ (the low mass slope of the SHMR) is observed
to remain remarkably constant at $\beta=0.46$ at $z<1$.

Given the simple arguments outlined above, we argue that the fact that
the pivot ratio remains constant may suggest that {\em star formation
  is fundamentally limited by a critical mass ratio}, $\eta_{\rm crit}
\equiv M_*/M_h \approx 0.04$. A very elegant and compelling
consequence of this model is that the observed downsizing trends in
$M_*^{\rm piv}$ and $M_h^{\rm piv}$ can be automatically explained
given the observation that $\lambda_{M_*}>\lambda_{M_h}$ below
$M_h\sim 10^{12}$ M$_{\odot}$. Previous work by \citet{Bundy:2006} on
the evolution of the galaxy stellar mass function has found a similar
downsizing trend for the ``transition mass'', $M_{tr}$, which is
defined in terms of $M_*$ by the declining fraction of star-forming
galaxies at the highest masses. The transition mass from
\citet{Bundy:2006} evolves from $M_{tr}\sim 9\times 10^{10}$
M$_{\odot}$ from $z\sim 0.9$ to $M_{tr}\sim 5\times 10^{10}$
M$_{\odot}$ from $z\sim 0.6$. This transition mass is similar to
$M_*^{\rm piv}$, reinforcing the notion that the pivot mass marks the
end of rapid star formation among halos. The obvious difference
between \citet{Bundy:2006} and this paper is that our work adds a key
missing ingredient, which is the evolution of the pivot halo mass and
the pivot ratio. If the quenching of star formation depends on a
critical $M_*/M_h$ ratio then the fact that low mass galaxies grow
more rapidly than dark matter below the pivot scale provides a {\em
  simple explanation for the observed downsizing in the sites of star
  formation} observed by studies such as \citet{Bundy:2006}.

\subsection{Physical mechanisms that might depend on $M_*/M_h$}

In the previous section, we demonstrated that a constant pivot ratio
provides important clues concerning the physical mechanisms that
quench star formation. We now discuss possible mechanisms that might
tie quenching to $M_*/M_h$.

The notion of a fixed maximum stellar-to-dark matter ratio, $\eta_{\rm
  crit}$, has been relatively unexplored in the
literature. Theoretical arguments tend to favor a relatively fixed (if
broad) critical halo mass \citep[see][]{Birnboim:2003} at $z \lesssim
1$, with significant modifications from cold-mode accretion occurring
mostly at higher redshifts. But, quenching at fixed halo mass alone is
not sufficient to reproduce the local stellar mass function and
red-sequence fraction, and also fails from simple arguments to produce
downsizing in the pivot masses, as shown by Row B of Figure
\ref{ms_mh_evol_illustration}. As a result, most semi-analytic models
include a quenching channel initiated by mergers or disk instabilities
among galaxies in halos below the critical halo mass threshold
\citep[e.g.,][]{Bower:2006,Croton:2006,Cattaneo:2006}. Above the
critical halo mass, once a quasi-static halo of hot gas has formed,
low-luminosity feedback (i.e., radio-mode AGN feedback) is often
invoked to prevent cooling and star-formation in massive halos at late
times. Using simple arguments, we have shown that mergers do not
provide a likely explanation for determining the scale of the pivot
mass. We therefore focus on AGN feedback and disk instabilities as
possible mechanisms.

It is common practice to consider low-luminosity AGN feedback only in
halos above a fixed halo mass. However, in practice, efficient AGN
feedback is more complex and requires at least two ingredients. First,
a quasi-static halo of hot gas must exist to which the AGN jets can
couple. However, AGN feedback also requires a sufficiently large
black-hole to produce jets powerful enough to initiate this
coupling. Given that gas cooling rates scale roughly with halo mass
and that black hole mass scales roughly with galaxy mass, via the
$M_*-\sigma$ relation \citep[][]{Gebhardt:2000,Ferrarese:2000}, it is
reasonable to speculate that AGN feedback efficiency would depend on
$M_*/M_h$.

Considering the galaxy population broadly, this scenario requires a
sufficiently large bulge component for AGN quenching to be effective.
While bulges may be built stochastically in galaxy mergers, a further
link may tie secular bulge formation via disk instabilities to the
value of $M_*/M_h$, thereby cementing the relationship between
quenching and $\eta_{\rm crit}$.  It has been shown that disks become
unstable to bar modes if the disk mass dominates the gravitational
potential \citep[][]{Efstathiou:1982,Mo:1998}. Semi analytic models
typically consider that disk instabilities occurs if:

\begin{equation}
V_{\rm max}/(G_N M_{\rm disk}/r_{\rm disk})^{0.5} \leqslant 1,
\label{disk_instability}
\end{equation}

\noindent where $M_{\rm disk}$ represents disk mass, $r_{\rm disk}$ is
the disk radius, and $V_{\rm max}$ is the maximum of the rotation
curve. Depending on the implementation of this criterion, $V_{\rm
  max}$ may be equated either to the halo virial velocity, or the disk
velocity at its half mass radius \citep[see discussion in
][]{Parry:2009}. An instability will cause either a partial or a total
collapse of the disk, leading to a burst of star formation at the
center, the formation of a spheroid, and also possibly fueling the
central black hole. The disk instability criterion in Equation
\ref{disk_instability} shows a dependence on $(M_h/M_{\rm disk})$
(relating $V_{\rm max}$ to $M_h$) which could be reflected in
$M_h/M_{*}$ as the instability converts cold gas into stars.  Disk
instabilities might therefore play a role in setting the pivot masses
and enforcing $\eta_{\rm crit}$.  The coincident fueling of the
central black hole during the instability may help initiate AGN
quenching and regulate the global decline in $M_*^{\rm cen}/M_h$
beyond the pivot mass.

Assuming the disk instability framework, we can test whether the
predicted sizes of stable galactic disks given our critical pivot
ratio, $\eta_{\rm crit} \equiv M_*/M_h \approx 0.04$, are consistent
with observations. Approximating the halo virial velocity for $V_{\rm
  max}$, we rewrite Equation \ref{disk_instability} to derive the
maximum size of stable disks:

\begin{equation}
r_{\rm disk} < R_h \times (M_*/M_h).
\label{disk_instability2}
\end{equation}

Applying this criteria to the pivot masses yields the condition that
$r_{\rm disk}<8$ kpc. Interestingly, this condition is very well satisfied
by the observed size distributions of disk galaxies which tend to fall
off rapidly just below $r_{\rm disk}=8$ kpc at both $z\sim 0$
\citep[see Figure 11 in][]{Shen:2003} and at $z>0$ \citep[see Figure
10 in][]{Sargent:2007}. We conclude that, via the growth of bulges and
initialization of AGN feedback, disk instabilities provide a promising
link between our observed critical pivot ratio and quenching, although
further investigation is clearly needed.

Finally, we note that $\eta_{\rm crit}$ might also be related to the
competition between the cooling and accretion of cold gas in central
star-bursts and the resulting feedback from either star formation
itself or a co-evolving quasar-mode AGN.  This Eddington-like limit
has been explored in the context of stellar systems by
\citet[][]{Hopkins:2010a} who derive a maximum stellar surface density
from simple arguments.  In the context of dark matter halos explored
here, similar arguments might naturally yield a fixed value for
$\eta_{\rm crit}$ similar to that obtained by our analysis.

%%%%%%%%%%%%%%%%%%%%%%%%%%%%%%%%%%%%%%%%%%%%%%%%%%%%%%%%%%%%%%%%%%%%%%%%%%%%%%
%     CONCLUSION
%%%%%%%%%%%%%%%%%%%%%%%%%%%%%%%%%%%%%%%%%%%%%%%%%%%%%%%%%%%%%%%%%%%%%%%%%%%%%%

\section{Conclusions}\label{conclusions}

The aim of this paper is to study the form and evolution of the
stellar-to-halo mass relation (SHMR) from $z=0.2$ to $z=1.0$. To
achieve this goal, we have performed a joint analysis of galaxy-galaxy
lensing, spatial clustering, and number densities of galaxies in
COSMOS. As a result, we have obtained a robust characterization of the
evolving relationship between stellar mass and halo mass over two to
three orders of magnitude in $M_*$.  The nature of this relationship
is not only a byproduct of cosmic mass assembly but is also shaped by
the physical processes that drive galaxy formation, ultimately
providing valuable constraints on both. A complete and comprehensive
interpretation of our results requires modelling and accounting for dark
matter accretion histories, galaxy merger rates, and star formation
rates as a function of redshift. This will be the focus of future
work. Nonetheless, we show how simple evolutionary models can already
provide a first interpretation of our results, setting the stage for a
more detailed treatment in future work. Using simple arguments, we
evaluate our results in the context of other observations and
theoretical work on galaxy formation models.

We have defined the pivot quantities ($M_{*}^{\rm piv}$, $M_{h}^{\rm
  piv}$, and $(M_h/M_*)^{\rm piv}$) as the location at which halos
reach their maximum accumulated stellar mass efficiency. The evolution
of the pivot quantities contain key clues about the physical processes
that are responsible for the quenching of star formation in halos
above $M_h>10^{12}$\msun. While we observe downsizing trends for both
$M_{*}^{\rm piv}$ and $M_{h}^{\rm piv}$, the co-evolution of these two
parameters leaves $(M_h/M_*)^{\rm piv}$ roughly constant with
redshift. We argue that this result raises the intriguing possibility
that the quenching of star formation may have a physical dependence on
$M_h/M_*$ and not simply on $M_h$ as is commonly assumed. If the
quenching of star formation indeed depends on a critical $M_*/M_h$
ratio then the fact that low mass galaxies grow more rapidly than dark
matter below the pivot scale provides a {\em simple explanation for
  observations of downsizing in the sites of star
  formation}. Additional and more precise measurements of the pivot
quantities would be highly interesting in order to confirm whether or
not the pivot ratio remains constant, or if instead it evolves mildly
with redshift. Interestingly, there are hints from the results of
\citet{Behroozi:2010} that the pivot mass might in fact remain
constant back to $z=4$.

We highlight four avenues of exploration that would be interesting to
pursue in the future and that would improve this analysis. Firstly, we
note that the comparison of our results with SDSS abundance matching
results \citep[e.g,][]{Behroozi:2010} is limited by systematic
differences between stellar mass estimates. In order to use the SDSS
data-set as a low-$z$ anchor point, a homogeneous analysis of SDSS and
COSMOS would be necessary. Secondly, our work in COSMOS is limited at
the high mass end by sample variance. Larger data-sets at higher
redshifts than SDSS such as the Baryon Oscillation Spectroscopic
Survey
(BOSS)\footnotemark[9]\footnotetext[9]{http://cosmology.lbl.gov/BOSS/}
and the Canada-France-Hawaii Telescope Legacy Survey
(CFHTLS)\footnotemark[10]\footnotetext[10]{http://www.cfht.hawaii.edu/Science/CFHLS/}
should provide interesting constraints on the evolution of the high
mass end of the SHMR. Thirdly, pushing the SHMR down to $10^8~{\rm
  M}_{\odot}<M_*<10^{9}~{\rm M}_{\odot}$ using techniques such as
described in this paper is clearly an exciting avenue to explore and
will be facilitated by upcoming very deep optical and IR surveys such
at UltraVista and the Hyper Suprime Cam (HSC) survey on the Subaru
telescope. Finally, we note that the stellar mass function is a
powerful tool for placing constraints on the SHMR. However, the
derivation of SMFs is clearly currently limited by systematic
uncertainties in stellar mass estimates. Working towards an improved
understanding of stellar mass estimates and towards reducing
systematic errors in the SMF will be the single most important avenue
for improving the type of analysis presented in this paper.

Finally, while our analysis demonstrates that the combination of
multiple and complementary dark matter probes is a powerful tool with
which to elucidate the galaxy-dark matter connection, we emphasize
that such probe combinations also hold great potential to constrain
fundamental physics, including the cosmological model and the nature
of gravity. Exploring the sensitivity of the combination of g-g
lensing, clustering, and the SMF to cosmological parameters will be
the focus of a follow-up paper.

%-----------------------------------------------------------------------------
%    ACKNOWLEDGMENTS
%-----------------------------------------------------------------------------

\acknowledgments
\noindent {\bf Acknowledgments}

We thank Uros Seljak, Beth Reid, Martin White, Surhud More, Rachel
Mandelbaum, Marcello Cacciato, Phil Hopkins, Charlie Conroy, Ian
McCarthy, and Frank van den Bosch for insightful discussions. We are
grateful to Rachel Mandelbaum, Benjamin Moster, Michael Blanton,
Surhud More, Henk Hoekstra, and Chris Bildfell for providing data in
electronic format. AL acknowledges support from the Chamberlain
Fellowship at LBNL and from the Berkeley Center for Cosmological
Physics. JPK acknowledges CNRS and CNES for support. JDR was supported
by JPL, operated under a contract by Caltech for NASA. This research
received partial support from the U.S. Department of Energy under
contract number DE-AC02-76SF00515.  RHW and PSB received additional
support from NASA Program HST-AR-12159.A, provided through a grant
from the Space Telescope Science Institute, which is operated by the
Association of Universities for Research in Astronomy, Incorporated,
under NASA contract NAS5-26555. MTB and RHW also thank their
collaborators on the LasDamas project for critical input on the
Consuelo simulation, which was performed on the Orange cluster at
SLAC. TS acknowledges support from the Netherlands Organization for
Scientific Research (NWO), NSF through grant AST-0444059-001, and the
Smithsonian Astrophysics Observatory through grant GO0-11147A. The HST
COSMOS Treasury program was supported through NASA grant
HST-GO-09822. We wish to thank Tony Roman, Denise Taylor, and David
Soderblom for their assistance in planning and scheduling of the
extensive COSMOS observations.  We gratefully acknowledge the
contributions of the entire COSMOS collaboration consisting of more
than 70 scientists.  More information on the COSMOS survey is
available at {\bf \url{http://cosmos.astro.caltech.edu/}}. It is a
pleasure the acknowledge the excellent services provided by the NASA
IPAC/IRSA staff (Anastasia Laity, Anastasia Alexov, Bruce Berriman and
John Good) in providing online archive and server capabilities for the
COSMOS data-sets.

\bibliographystyle{apj}
%\bibliography{all_refs}

 %-----------------------------------------------------------------------------

\end{document}